\documentstyle[12pt,epsf]{article}

\newcommand{\eq}{\begin{equation}}
\newcommand{\en}{\end{equation}}
\newcommand{\bd}{\begin{displaymath}}
\newcommand{\ed}{\end{displaymath}}
\newcommand{\minimo}[2]{\mbox{\emph{min}$_{_{_{\hspace*{-11pt}#1\hspace*{6pt}}}}\{#2\}$}}

\newcommand{\Integer}{\mbox{$\bf \cal{Z}$}}

\newcommand{\ignorar}[1]{}

\title{{\bf 
Scene dependence of the non-gaussian scaling properties of natural images
} }

\author{
{\bf\sc \'Angel Nevado}\thanks{E-mail: anevado@delta.ft.uam.es},
{\bf\sc Antonio Turiel}\thanks{E-mail: amturiel@delta.ft.uam.es} 
and {\bf\sc N{\'e}stor Parga}\thanks{To whom correspondence should be 
addressed. E-mail: parga@delta.ft.uam.es}
\\
\it Departamento de F{\'{\i}}sica Te{\'o}rica \\
\it Universidad Aut{\'o}noma de Madrid \\
\it Canto Blanco, 28049 Madrid, Spain
}

\date{}

\begin{document}

\maketitle

\begin{abstract}

We report results on the scaling properties of changes in contrast of
natural images in different visual environments.  This study confirms
the existence, in a vast class of images, of a multiplicative process
relating the variations in contrast seen at two different scales, as
was found in \cite{NIPS97,TuMaPaNa98}. But it also shows that the scaling
exponents are not universal: Even if most images follow the same type
of statistics, they do it with different values of the distribution
parameters. Motivated by these results, we also present the analysis
of a generative model of images that reproduces those properties and
that has the correct power spectrum.  Possible implications for visual
processing are also discussed.

\end{abstract}

{\em NETWORK} {\bf 11} 131-152 (2000)

\pagebreak

\section{Introduction}
\label{section:introduction}

As it has been suggested a long time ago \cite{Ba61}, the early stages of
the visual system must have evolved by adaptation to the statistics of
the external stimuli. During this process the neurons in
the visual pathway have developed their receptive fields in such a way
that information about visual scenes is represented internally in an
efficient way. The large amount of redundancy present in the external
world is, at least in part, eliminated from the internal sensory code.

In order to achieve this quasi-optimal representation the visual system must
have learnt the regularities present in the environment where the
organism lived. If a given conjunction of some elementary features
tend to appear together, a cell responding optimally to the
combination of features is rather likely to exist.

Carrying out this program requires, as a first step, to perform an 
analysis of the statistical properties of the environment. In the case
of visual scenes, a systematic study of this matter has began rather
recently
\cite{RuBi94,Ru94,NIPS97,TuMaPaNa98,Singularities,TuPaRuCr99}. 
Although the relevance of the second order statistics has been pointed
out some time ago \cite{BuMo87,Fi87}, and a gaussian distribution for
images has very often been used \cite{Li88,At92,vHa92,LiAt94} to make
predictions on the receptive fields of cells in V1 and in previous
stages of the visual pathway, there are many reasons to believe that
this statistics leaves aside a vast number of qualitatively important
properties. An indication of this is that once the image is
decorrelated (i.e. the correlations between pairs of pixels are
eliminated) the scene can still be recognized
\cite{Fi93}, something that is mainly due to the fact that the borders
of the objects are still present.

As it was emphasized in \cite{Ru94,TuMaPaNa98,Singularities}, a better
understanding of the statistics of images should be achieved before
making predictions on the visual system. In fact, an analysis of the
properties of changes in contrast in natural scenes revealed the
existence of multiscaling properties: images do not have uniform scale
properties, but they can be decomposed in sets of pixels such that
only those pixels in a given set have similar scale properties. 
\footnote{Image structure in scale-space has been considered by
several authors, although from a perspective different from ours 
\cite{Li94,MaZh91}.}
Interestingly enough, these properties can be explained
\cite{NIPS97,TuMaPaNa98} by means of a simple model, that obtains the
statistics of changes in contrast at a scale $r$ in terms of an
independent multiplicative process applied to the changes occurring a
a larger scale $L$.

The multiplicative process is a log-Poisson distribution.  The events
it generates represent sharp changes, or modulations, of the contrast
gradient. It contains two parameters: the average number of events per
unit of scale, $\bar s$, and the strength of each of these elementary
changes, $\beta$.  For the ensembles considered in
\cite{TuMaPaNa98,Singularities} these parameters took the values  
$\bar s \approx 1.$ and $\beta \approx 0.5$. \footnote{Similar
processes also occur in the physics of turbulent flows
\cite{ShLe94,ShWy95,CaGaHo90,Ca96}.}

The set of images used for those studies was rather uniform, they were
generally forest scenes, and their contrast distribution was also
similar from image to image. This fact made the statistical analysis
very stable, and the existence of the multiplicative relation between
different scales was clear. It left however open the question of how
much dependent those properties are on the chosen image ensemble. Their
possible relevance for the development of the early visual system
would be greatly diminished if they only held in a restricted set
of images. On the other hand, if the same statistical properties of
changes in contrast could be found in a large set of visual scenes,
their importance would be more clear.

In this paper we address the issue of robustness of the properties of
edges and softer textures analyzed in \cite{NIPS97,TuMaPaNa98}. In order
to establish their validity we use a larger data set (more than a
hundred times larger than the one used in our first analysis) 
and more heterogeneous images, containing many
different environments. 

Our result is both surprising and encouraging. The statistical
analysis can be performed on an image by image basis. When this is
done, one finds that most of them (375 out of 400) exhibit the scale
properties found in the smaller data set. Besides, the same log-Poisson
model can be applied to explain the data. That is, for most images,
contrast changes at different scales are related by the same type of
multiplicative process. Nevertheless, the model parameter $\beta$ depends
on the image.

We have also carried out a similar statistical analysis for ensembles
of images, where each ensemble was characterized by the type of
scenes. Again, each of the ensembles analyzed presents the same
multiscaling properties, although with different values of the
parameter $\beta$. Finally, we have done the study over the whole set
of images, and have again found that the same model is able to explain
the observed quantities.

Given these results, our conjecture is that the visual system has
adapted, during a long period of time, to the existence of these
multiscale properties and, in particular, that its architecture has
captured the existence of the multiplicative process.  On the other
hand, since different visual scenes differ in the value of the
parameter $\beta$, the adaptation to this should occur more rapidly,
and could be implemented, for instance, by gain control mechanisms.  

Stochastic generative models of images can be useful to understand the
role played by the statistical properties of images in the structure
of receptive fields. These models become very simple if it is assumed
that only the second order statistics is relevant; in this case the
contrast distribution is just a gaussian with the correct correlation
function between pairs of pixels. But the definition of a model able
to describe correctly the statistics of changes in contrast requires a
different approach.  Here we have studied a model that reproduces the
properties exhibited by the natural images analyzed in the first part
of this work. As we will see, it also possesses the correct power
spectrum.  This generative model will also allow us to see more
clearly the role of the log-Poisson parameters on the geometrical
structure of the images.

The layout of this paper is as follows. The next section contains a
brief theoretical background about the multiplicative log-Poisson
process and multifractality of images. The image data sets used in
this work are described in Sec.
\ref{section:environments}, as well as the criteria that have been
used to select them. The results about the statistics of changes in
contrast and the existence of the multiplicative process in the
various data sets are presented in Sec. 
\ref{section:results}.  Sec. \ref{section:generative_model}
is devoted to the presentation and analysis of a generative model of
images that follows the non-gaussian statistical properties discussed
in the previous sections.  The last section is devoted to discussions,
where the relevance of this work for the visual system is briefly
discussed.


\section{Multiscaling properties of images and the multiplicative process}
\label{section:theoretical_background}

In this section we give the basic theoretical background necessary for
our numerical analysis. For further details the reader should consult 
the original work \cite{NIPS97,TuMaPaNa98,Singularities,TuPaRuCr99}

\subsection{The local linear edge variance}
\label{subsection:the_measure}

The contrast $C(\vec{x})$ is defined as $C(\vec{x})=I(\vec{x})-\langle
I \rangle$, where $I(\vec{x})$ is the field of luminosities and
$\langle I \rangle$ its average value across the image ensemble.
Since we are interested in quantifying changes of the contrast $C({\bf
x})$ at the point $\bf x$ and at a given scale $r$, it is natural to
consider the following quantities \cite{NIPS97,TuMaPaNa98}:

\begin{equation}
\epsilon_{h,r}({\bf x}) = \frac{1}{r} \int_{x_1}^{x_1+r}
\left. \left( \frac{\partial C({\bf x}')}{\partial x'_1}\right)^2 \right|_{{\bf x}'=\{x'_1,x_2\}} dx'_1
\label{eq:edge-horiz}
\end{equation}

\noindent
and

\begin{equation}
\epsilon_{v,r}({\bf x}) = \frac{1}{r} \int_{x_2}^{x_2+r}
\left. \left( \frac{\partial c({\bf x}')}{\partial x'_2}\right)^2 \right|_{{\bf x}'=\{x_1,x'_2\}} dx'_2
\label{eq:edge-vert}
\end{equation}

\noindent
The variables $\epsilon_{h,r}({\bf x})$ and $\epsilon_{v,r}({\bf x})$
are defined at the position ${\bf x}$ and at a scale $r$. 
While the first takes contributions from edges
transverse to a horizontal segment of size $r$, the second 
accumulates changes in contrast transverse
to a vertical segment of the same size.

These variables have been interpreted as the {\em local linear edge
variance} along the corresponding direction and at the scale $r$.
They describe how far the changes in contrast are from being uniformly
distributed in a segment, either horizontal or vertical, of size $r$.
If these changes were constant, the estimators would be independent of
$r$ and, on the contrary, a dependence on $r$ indicates
the existence of geometrical structure. 
Although in this work we will
discriminate between the vertical and horizontal directions, we
mention that the behaviour of contrast changes at a scale $r$ can also
be studied by means of a bi-dimensional integral \cite{Singularities}:

\eq
\epsilon_r({\bf x})  \:=\: 
 \frac{1}{r^2} 
\int_{x_1 - \frac{r}{2}}^{x_1 + \frac{r}{2}} dx_1'
\int_{x_2 - \frac{r}{2}}^{x_2 + \frac{r}{2}} dx_2'  \;\;  
\mid \nabla C({\bf x'}) \mid   \;\;   .
\label{eq:square}
\en

\noindent
Similarly to the other two variables, $\epsilon_r(\vec{x})$ quantifies
the deviation from uniform of the distribution of $|\nabla C|$.

\subsection{The multiplicative process}
\label{subsection:multiplicative_process}

\indent
The question now arises of how the statistics of $\epsilon_{j,l}({\bf
x_0})$ (where $j=h,v$) at the scale $l$ is related to the distribution of
the variables at a larger scale $L$. This was discussed in
\cite{NIPS97,TuMaPaNa98,Singularities} for the marginal distribution of
$\epsilon_{j,r}$. The answer is simple:  they are related by a
multiplicative process. Denoting by $\alpha_{lL}$ its associated
stochastic variable, this means that

\eq
\epsilon_{j,l} \; \doteq \; \alpha_{lL} \: \epsilon_{j,L}
\label{eq:alpha} 
\en

\noindent
The random variable $\alpha_{lL}$ is independent of $\epsilon_L$. The
symbol $\doteq$ indicates that the equality holds in the
distributional sense, that is, both sides of the equality have the
same distribution.  This relation implies a linear relation between
the logarithms of the variables at two different scales. This property
has also been discussed in \cite{BuSi99}, although the existence of a
multiplicative process was not noticed.

\indent
The process for arbitrary changes in scale was derived in
\cite{TuMaPaNa98} where it was shown that the random variable
$\alpha_{lL}$ follows a log-Poisson process \cite{ShWy95}.  This
process can be justified as follows.  As the scale is gradually
reduced from $L$ to $l$ some changes in contrast are left outside the
segment of size $L$. But some of these changes are special in that
they give rise to a singular behaviour of the derivatives of the
contrast, and in turn to a discontinuity in the $\epsilon$'s, which
acquire a factor $\beta$ each time that this occurs ($0 \leq \beta <
1$). We will refer to this effect by saying that a modulation has
occurred.

\noindent
It will be assumed that:

\begin{itemize}

\item
modulations are independent events. The average number of them
contained in the change of scale $(\ln L - \ln l)$) is denoted by $s$,

\item
the image ensemble is translational invariant, 

\item
the multiplicative process is scale invariant. 

\end{itemize}

Under a finite change from $L$ to $l$, there will be $n$ 
modulations with probability $p_n$ that, given the first 
assumption, is Poisson:

\eq
p_n = \frac{s^n}{n!}{e^{-s}} 
\label{eq:Poisson}
\en

\noindent
The value of $\alpha_{lL}$ after $n$ modulations is proportional to
$\beta^n$: $\alpha_{lL}=\beta^n M(l,L)$.  The fact that the
proportionality constant is not one can be understood by noticing that
if no modulations occur (i.e., $n=0$), $\epsilon_{j,l}$ keeps as much
singular structure in the interval $l$ as that present in
$\epsilon_{j,L}$, which is distributed in the larger interval $L$.

The dependence of $M(l,L)$ with the scales $l$ and $L$ can be obtained by
invoking a scale invariant multiplicative process. In this case, it
can only depend on the ratio $l/L$, that is $M(l,L)=M(\frac{l}{L})$.
A standard argument shows that it has to be a power law: if the change
from $L$ to $l$ is done through an intermediate scale $r$, then
because of the multiplicative character of the process we have
$M(\frac{l}{L}) \; =\; M(\frac{l}{r}) M(\frac{r}{L})$, which can
only be satisfied if $M(\frac{l}{L}) \; =\;
\left(\frac{l}{L} \right)^{- \Delta}$. The exponent $\Delta$ is 
another parameter of the model.

\indent
Taking all these arguments into account, the distribution of
$\ln \alpha_{lL}$ can be expressed as:

\eq
\rho_{\alpha_{lL}}(\ln \alpha_{lL}) \; =\; e^{-s} \: 
\sum_{n=0}^{\infty}\frac{s^n}{n!} \: \delta \left( \ln \alpha_{lL} 
- n \ln \beta \: - \Delta \ln \left( \frac{L}{l} \right) \right)  \;\; .
\label{eq:Log-Poisson}
\en

\indent
Up to now we have considered $s$, $\beta$ and $\Delta$ as independent
parameters. However translational invariance fixes one of them. In
fact, from the definition of the $\epsilon_{j,r}$'s one easily checks
that its average over a translational invariant ensemble of images
does not depend on $r$. In turn, taking the average on both sides of
eq. (\ref{eq:alpha}), this implies that

\eq
\langle \alpha_{rL} \rangle \; =\; 1 \;\; , 
\label{eq:average_alpha}
\en

\noindent
where $\langle \ldots \rangle$ indicates the average over an ensemble
of images. But then, imposing this condition over the value of this
average obtained from the distribution in eq. (\ref{eq:Log-Poisson})
one has:

\eq
s \; =\; \frac{\Delta}{1 - \beta}  \ln \frac{L}{r}   \;\; , 
\label{eq:s}
\en

\noindent
from where the average number of modulations per unit of change in 
scale is $\bar s = \frac{\Delta}{1 - \beta}$.

\indent
The existence of a multiplicative process has direct consequences on
the scaling properties of the moments of $\epsilon_{j,r}$.  Let us
denote these moments by $\langle \epsilon_{j,r}^p \rangle$.  If the
log-Poisson model holds, then from eq. (\ref{eq:Log-Poisson}) it is
easy to check that the moments have a property called 
Self-Similarity (SS),  

\eq
\langle \epsilon_{j,r}^p \rangle \; =\; A_p\: r^{\tau_p} \;\; ,
\label{eq:SS}
\en

\noindent
where the $\tau_p$'s are the SS exponents. 
Notice that this also implies that any
moment can be expressed as a power of the moment of order $q$. 
Choosing $q=2$ this means that 

\eq
\langle \epsilon_{j,r}^p \rangle \; =\; A(p,2)
\left[ \langle \epsilon_r^2 
\rangle\right]^{\rho(p,2)}\;\;,
\label{eq:ESS}
\en

\noindent
This relation could hold even when SS is not true. It is called
Extended Self-Similarity (ESS) \cite{ESS}. The $\rho(p,2)$'s are the
ESS exponents and the $A(p,2)$'s are geometrical factors.  The 
exponents $\rho(p,2)$ can be predicted using the distribution of the
multiplicative process, eq. (\ref{eq:Log-Poisson}), to evaluate the
moments of order $p$ in eq. (\ref{eq:alpha}). This computation yields

\eq
\rho(p,2) \; =\; \frac{p}{1-\beta}\: -\: \frac{1-\beta^p}{(1-\beta)^2} \;\; .
\label{eq:rho_S-L}
\en

\noindent
Let us notice that although the model has two parameters, $\beta$ and
$\bar s$, the ESS exponents $\rho(p,2)$ depend only on the modulation
parameter $\beta$.  There is a simple relation between $\tau_p$ and
$\rho(p,2)$:

\eq
\tau_p\; =\; - \bar s  (1-\beta)^{2}  \: \rho(p,2)  \;\; ,
\label{eq:rel_SS-ESS}
\en

\noindent
Conversely, it can be proven that if SS and ESS hold and the exponents
$\rho(p,2)$ verify eq.~(\ref{eq:rho_S-L}), then $\epsilon_r$ can be
described in terms of a multiplicative process (eq.~(\ref{eq:alpha}))
of the log-Poisson type (eq.~(\ref{eq:Log-Poisson})). It is then
enough to check eq.~(\ref{eq:rho_S-L}), from where the existence of a
log-Poisson process is derived.

\subsection{Multifractality of images}
\label{subsection:multifractality}

\noindent
The process just described has an interesting geometrical
interpretation, as has been discussed in \cite{Singularities}. In
fact, the power law behaviour of the moments $\langle \epsilon_{j,r}^p
\rangle$, given in eq.~(\ref{eq:SS}), can be traced back to the existence of a
very irregular behaviour of $\epsilon_r(\vec{x})$, which can be
expressed as:

\eq
\epsilon_r(\vec{x})\: =\: \alpha(\vec{x})\; r^{h(\vec{x})}\;\; ,
\label{eq:exponent_epsilon}
\en

\noindent
where the exponent $h(\vec{x})$ depends on the site of the image.
This property is, in turn, related to a singular behaviour of $|\nabla
C|(\vec{x})$: $h<0$ indicates a divergence of $|\nabla C|(\vec{x})$ to
infinity, while $h>0$ indicates finiteness and continuous
behavior. The greater the exponent the smoother is the the change in
contrast around that point.  Eq.~(\ref{eq:exponent_epsilon}) expresses
that all the points are singular (in this wide sense), and that the
singularity exponent is not uniform. This property allows to classify
the pixels in a given image: the set of points with a singularity
exponent contained in the interval $[h -\Delta_h, h + \Delta_h]$
(where $\Delta_h$ is a small positive number) define a class $F_h$.
These classes are the
\emph{fractal components} of the image 
(the notion of fractality is discussed in \cite{Ma77,Fa90}).  The
smaller the exponent the more singular is the class, and the most
singular component is the one with the smallest value of $h$.  A
mathematical object with this structure is said to be a
\emph{multifractal}, a concept originally introduced in the context
of fully developed turbulence
\cite{PaFr85}.

\indent 
The irregular arrangement of pixels in a fractal component $F_h$ can be
characterized by counting the number of pixels contained inside a
given ball of radius $r$, $N_r(h,\Delta_h)$. As $r \rightarrow 0$ it
is verified that

\eq
N_r(h, \Delta_h) \:\sim\: r^{D(h)}  \;\;.
\label{eq:component_num}
\en

\noindent
This exponent $D(h)$ quantifies the size of the set of pixels with
singularity $h$ as the image is covered with small balls of radius
$r$. It is the fractal dimension of the associated fractal component
$F_h$, and the function $D(h)$ is called the \emph{singularity
spectrum} of the multifractal \cite{PaFr85}.

\indent
There is an important connection between the local singularity
analysis and the statistical description of the image in terms of the
moments $\langle \epsilon_{j,r}^p \rangle$. In fact, the ESS exponents
$\tau_p$ are the Legendre transform of the singularity spectrum $D(h)$
(\cite{PaFr85}, see also \cite{Singularities}),

\eq
D(h)\; =\; \minimo{p}{ph +d -\tau_p} \;\; , 
\label{eq:sing_spectrum}
\en

\noindent
where $d=2$ is the dimensionality of the images. Given the $\tau_p$'s,
eqs.~(\ref{eq:rel_SS-ESS}) and (\ref{eq:rho_S-L}), the dimension
spectrum can be predicted:

\eq
D(h)\; = \; D_{\infty} - \frac{h+\Delta}{\ln\beta}
\left[1-\ln\left( -\frac{h+\Delta}{(d-D_{\infty})\ln\beta}\right) \right]\;\; , 
\label{eq:Dh_S-L}
\en

\noindent
where $D_{\infty}$ is defined as $D_{\infty} = d - \bar s$. From here
one notices that there is a minimum value of the singularity exponent
$h$, given by $h = -\Delta$ which then is the singularity of the most
singular fractal component. Its fractal dimension can be computed from
eq. (\ref{eq:Dh_S-L}) which yields $D(-\Delta)=D_{\infty}$.  Then
$D_{\infty}$ is the dimension of the most singular set; it can be
expressed in terms of the parameters of the model, e.g. $\beta$ and
$\Delta$, as

\eq
 d - D_{\infty} = \bar s = \frac{\Delta}{1 - \beta} \;\; .
\label{eq:beta_vs_delta}
\en

\indent
The points with the strongest singularity ($h = -\Delta$) are those with
the sharpest changes in contrast (edges). This can be explicitly verified by
evaluating the singularity exponents of all the pixels in the image
and extracting those with the minimum value of $h$ (within a
resolution $\Delta_h$) \cite{Singularities}.  Let us remark that the
most singular points could have a rather weak singularity, but even in
these cases they are the points where the most important changes in
contrast occur. This happens, in particular, when there are no
modulations ($\bar s = 0$ or, equivalently, $D_{\infty} = 2$), in which
case $\Delta = 0$.

\indent
The parameter $\beta$ was defined as the strengh of a modulation in
the multiplicative process, but it can also be interpreted in terms of
the geometrical properties of the edges in the image.  In fact,
eq. (\ref{eq:beta_vs_delta}) relates $\beta$ with $D_{\infty}$ and
$\Delta$, which describe properties of the set of pixels where the
changes in contrast are most singular: $\beta \; = \; \frac{-\Delta}{2
- D_{\infty}} + 1$.  For fixed $D_{\infty}$, as $\beta$ decreases, the
singularity exponent of the edges, $-\Delta$, decreases.  In order to
gain some intuition about this relation, let us consider an isolated
jump in contrast. Its dimensionality is $D_{\infty}=1$, and it can be
proved \footnote{see, e.g., \cite{Singularities} } that $\Delta = 1$,
what gives $\beta = 0$. By contrast, images with statistics dominated
by smooth changes of contrast have $\beta = 1$. This is so because
$\Delta = 0$, but this result can also be understood by noticing that
the way to obtain smooth images with the log-Poisson process is to set
$\beta = 1$.

\indent
The measurable quantities are the moments $\langle \epsilon_{j,r}^p
\rangle$, and from them one can obtain the SS exponents $\tau_p$ and
the ESS exponents $\rho(p,2)$.  Given the experimental ESS exponents
$\rho(p,2)$, it is easy to obtain the value of $\beta$ by performing a
least squares regression on eq.~(\ref{eq:rho_S-L}). To define the
log-Poisson process completely one should still estimate $D_{\infty} =
2 - \bar s$, which can be done for instance using the measured value of
$\tau_2$. This can be seen from eq.~(\ref{eq:rel_SS-ESS}) which after  
setting $p=2$ yields

\eq
D_{\infty} = 2 - \frac{\tau_2}{(1-\beta)^2}  \;\;\; .
\label{eq:D_vs_beta}
\en

\noindent
Let us remark that this expression contains a dependence in
$1/(1-\beta)^2$ which makes the propagation of errors important as
$\beta$ approaches the value one, even if the uncertainty in the
values of $\beta$ and $\tau_2$ are small.


\section{Choice of environments}
\label{section:environments}

In the present study we used a data set of $200$ images, having $1536
\times 1024$ pixels and 16 bits in luminance depth. These images have been
kindly provided to us by Hans van Hateren \cite{vHavSc98} and were
selected from his data set of about $4200$ images.  The selected
images can be observed and downloaded from the URL address
http://hlab.phys.rug.nl/archive.html.  A list of them is provided in
Table \ref{tab:selected_images}. The files are originally named as 
``IMK\#.IMC''. In the table only the number (\#) of the corresponding
image is specified.

To compare with previous studies we notice that the images used in 
\cite{TuMaPaNa98} contained $45$ natural scenes with a size of $256 \times 256$
pixels and a luminance depth of 13 bits.  This means that our present
statistical database has 130 times as many bits as the previous one.

The data set has been considered either complete or divided into four
different subsets with $50$ images in each of them.  These four ensembles
were selected to meet the following requirements:

\indent 
\begin{itemize}

\item  Ensemble A:

It contains natural scenes of trees and woods. The images do not have
artificial objects or open skies. The woods are dense and neither
shadows nor direct light rays are allowed.

\item  Ensemble B:

It contains natural scenes. The rest of conditions imposed in the 
definition of ensemble A have not been required here.

\item  Ensemble C:

Images containing both artificial and natural objects are included. We
chose images that in a visual scan appear to have predominantly
horizontal structures, i.e, the horizontal edges seem to be longer
than the vertical ones. We are interested in testing whether this
implies a difference between the horizontal and vertical statistics.

\item  Ensemble D:

Out of the whole database of $4212$ images in \cite{vHavSc98} we
picked up $50$ randomly. No image belonging to any of the other three
ensembles was included here. The intention is to have a data set
as varied as possible.

\end{itemize}

Each ensemble is used to study the vertical and the horizontal
statistics (that is the marginal statistics of $\epsilon_{h,r}({\bf
x})$ and $\epsilon_{v,r}({\bf x})$, respectively), what gives a total
of eight different elements.

\begin{table}[hbt]
\begin{center}
\begin{tabular}{|c|c|}
\hline
\bf{Ensemble} & $\bf{Image~number}$ \\ \hline
   & 
 0034, 0211, 0263, 0478, 0586, 0605, 0662, 0683, 0801, 0808 \\ \cline{2-2}
   &
 0881, 1017, 1031, 1164, 1406, 2035, 2263, 2280,  2411, 2417\\ \cline{2-2}
A   &
 2603, 2605, 2606, 2626, 2649, 2935, 3002, 3134,  3491, 3514\\ \cline{2-2}
   &
 3536, 3789, 3807, 3830, 3842, 3940, 4010, 4031, 4037, 4042\\ \cline{2-2}
   &	
 4044, 4056, 4069, 4078, 4094, 4098, 4101, 4123, 4133, 4134\\ \hline
   
    &   
0037, 0071, 0222, 0265, 0511, 0758, 0774, 0807, 0836, 0877 \\ \cline{2-2}
    &
1206, 1341, 1375, 1814, 1831, 1836, 1852, 1867, 1889, 1926 \\ \cline{2-2}
B    &
1976, 1992, 2071, 2159, 2214, 2409, 2437, 2619, 2623, 2805 \\ \cline{2-2}
    &	
2806, 3236, 3269, 3318, 3414, 3415, 3611, 3615, 3635, 3913 \\ \cline{2-2}
    &
4028, 4066, 4087, 4089, 4107, 4121, 4122, 4139, 4184, 4193 \\ \hline

   &
0001, 0038, 0046, 0052, 0085, 0090, 0127, 0145, 0146, 0147 \\ \cline{2-2}
   &
0168, 0177, 0210, 0403, 0423, 0459, 1282, 1357, 1400, 1411 \\ \cline{2-2}
C   &
1423, 1427, 1431, 1432, 1471, 1489, 1529, 1562, 1566, 1600 \\ \cline{2-2}
   &
1878, 1879, 1934, 1946, 1950, 1964, 3029, 3054, 3065, 3082 \\ \cline{2-2}
   &
3099, 3137, 3154, 3179, 3223, 3225, 3286, 3322, 3332, 3384 \\ \hline

   &   
0288, 0325, 0342, 0446, 0609, 0668, 0685, 0695, 0850, 0879 \\ \cline{2-2}
   &
0914, 0920, 1092, 1171, 1191, 1201, 1249, 1322, 1414, 1422 \\ \cline{2-2}
D   &
1462, 1645, 1691, 1775, 1945, 2231, 2235, 2499, 2582, 2749 \\ \cline{2-2}
   &
2750, 2770, 3079, 3123, 3181, 3214, 3356, 3474, 3496, 3540 \\ \cline{2-2}
   &
3546, 3680, 3752, 3784, 3786, 3894, 3952, 3988, 4058, 4086 \\ \hline

\end{tabular}
\end{center}
\caption{The 200 selected images and its classification in the four ensembles A-D}
\label{tab:selected_images}
\end{table}

\section{Results}
\label{section:results}

The main purpose of this section is to establish whether the
statistical properties of SS and ESS, eqs.(\ref{eq:SS},
\ref{eq:ESS}), are present in the ensembles considered, and whether
the ESS exponents can be predicted by the log-Poisson model,
eq.(\ref{eq:Log-Poisson}).  This will be done taking three different
categories of ensembles. In Sec. \ref{subsection:results_ensembles}
we consider the four ensembles defined in
Sec. \ref{section:environments} for the horizontal and vertical
directions.  Next, in Sec.
\ref{subsection:results_single_images} we take as ensembles the individual
images themselves, again for both directions. Finally in Sec. 
\ref{subsection:results_whole} we regard the
$200$ images as a single ensemble and we accumulate the horizontal and
vertical statistics in the same distribution.  Notice that the moments
are self-averaging quantities but the ESS exponents, and therefore the
parameter $\beta$ of the probability distribution of
eq. (\ref{eq:Log-Poisson}) are not, so we do not expect the parameter
$\beta$ of an ensemble to be the average of the $\beta$'s of its
images.

\subsection{Methods}
\label{subsection:results_methods}

Here we describe the methods used to check the presence of SS, ESS
and the log-Poisson model for given $p$-moments of $\epsilon_r$. The
statistical moments will be computed over the three different
categories we are going to analyze, i.e., for each of the eight
orientational ensembles, for single images and for the whole
data set. The items of a category giving rise to the different moments
$\langle \epsilon_r^p\rangle$ will be referred to as elements. (Hence, we
have eight elements in the first category, 400 in the second and just
one in the third).

\indent
The existence of SS will be checked by verifying that $\ln
<\epsilon_{j,r}^{p}>$ is linear in $\ln r$ for each element (the symbol
$<.>$ denotes the ensemble average over the images belonging to that
element). In order to present the results for different elements in a
single graph and to evaluate the standard deviation of the data, we
have proceeded as follows. Since each element has its own scaling
exponents the curves have to be normalized; we first define,

\eq
S_{p}(x) = \ln <\epsilon_{j,r}^{p}>  \;\; ,
\en

\noindent
where $x = \ln r$. When $S_{p}(x)$ vs $x$ is fitted by a least-squares
regression it yields the fitting line \footnote{Notice that the
fitting parameter $b$ gives an estimate of the SS exponent $\tau_p$.
The linear regression is performed in the range $r = [8:25]$, 
where the SS property is best realized}:
$\hat{S}_p(x) = a + bx$. We now perform a linear transformation
$\hat{S_p} \rightarrow {\hat{S}_p}'$ such that ${\hat{S}_p}' = x$ for
each ensemble.  The same transformation is applied to $S_{p}(x)$ to
give

\eq
S_p(x) \rightarrow {S_p}'(x) = \frac{S_p(x)-a}{b}  \;\; .
\label{eq:S_p}
\en

\noindent
The average over different ${S_p}'$ of a given category can now be
carried out, yielding the quantities $\bar{S'_p}$. ESS is analyzed in a
similar way; now we define

\eq
E_{p}(y) = \ln <\epsilon_{j,r}^{p}>
\en

\noindent
where $y = \ln <\epsilon_{j,r}^{2}>$.  The fitting line for each
element is expressed as $\hat{E}_p(y) = c + d \; y$. However now not
only the fit parameters $c$ and $d$ depend on the element considered,
but also the variable $y$, for fixed $r$, varies from element to element.
Therefore, before transforming all the fitting lines onto a common
function, one has to normalize $y$ properly. This is done as:

\eq
y \rightarrow y' = \frac{y-y_{min}}{y_{max}-y_{min}} 
\label{eq:y}
\en

\noindent
where $y_{max}$ and $y_{min}$ are respectively the maximum and minimum
values of $y$ for a given element. This maps $y$ into the
interval [0:1]. As before we apply to $E_p$ the transformation that yields
$\hat{E}_p\rightarrow\hat{E}_p^{\prime}=y^{\prime}$:

\eq
{E}_p \rightarrow {E'}_p(y') = \frac{ E_p(y) -
(a+by_{min})}{b(y_{max}-y_{min})}  \;\; .
\label{eq:E_p}
\en

\indent
The experimental ESS exponents $\rho(p,2)$ for each element in the
category are fitted by a least-squares regression. An important result
is that each element follows the log-Poisson model although the optimal
value of $\beta$ varies from element to element. To represent the data
from all the elements together we define a normalized $\rho'(p,2)$ as:

\eq 
\rho'(p, 2) = \frac{\rho(p, 2)}{\rho^{\beta}(p, 2)}
\rho^{\beta_{0}}(p, 2) 
\label{eq:rho}
\en

\noindent
where $\rho(p, 2)$ denotes the computed ESS exponents 
\footnote{These exponents
are obtained with a linear regression of the ESS curves for each
moment. This regression is performed in the range $r =
[8:25]$, where also the ESS property is best realized.  The function
$\rho(p,2)$ is computed up to $p=7$. For higher order moments the properties
of SS and ESS begin to appear less clearly. The reason for this is
that the main contribution to these moments comes from the tail of the
distribution of $\epsilon_{r,l}$, where the sampling is worse.}
obtained from each element, $\rho^{\beta}$ denotes the exponents
predicted by the log-Poisson model (\ref{eq:rho_S-L}) using the value
of $\beta$ that best fits the data and $\rho^{\beta_{0}}(p, 2)$ are
the exponents predicted by the log-Poisson model with a certain
$\beta_{0}$ conventionally set equal to $0.5$.  (i.e., half of its
maximum value).

\subsection{Ensembles}
\label{subsection:results_ensembles}

\indent

\begin{figure}[hbt]
\hspace*{5cm}$\bf{a}$ \hspace*{6cm} $\bf{b}$ \\
$I)$ \\
\hspace*{0.5cm} $\bar{{S'}_2}(x)$ \hspace*{5.5cm}$\bar{{S'}_7}(x)$
\begin{center}
\leavevmode
\epsfysize=4cm
\epsfbox[50 61 388 291]{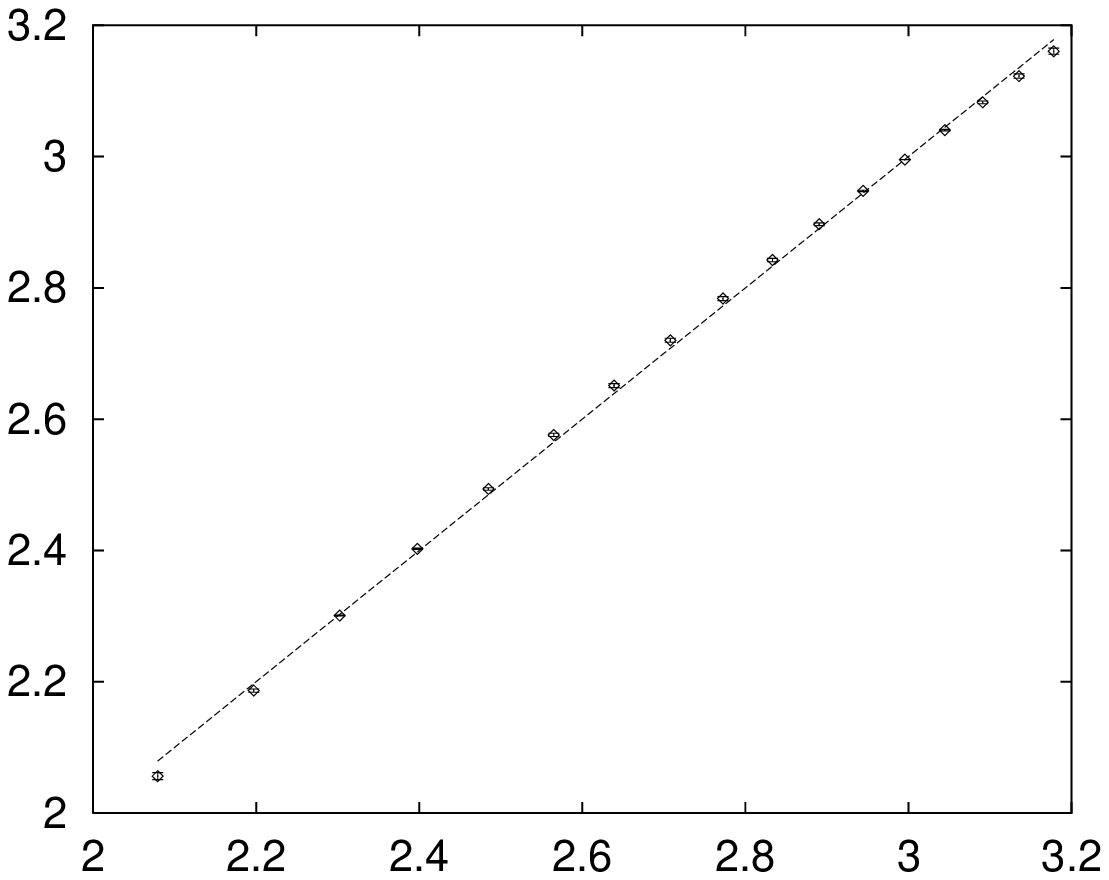}
\hspace*{1cm}
\epsfysize=4cm
\epsfbox[72 61 410 291]{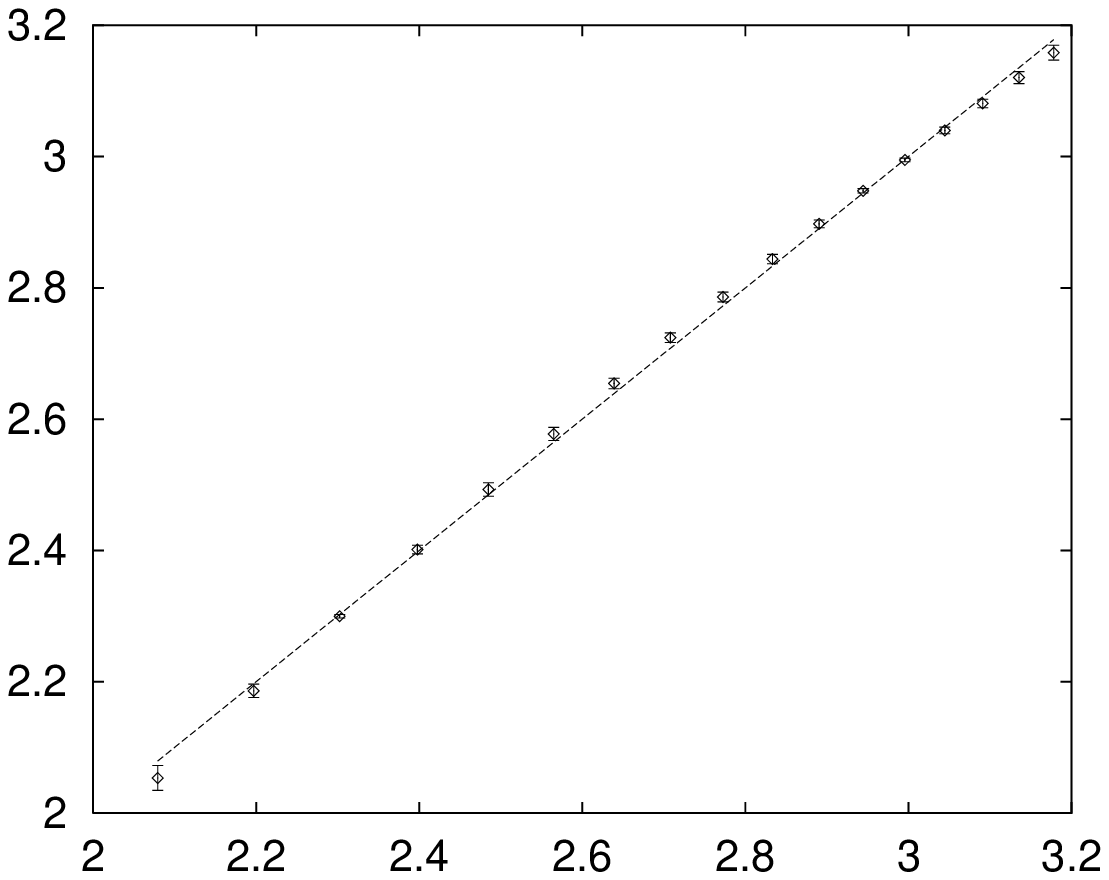}
\\
\hspace*{1cm}$x$
\hspace*{6.5cm}$x$
\\
\end{center}


$II)$ \\
\hspace*{0.5cm}${E'}_3(y')$ 
\hspace*{5.5cm}${E'}_7(y')$
\begin{center} 
\leavevmode 
\epsfysize=4cm 
 \epsfbox[50 61 388 291]{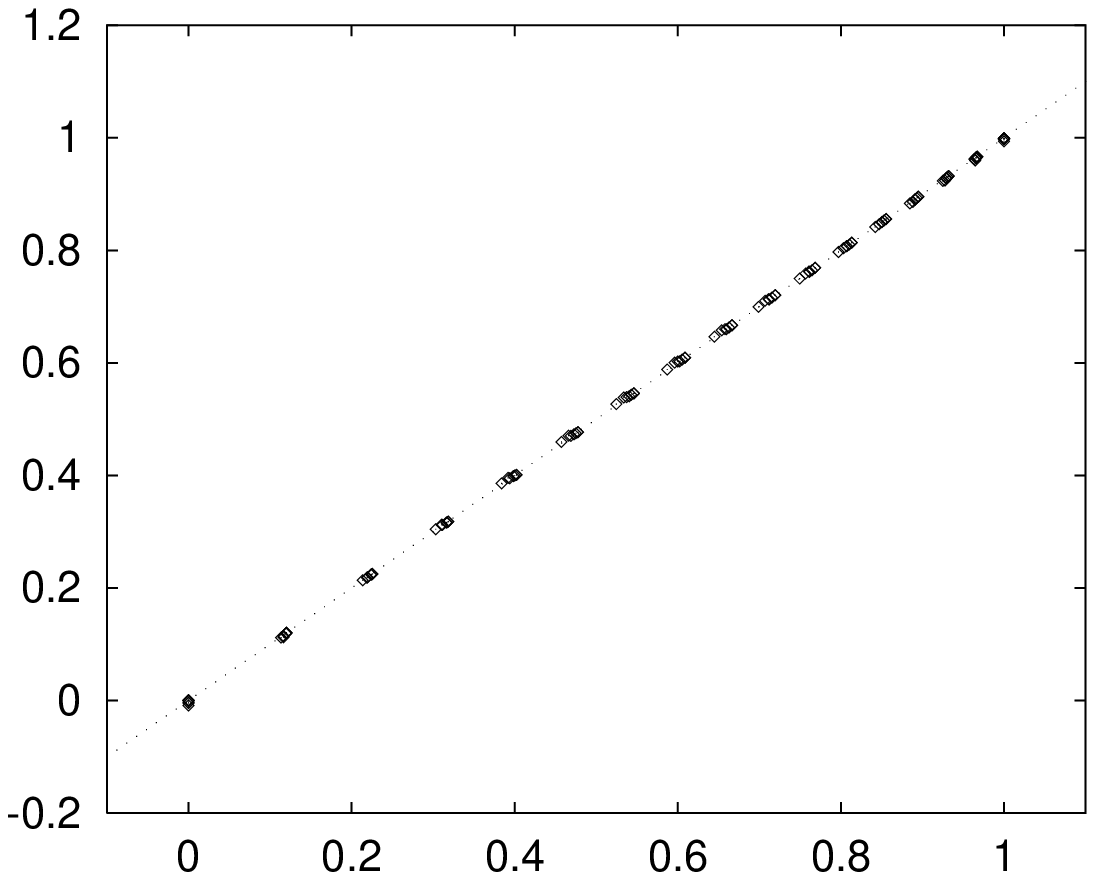}
\hspace*{1cm} 
\epsfysize=4cm 
\epsfbox[72 61 410 291]{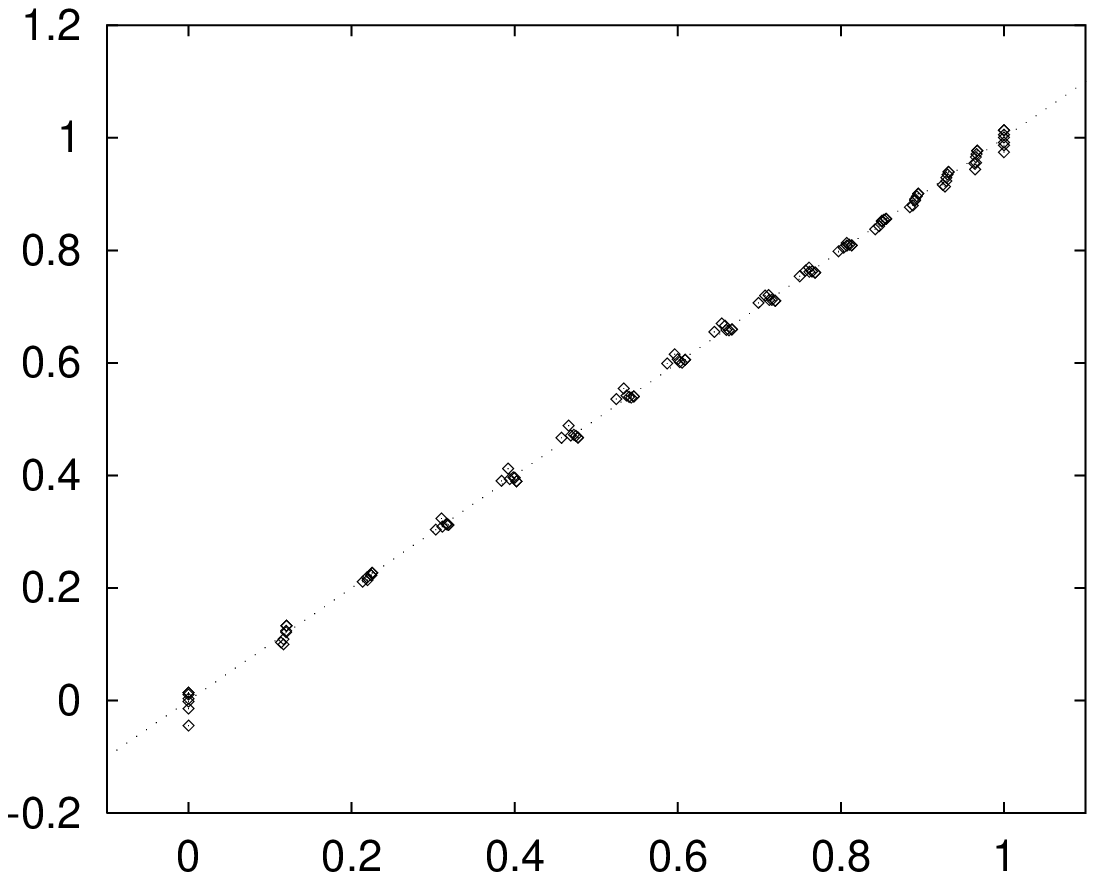}   
\\ 
\hspace*{1cm}$y'$
\hspace*{6.5cm}$y'$
\\ 
\end{center} 


\caption{ 
Test of SS and ESS for the eight ensembles. 
I) SS for the moments of order 2 (a) and 7 (b). The dashed line is
$f(x)=x$. The diamonds represent the
mean and the (small) bars the standard deviation over the eight elements. 
II) ESS for the moments of order 3 (a) and 7 (b). The dotted line is
$f(y')=y'$
}
\label{fig:8_colec}
\end{figure}

We study the properties of the four ensembles in the two directions, which
gives us eight elements altogether.  Figs. \ref{fig:8_colec}Ia and
\ref{fig:8_colec}Ib show the mean $\bar{S'_p}(x)$ and the standard
deviation $\sigma_{{S_p}'(x)}$ for $p=2$ and $7$. It is observed that SS
holds rather well, although there appear small deviations for the lowest
and highest values of $x$.

\indent
The values of ${E'}_p(y')$ are plotted together for the eight elements as
a function of $y'$. This time we cannot average over the values of
$E_p^{\prime}(y^{\prime})$, since $y^{\prime}$ varies from element to 
element. Figs. \ref{fig:8_colec}IIa and \ref{fig:8_colec}IIb
for $p=2$ and $7$, respectively, show that ESS holds better than SS.


\begin{figure}[hbt]
\begin{center}  
\hspace*{0.0cm}
\makebox[7cm]{\makebox[1cm]{$\bar{\rho'}(p, 2)$}\makebox[6cm]{}}
\\
\vspace*{.5cm}
\makebox[7cm]{\makebox[1cm]{}
\makebox[6cm]{
\epsfysize=6cm
\epsfbox[50 61 388 291]{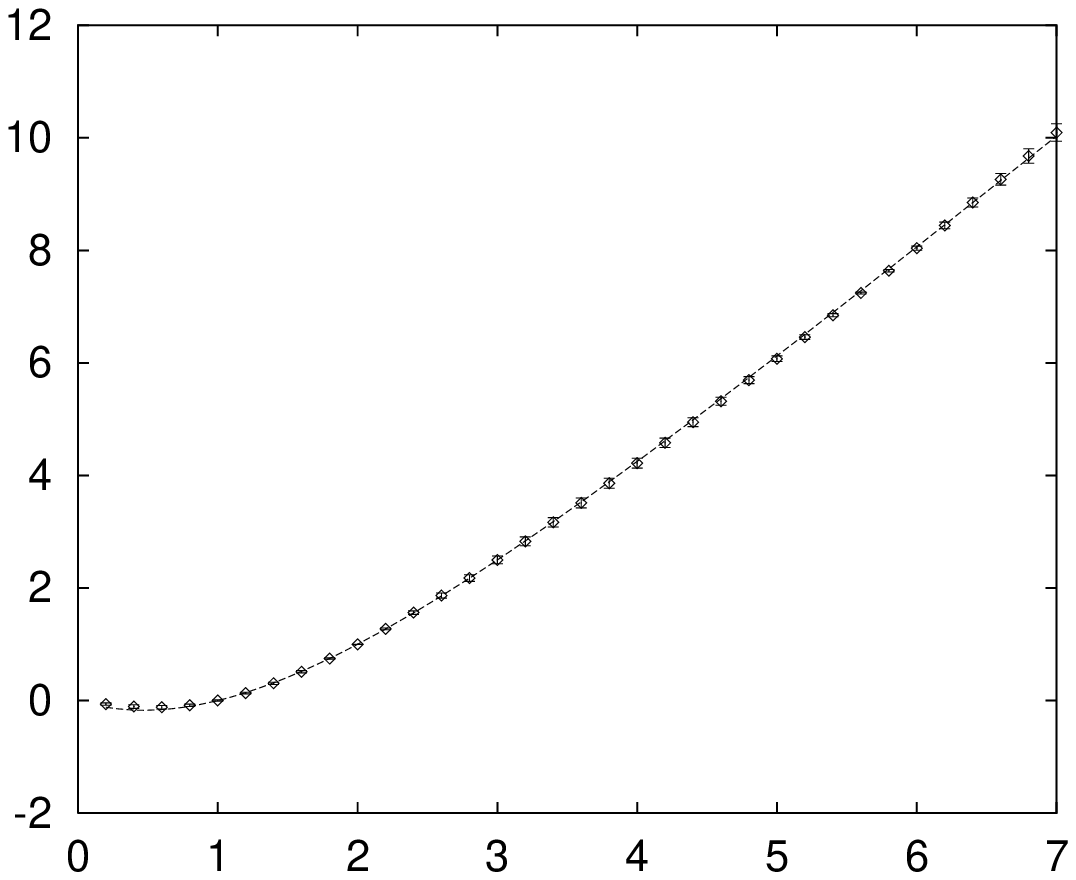}
}}
\\
\makebox[7cm]{\makebox[1cm]{}\makebox[6cm]{\hspace*{2cm}$p$}}  
\end{center} 
\caption{ Test of the log-Poisson model for the eight ensembles. 
The diamonds represent the mean of $\rho'(p, 2)$ over the eight  
ensembles as a function of $p$ and the error bars are twice the 
standard deviation. 
These error bars are small and difficult to appreciate. 
The dashed line represents $\rho^{0.5}(p,2)$, the log-Poisson 
$\rho(p,2)$ with $\beta=0.5$. }
\label{fig:8_colec_rho}
\end{figure}

\indent
Fig. \ref{fig:8_colec_rho} represents the mean value $\bar{\rho'}(p)$ over
the eight ensembles as a function of $p$ and its standard deviation. It is
observed that the log-Poisson model is very well satisfied. In Table
\ref{tab:8_colec_beta}, the observed values of $\beta$ are given. There
are important numerical differences among them. As we will see in the next
section this also occurs when the images are analyzed individually.

\begin{table}[hbt]
\begin{center}
\begin{tabular}{|c|c|}
\hline
\bf{Ensemble and orientation} & $\bf{\beta}$ \\ \hline
A horizontal &       0.17  \\ \hline
A vertical   &       0.022 \\ \hline
B horizontal &       0.27  \\ \hline
B vertical   &       0.14  \\ \hline
C horizontal &       0.10  \\ \hline
C vertical   &       0.10  \\ \hline
D horizontal &       0.28  \\ \hline
D vertical   &       0.030 \\ \hline
\end{tabular}
\end{center}
\caption{Values of $\beta$ for the four ensembles A-D and two orientations}
\label{tab:8_colec_beta}
\end{table}


\indent
In most of the entries of Table \ref{tab:8_colec_beta} there is an
anisotropy between the observed values of $\beta$ along the horizontal
and vertical directions. For instance, in ensemble A the horizontal
$\beta$ is much larger than the vertical one. A difference between the
two $\beta$'s is still present in ensemble B, although it is smaller.
Both are dominated by vertical statistics (they are made up of images
of woods), which somehow is reflected in the different values observed
for $\beta$ along each direction.  However, ensemble B is more
isotropic. This can be due to the existence in this ensemble of
structures not present in A such as shadows, open skies and clouds,
and to the fact that the woods are not as dense as in A.

Naively, one would expect a difference of the opposite sign in
ensemble C, which was defined to be dominated by horizontal
statistics. However, this particular ensemble reveals to be the most
isotropic of the four. This apparent contradiction is explained by the
fact that even when changes in contrast along the horizontal direction
are normally smooth, they present very localized fluctuations that
contribute to $\epsilon_{h,r}$.  In fact, a closer inspection of the
images in ensemble C shows that although the contrast is more
correlated along the horizontal direction, the edges appear
isotropically distributed.
Finally, ensemble D behaves similarly
to A, except for the numerical values of $\beta$.

The values of $\beta$ in Table~\ref{tab:8_colec_beta} also 
inform of the singularity exponents of the edges, $-\Delta$ (cf. end
of Sec. (\ref{subsection:multifractality})). 
For instance, and assuming that $D_{\infty} \sim 1$, 
the small values of $\beta$ in the table indicate that 
the corresponding statistics is dominated by 
sharp contrast changes. This effect is more noticeable in the 
vertical variable than in the horizontal one.

\subsection{Single images}
\label{subsection:results_single_images}

\indent

Again, our aim is to establish whether the statistical properties of
SS and ESS are present but this time in single images, and whether the
ESS exponents can be predicted by a log-Poisson model for each of
them. Now the moments $<\epsilon_{r,l}^{p}>$ are obtained by averaging
over the pixels of each individual image. Note that the resulting
exponents $\rho(p, 2)$ of the images of a certain ensemble do not
necessarily average to the ensemble exponents.  This is checked for
the set of $200$ images for both the horizontal and vertical
direction. From now on we shall refer to this set as containing $400$
``images'', implying that we consider the statistics from the horizontal and the
vertical directions.

\begin{figure}[hbt]
\hspace*{5cm}$\bf{a}$ \hspace*{6cm} $\bf{b}$ \\
$I)$ \\
\hspace*{0.5cm} $\bar{{S'}_2}(x)$ \hspace*{5.5cm}$\bar{{S'}_7}(x)$
\begin{center}
\leavevmode
\epsfysize=4cm
\epsfbox[50 61 388 291]{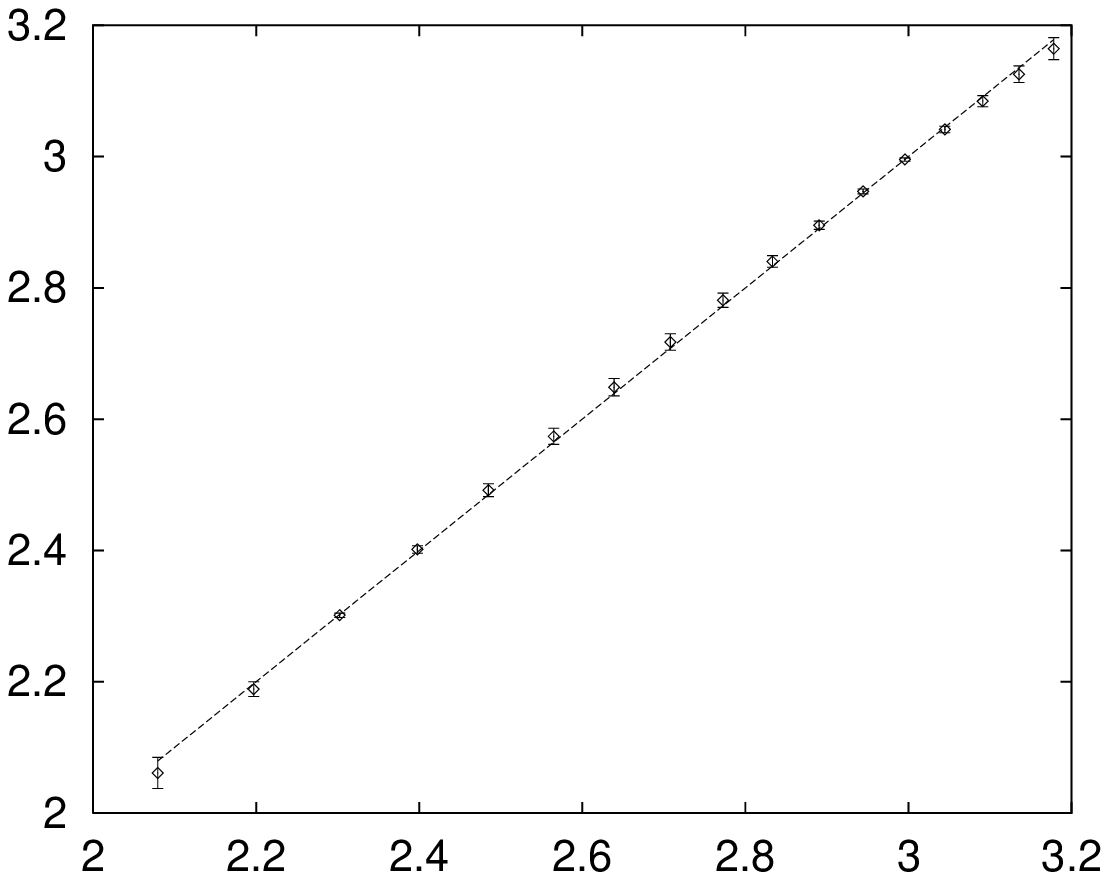}
\hspace*{1cm}
\epsfysize=4cm
\epsfbox[72 61 410 291]{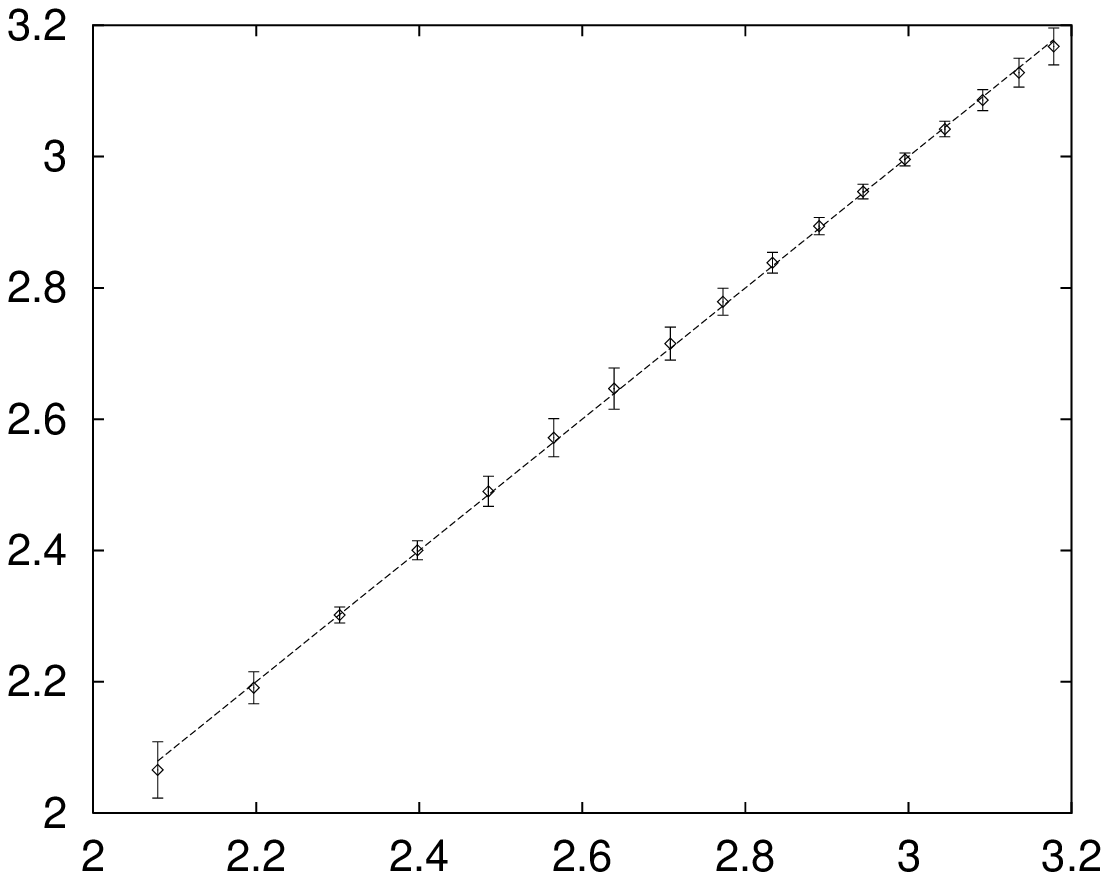}
\\
\hspace*{1cm}$x$
\hspace*{6.5cm}$x$
\\
\end{center}


$II)$ \\
\hspace*{0.5cm}${E'}_3(y')$ 
\hspace*{5.5cm}${E'}_7(y')$
\begin{center} 
\leavevmode 
\epsfysize=4cm 
 \epsfbox[50 61 388 291]{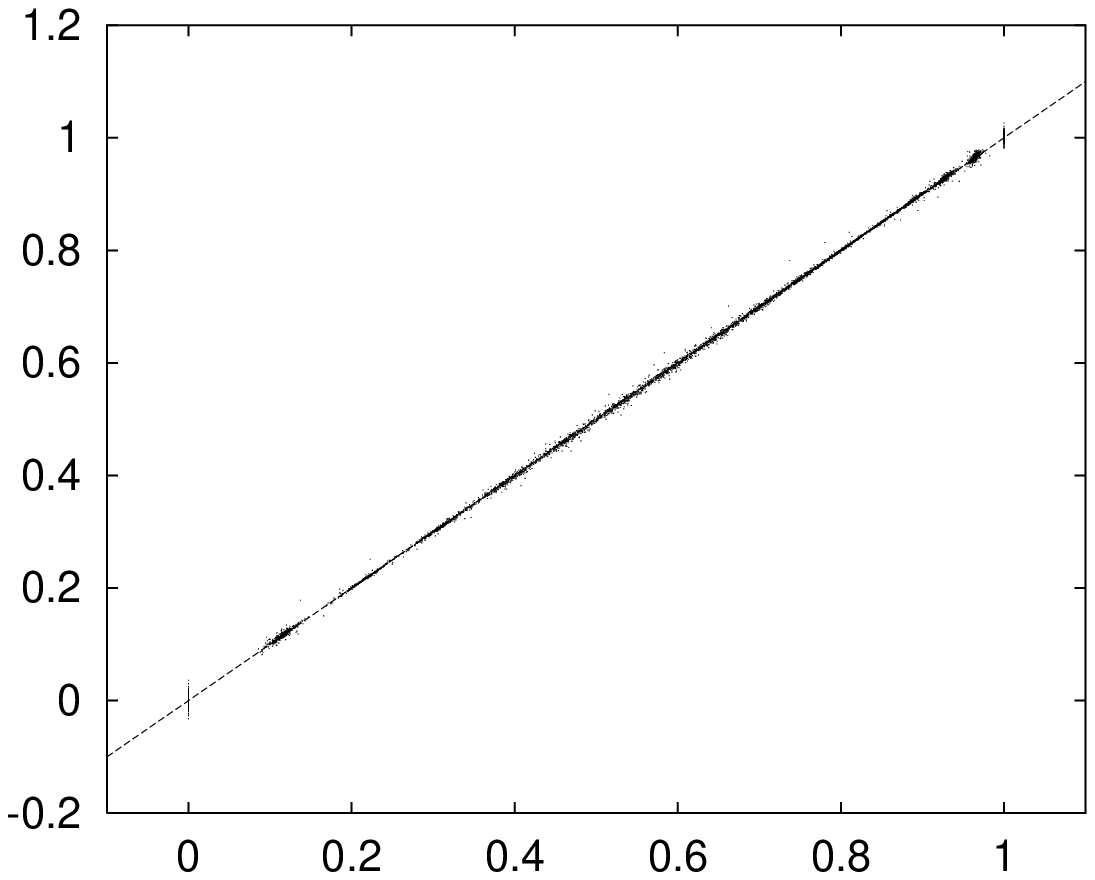}
\hspace*{1cm} 
\epsfysize=4cm 
\epsfbox[72 61 410 291]{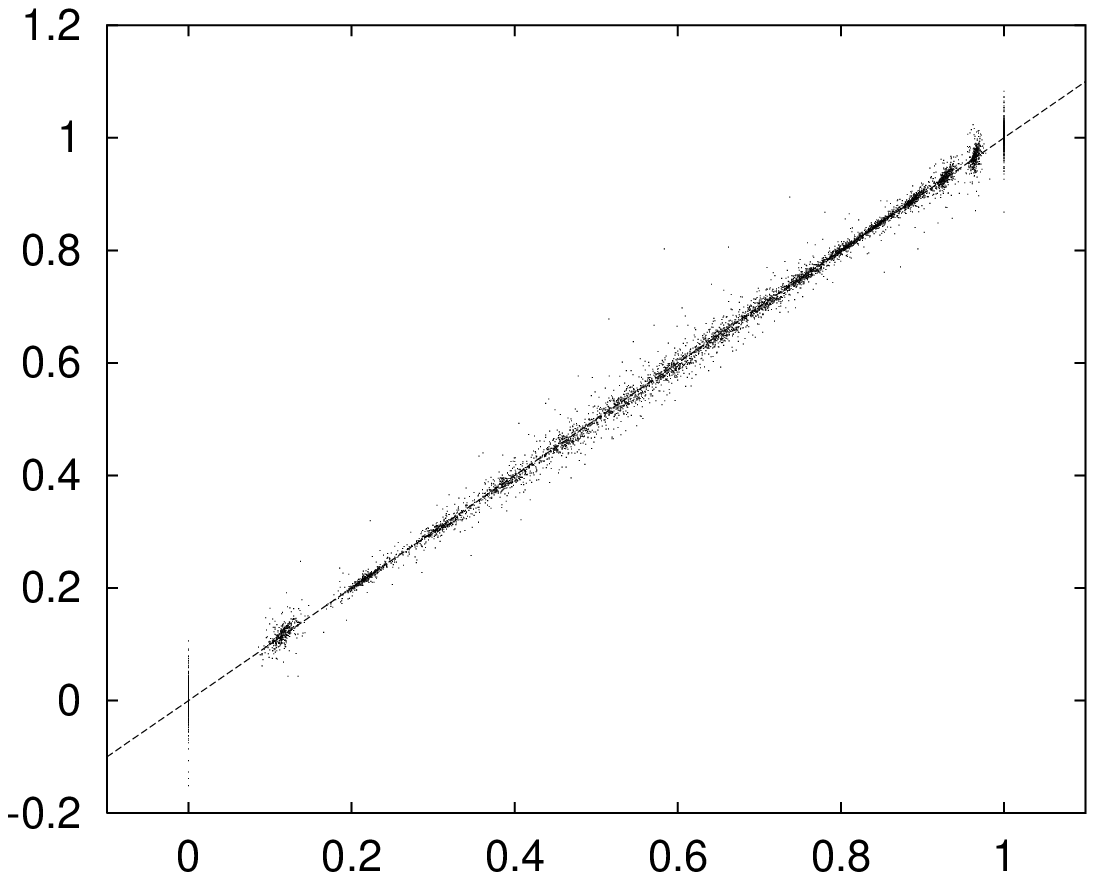}
\\ 
\hspace*{1cm}$y'$
\hspace*{6.5cm}$y'$
\\ 
\end{center} 


\caption{ Test of SS and ESS for the $375$ out of $400$ chosen images. 
I) SS for the moments of order 2 (a) and 7 (b). The dashed line is
$f(x)=x$.  
II) ESS for the moments of order 3 (a) and 7 (b). The dashed line is
$f(y')=y'$}
\label{fig:400_imag_SS}
\end{figure}

\begin{figure}[htb]
\hspace*{5cm}$\bf{a}$ \hspace*{6cm} $\bf{b}$ \\
$I)$ \\
\hspace*{0.5cm} $\bar{\rho'}(p, 2)$ \hspace*{5.5cm}$\bar{\rho'}(p, 2)$
\begin{center}
\leavevmode
\epsfysize=4cm
\epsfbox[50 61 388 291]{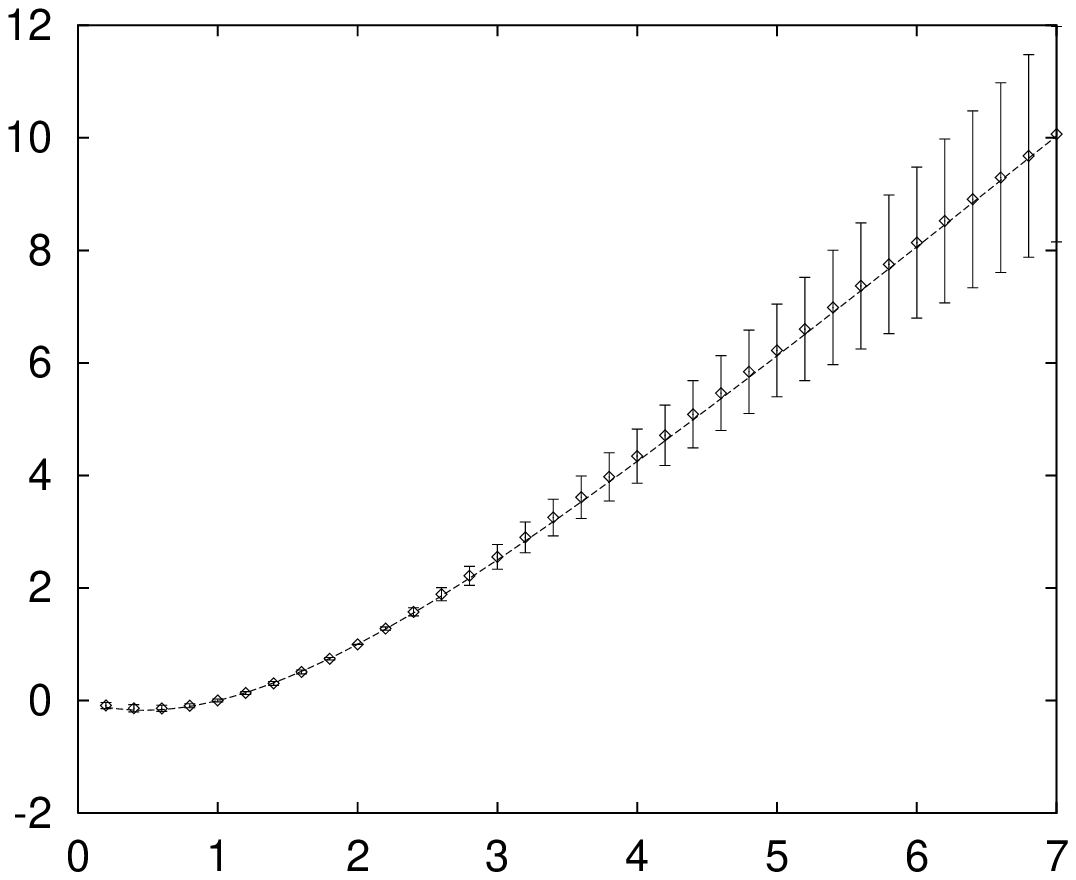}
\hspace*{1cm}
\epsfysize=4cm
\epsfbox[72 61 410 291]{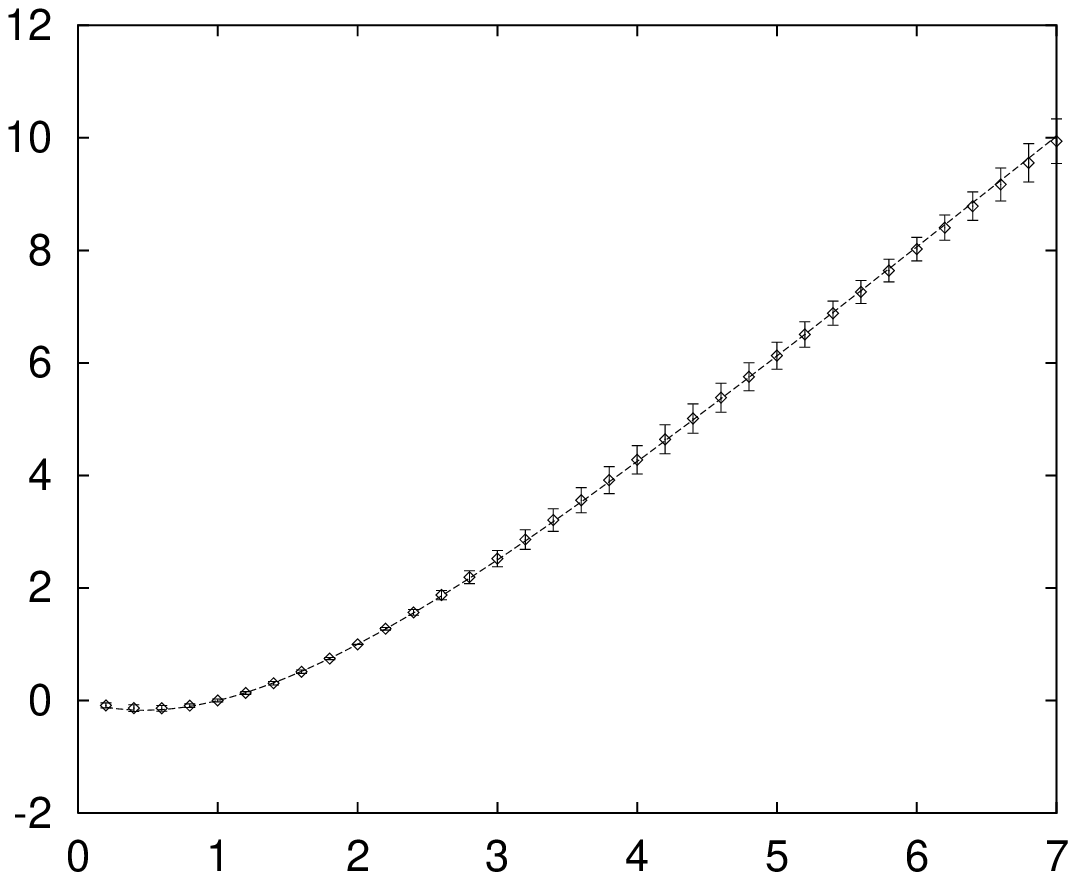}
\\
\hspace*{1cm}$p$
\hspace*{6.5cm}$p$
\\
\end{center}

$II)$ \\
\hspace*{0.5cm} $P(\beta)$ \hspace*{5.5cm}$P(D_{\infty})$
\begin{center}
\leavevmode
\epsfysize=4cm
\epsfbox[50 50 410 302]{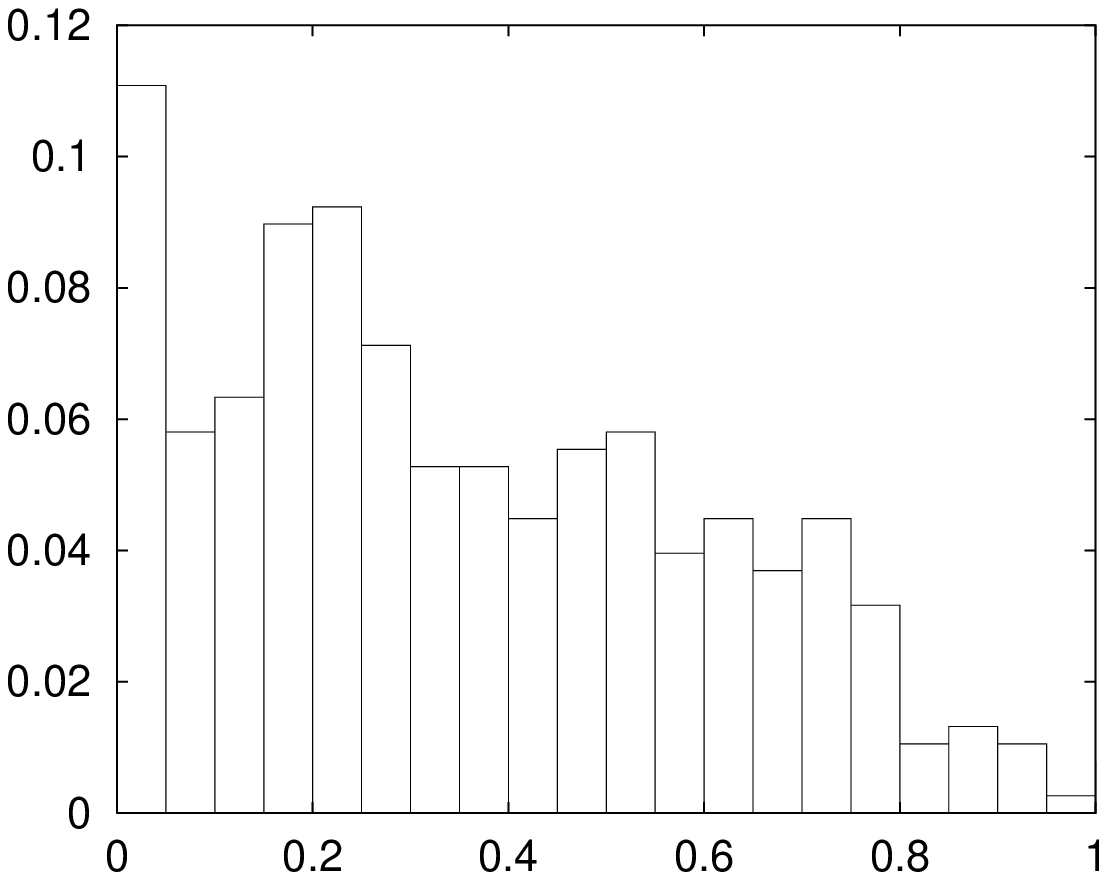}
\hspace*{1cm}
\epsfysize=4cm
\epsfbox[50 50 410 302]{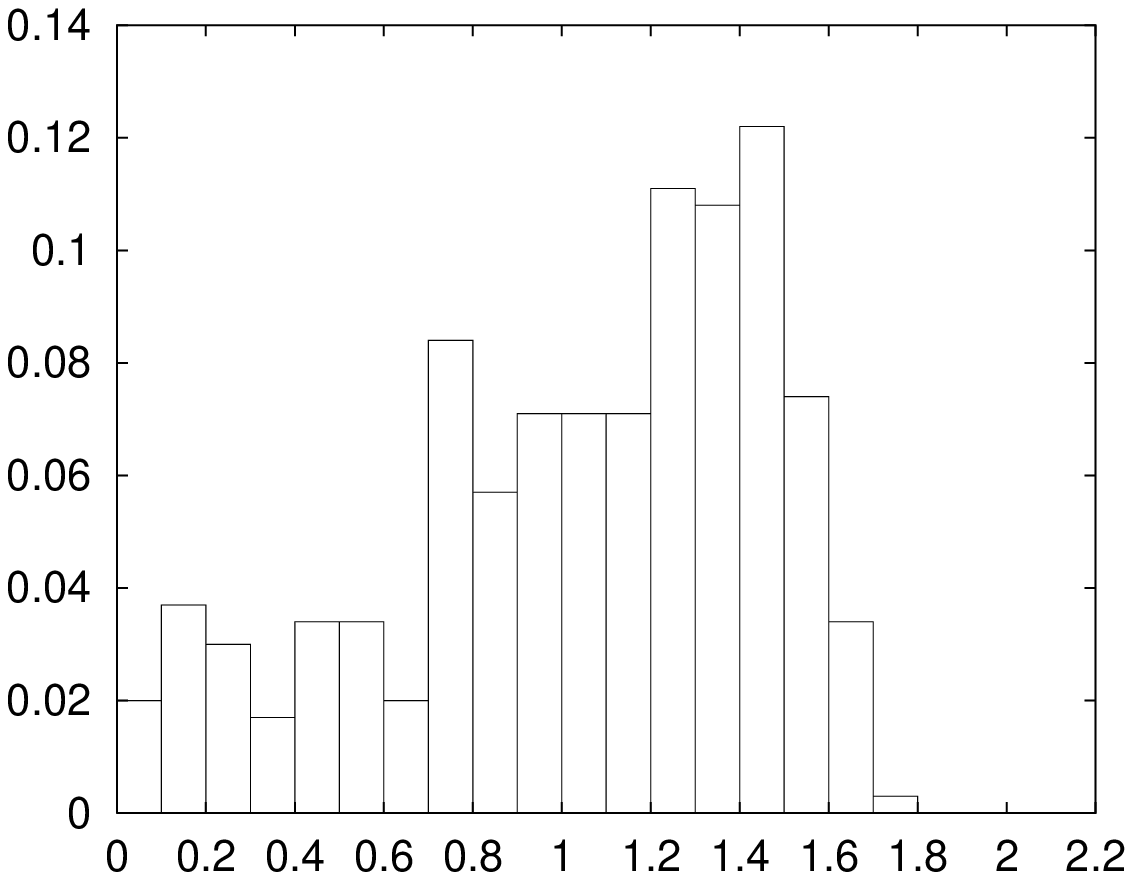}
\\
\hspace*{1cm}$\beta$
\hspace*{6.5cm}$D_{\infty}$
\\
\end{center}

\caption{ 
Test of the log-Poisson model for the $400$ images. 
Ia) The mean of the $\rho'(p, 2)$'s over the $400$ images. 
Ib) The mean of the $\rho(p, 2)$'s for the $375$ best images. 
Again, for a) and b) the diamonds represent the mean of 
$\rho'(p, 2)$ and the error bars are twice the standard deviation. 
The dashed line represents $\rho^{0.5}(p,2)$.
IIa) Histogram of values of $\beta$ for the $375$ selected images.
IIb) Histogram of values of $D_{\infty}$ for the $290$ 
images with $D_{\infty}$ in the range [0:2]}
\label{fig:400_imag_rho}
\end{figure}

The results for SS, ESS and the ESS exponents are presented applying
the procedures described in Sec. \ref{subsection:results_methods}.
Fig. \ref{fig:400_imag_SS}) shows the data for SS and ESS while 
Fig. \ref{fig:400_imag_rho} includes the verification that the
log-Poisson model is satisfied and the corresponding distribution of
values of $\beta$ and $D_{\infty}$.

The means $\bar{{S'}_p}$ and $\bar{\rho}(p,2)$ are not over the $400$
images but over the 375 images that best satisfy the log-Poisson
model. Of the other $25$ images ($6.2\%$), $6$ of them ($1.5\%$),
have been discarded because their $\beta$ exceeds $1$, the maximum
admissible one. The rest have been discarded because the
log-Poisson model fits their $\rho(p, 2)$ significantly less
accurately than that of the $375$ selected images. This can be seen in
Figs. \ref{fig:400_imag_rho}Ia and \ref{fig:400_imag_rho}Ib
The $375$ images with smallest $\chi^2 = 
\sum_{p=1}^{n} {(\rho(p, 2)-\rho^{\beta}(p, 2))}^2$ are a
homogeneous set in the sense that discarding a few of the worst of
them produces a small improvement of the overall fit represented in
Fig. \ref{fig:400_imag_rho}Ib. On the contrary, discarding the $25$
worst images of the complete set produces a significant improvement of
the overall fit.
The discarded images have some features in common. Almost all of them
either are obscure and therefore lack definition or are scenes with
little structure: sea, rivers, ground with little more than grass,
buildings with simple repeating patterns.

\indent
Fig. \ref{fig:400_imag_rho}IIa shows that the probability of $\beta$
decreases approximately linearly with $\beta$. The corresponding
histograms for the eight ensembles considered in
Sec. (\ref{subsection:results_ensembles}) have a similar shape, though
more irregular. They extend also over the whole range [0:1]. The
histograms of the vertical ensembles are peaked at a lower value of
$\beta$ ($\sim 0.05$) than the horizontal ones ($\sim 0.25$), except
for ensemble C, which is in accordance with the global values of
$\beta$ for these ensembles, shown in Table~\ref{tab:8_colec_beta}.

\indent 
The fractal dimension of the set of edges, $D_{\infty}$, can be
estimated from eq.(\ref{eq:D_vs_beta}) by measuring $\beta$ and
$\tau_2$. As has already been mentioned in section
(\ref{subsection:multifractality}), the
evaluation of this fractal dimension becomes difficult for large
$\beta$.  In fact, $D_{\infty}$ has  unrealistic negative values for
about 23\% of the images which satisfy the log-Poisson model (all of
them have $\beta>0.5$). This can be explained by the fact that, for
large $\beta$, Eq.(\ref{eq:D_vs_beta}) propagates errors in the
estimation of $\beta$ and $\tau_2$ as very large errors in the
estimation of $D_{\infty}$, due to the factor
$\frac{1}{(1-\beta)^2}$. Furthermore, even if errors in $\beta$ are
not biased, eq.(\ref{eq:D_vs_beta}) gives an estimation of
$D_{\infty}$ biased towards lower values \footnote{This can be seen by
calculating the propagated error in $D_{\infty}$ to second order in
the error in $\beta$. For large enough $\beta$ this order becomes
relevant: $\delta D_{\infty} = \frac{\tau_2}{(1-\beta)^3} [2 \delta
\beta+\frac{3}{(1-\beta)}(\delta \beta)^2]$}, which may explain why, 
for some images, the dimension of the most singular fractal component
(the set of edges) is negative.

\indent
The result is shown in Fig. \ref{fig:400_imag_rho}IIb. Notice that it
is consistent with the expectation that the fractal dimension of the
set of pixels with the largest changes in contrast should be close to
one.  Even if this histogram seems to be rather broad, a visual
inspection of the fractal component corresponding to an image with,
say, $D_{\infty} = 1.3$ looks like a not too thick line.


\subsection{Whole data set}
\label{subsection:results_whole}

\begin{figure}[hbt]
\hspace*{5cm}$\bf{a}$ \hspace*{6cm} $\bf{b}$ \\
$I)$ \\
\hspace*{0.5cm} ${S'}_2(x)$ \hspace*{5.5cm}${S'}_7(x)$
\begin{center}
\leavevmode
\epsfysize=4cm
\epsfbox[50 61 388 291]{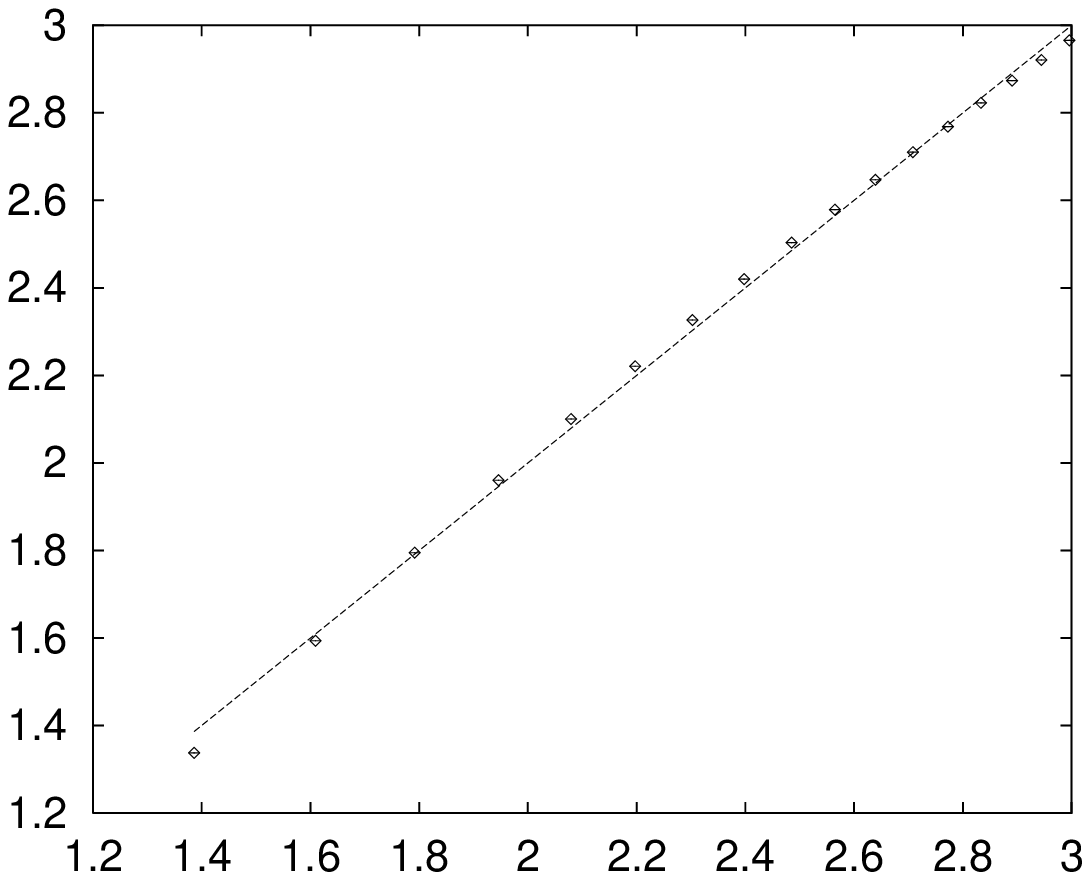}
\hspace*{1cm}
\epsfysize=4cm
\epsfbox[72 61 410 291]{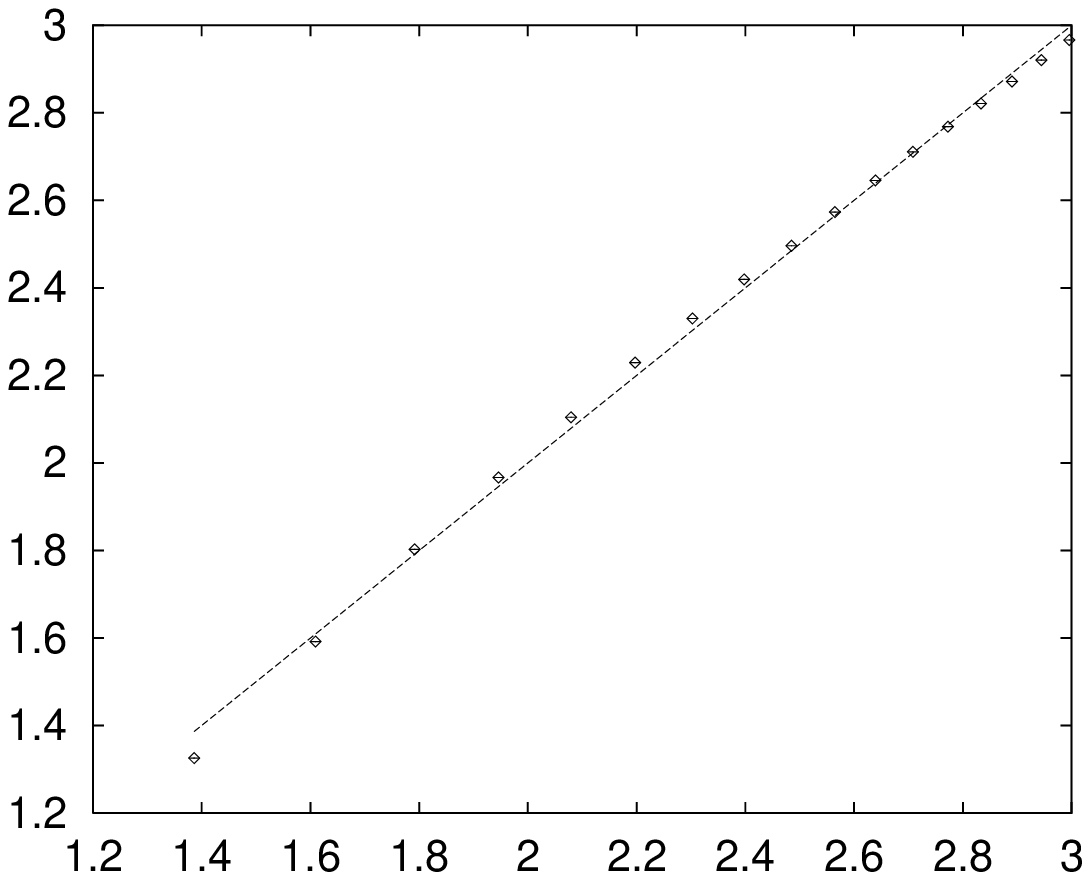}
\\
\hspace*{1cm}$x$
\hspace*{6.5cm}$x$
\\
\end{center}


$II)$ \\
\hspace*{0.5cm}${E'}_3(y')$ 
\hspace*{5.5cm}${E'}_7(y')$
\begin{center} 
\leavevmode 
\epsfysize=4cm 
 \epsfbox[50 61 388 291]{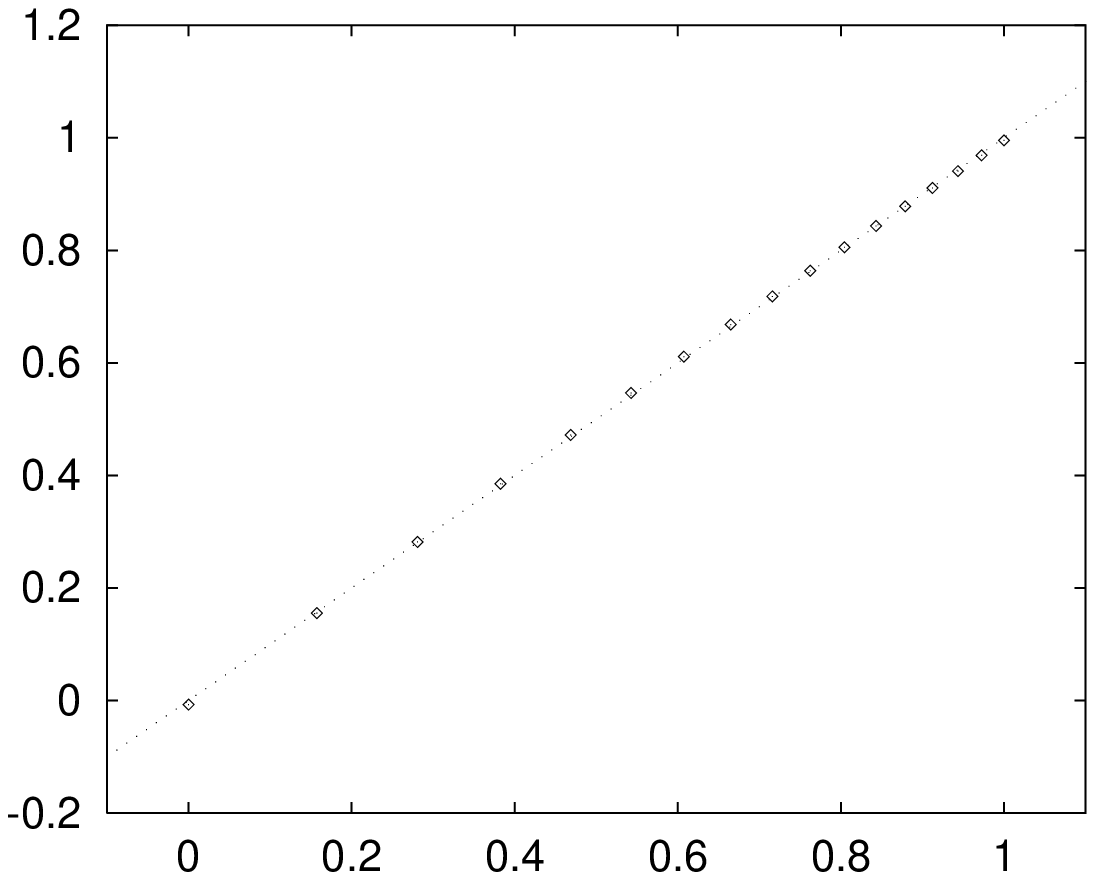}
\hspace*{1cm} 
\epsfysize=4cm 
\epsfbox[72 61 410 291]{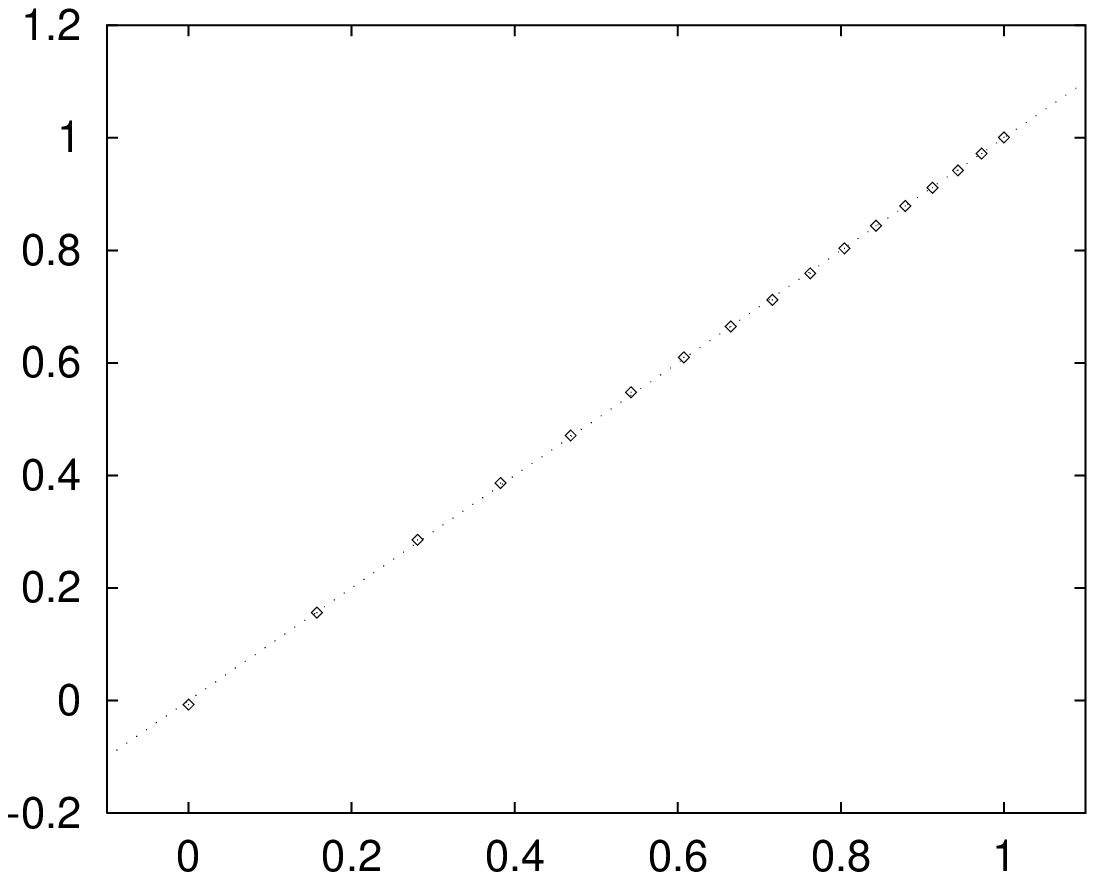}   
\\ 
\hspace*{1cm}$y'$
\hspace*{6.5cm}$y'$
\\ 
\end{center} 


\caption{ 
Test of SS and ESS for the whole image data set (400 images).  I) SS
test for the moments of order two (a) and seven (b). The dashed line
represents $f(x)=x$.  II) ESS test for the moments of order three (a)
and seven (b). The dotted line represents f(y')=y'.}
\label{fig:prom_8_colec}
\end{figure}

\begin{figure}[hbt]
\begin{center}  
\hspace*{0.0cm}
\makebox[7cm]{\makebox[1cm]{$\rho'(p, 2)$}\makebox[6cm]{}}
\\
\vspace*{.5cm}
\makebox[7cm]{\makebox[1cm]{}
\makebox[6cm]{
\epsfysize=6cm
\epsfbox[50 61 388 291]{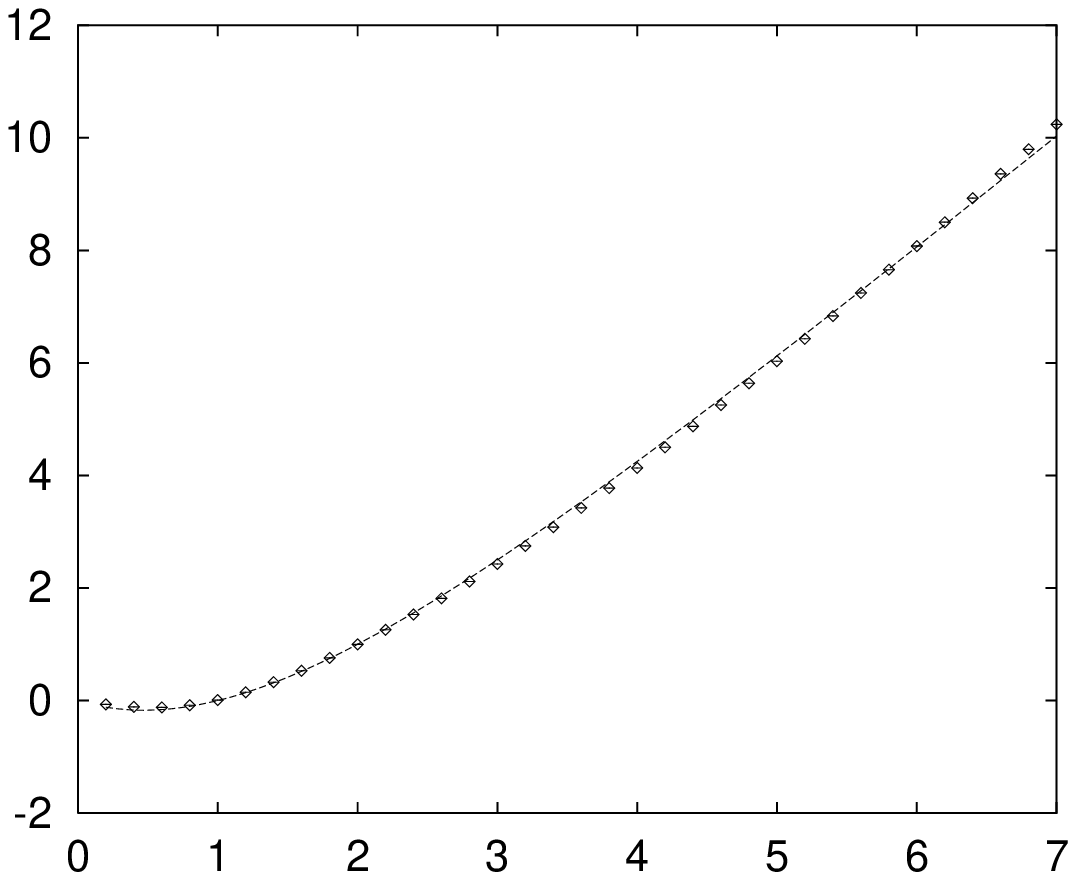}
}}
\\
\makebox[7cm]{\makebox[1cm]{}\makebox[6cm]{\hspace*{2cm}$p$}}  
\end{center} 
\caption{ Test of the log-Poisson model for the whole data set. }
\label{fig:prom_8_colec_rho}
\end{figure}

\indent
We now perform the same analysis on the global data set containing the
whole set of $200$ images. These are again treated as $400$ ``images''
by averaging together the moments for the horizontal and vertical
variables $\epsilon$. The results are presented in terms of the same
quantities (eqs.~(\ref{eq:S_p}) to (\ref{eq:rho})) 
used in Sec.~\ref{subsection:results_methods} to allow for direct
comparison. The value of $\beta$ obtained from this set is $0.17$. Fig. 
\ref{fig:prom_8_colec} 
shows the SS test for the moments of order two and seven and the ESS test
for the moments of order three and seven. The fit of the ESS exponents 
with a log-Poisson curve is presented in
Fig. \ref{fig:prom_8_colec_rho} (Notice that, following the convention
defined in eq. (\ref{eq:rho}) the curve is referred to $\beta_0 =
0.5$). It can be seen that also for the whole data set all these
properties are well confirmed. Again ESS holds better than SS.



\section{Generative model of images}
\label{section:generative_model}

\indent

The fact that the SS exponents $\tau_p$ do not behave linearly with
$p$ implies that scale-invariance of natural images is rather
subtle. It is related to the fact that for a given image not all the
points transform in the same way under scale transformations
\cite{Singularities}. As was shown in Sec. (\ref{subsection:multifractality}), the log-Poisson model, which explains the
non-linear behaviour of $\tau_p$, also predicts the existence of a
multifractal structure: the image pixels can be decomposed in fractal
components such that points in the same component have the same
scaling properties.

Generative models of natural images based on their scale properties 
should take this important property into account. 
The intuitive model proposed in  \cite{Ru97} 
puts the emphasis in reproducing the correct power spectrum, but it 
does it in such a way that all the points have the same scale properties.
This is because all the changes in contrast are sharp, and equally singular, 
what gives rise to a single fractal component. An evaluation of 
the SS exponents $\tau_p$ would show that these are linear in $p$.

\indent
Here we address the issue of how to define a simple stochastic model
for image generation which takes into account the properties discussed
in the previous sections and that has the correct power spectrum.  The
model described below is based on a wavelet expansion of the contrast
$C(\vec {x})$ itself. 
Although generative models of images based 
on a wavelet expansion have been proposed before (see, e.g.,
\cite{Fi94}), the existence of a multiplicative process associated with
the wavelet coefficients was noticed more recently in 
\cite{Singularities}. According to this result, the multifractal
structure described in section (\ref{subsection:multifractality}) is
produced through a convenient stochastic process for the wavelet
coefficients. \footnote{Although in a different context, the model has
been used before in \cite{synthetic_turbu96} to study synthetic
turbulence.}

\indent
We will express the contrast $C(\vec {x})$ of the image in terms of a
dyadic wavelet set \cite{Dau92} $\Psi_{j,\vec{k}}(\vec{x})\equiv
\Psi(2^j \vec{x} -\vec{k})$, where the scale $r$ is given by $r =
2^{-j}$ ($j \in \Integer$), and the space is sampled at the points
$\vec{x}_0=2^{-j} \vec{k}$, with $\vec{k}\equiv(k_1,k_2)$ ($k_1$ and 
$k_2$ are integer numbers taking values from 0 to $2^{j}-1$).

\begin{equation}
C(\vec{x})\; =\; \sum_{j=0}^{N} \;\; \sum_{k_1,k_2=0}^{2^{j}-1} 
\gamma_{j, \vec{k}} \Psi_{j,\vec{k}}(\vec{x}) \;\; .
\label{eq:Benzi}
\end{equation}

\noindent
where $N$ is the number of scales considered; thus, the number of
pixels of the image is $2^N\times 2^N$ (each change in scale is of
size $2$).  Synthetic images will be generated with $N=8$, which gives
rise to images with $256\times 256$ pixels. The mother wavelet
$\Psi(\vec{x})$ will be taken as the Laplacian of a gaussian with 
zero mean and standard deviation one. \footnote{The interesting 
question of how to choose a convenient  mother wavelet for a given 
image ensemble is not addressed here. This problem is discussed in 
\cite{TuPa99}}

The wavelet coefficients $\gamma_{j, \vec{k}}$ are then generated
following a hierarchical procedure which starts at the scale with the
poorest resolution ($j=0$), which is described by the mother wavelet.
Its coefficient, $\gamma_{0 \vec{0}}$, will appear just as a
normalization of the contrast level.  It is then used to generate the
coefficients $\gamma_{1, \vec{k}}$ ($k_1,k_2 = 0,1$) of the wavelets
located at the four points used to sample the next scale ($j=1$). They
are of the form $\gamma_{1,\vec{k}} \;=\; \alpha_{1 \vec{k}}
\gamma_{0,
\vec{0}}$, where the $\alpha_{1 \vec{k}}$'s are independent, identically
distributed random variables which follow a given probability distribution
(to be given below). This operation is repeated recursively at every new
scale. At the scale $j-1$ each $\gamma_{j-1, \vec{k}}$ generates, by
multiplication by the corresponding independent $\alpha$'s, four new
coefficients at the scale $j$:

\begin{equation}
\gamma_{j \vec{k}}\; =\; \frac{1}{2}\:\alpha_{j \vec{k}} \gamma_{j-1, 
[\frac{\vec{k}}{2}]}
\label{eq:reletaalpha}
\end{equation}

\noindent
where $[\frac{\vec{k}}{2}]$ means the vector with components
equal to the integer part, rounding down, of those of $\vec{k}$ divided by
2.

\indent
The projections of such $C(\vec{x})$ over appropriate wavelets of size $r$
possess the property of SS. Let us consider first the dual wavelet
$\tilde{\Psi}$ (that is, $\langle \tilde{\Psi}_{j\vec{k}} |
\Psi_{j^{\prime}\vec{k}^{\prime}}\rangle = 2^{-2j} \delta_{jj^{\prime}}
\delta_{\vec{k}\vec{k}^{\prime}}$). Defining the wavelet projection  
$T_{\tilde{\Psi}}^r C(\vec{x})$ as:

\begin{equation}
T_{\tilde{\Psi}}^r C(\vec{x})\; =\; \frac{1}{r^2} \int d\vec{y} \;\;
\tilde{\Psi}(\frac{\vec{x}-\vec{y}}{r}) C(\vec{y})
\end{equation}

\noindent
It follows from eq.~(\ref{eq:Benzi}) that:

\begin{equation}
\langle |T_{\tilde{\Psi}}^r C|^p\rangle \; =\; A_p^C\: r^{\tau_p^C}
\label{eq:multiaffine}
\end{equation}

\noindent
where

\begin{equation}
\tau_p^C \; =\; p\:-\: \log_2 \overline{\alpha^p}  \;\;, 
\label{eq:reletatau}
\end{equation}

\noindent
where $\overline{\alpha^p}$ is the moment of order $p$ of the
distribution of $\alpha$. Then, there exists at least the wavelet
$\tilde{\Psi}$ for which the wavelet projections of $C$ have SS,
eq.~(\ref{eq:multiaffine}) (in fact, all the wavelets vanishing the
same number of moments define wavelet projections of $C$ with SS of
the same exponents $\tau_p^C$). As shown in \cite{Singularities}, this
implies that also $\epsilon_{\vec{r}}$ verifies SS (i.e. it follows 
eq. (\ref{eq:SS})) and there is a simple relation between the SS
exponents $\tau_p^C$ and $\tau_p$ given by:

\begin{equation}
\tau_p \; = \;\tau_p^C \: -\: p \;\;  
\label{eq:rel_taus}
\end{equation}

\noindent
and using eq.~(\ref{eq:reletatau}) we obtain 
$\tau_p=-\log_2\overline{\alpha^p}$. It follows that in order to obtain 
SS exponents $\tau_p$ according to the log-Poisson model the random
the $\alpha_{j\vec{k}}$ are computed according the log-Poisson 
distribution eq.~(\ref{eq:Log-Poisson}) where the ratio $\frac{L}{r}$ is $2$.

\begin{figure}[htb]
\begin{center}
\hbox{
        \makebox[0.25cm]{$S(f)$}
        \makebox[10cm]{\hspace*{3cm}a\hspace*{6cm}b}
}
\vspace{0.5cm}
\hbox{
        \makebox[0.25cm]{}
        \leavevmode
        \epsfysize=5cm
        \epsfbox[50 50 379 302]{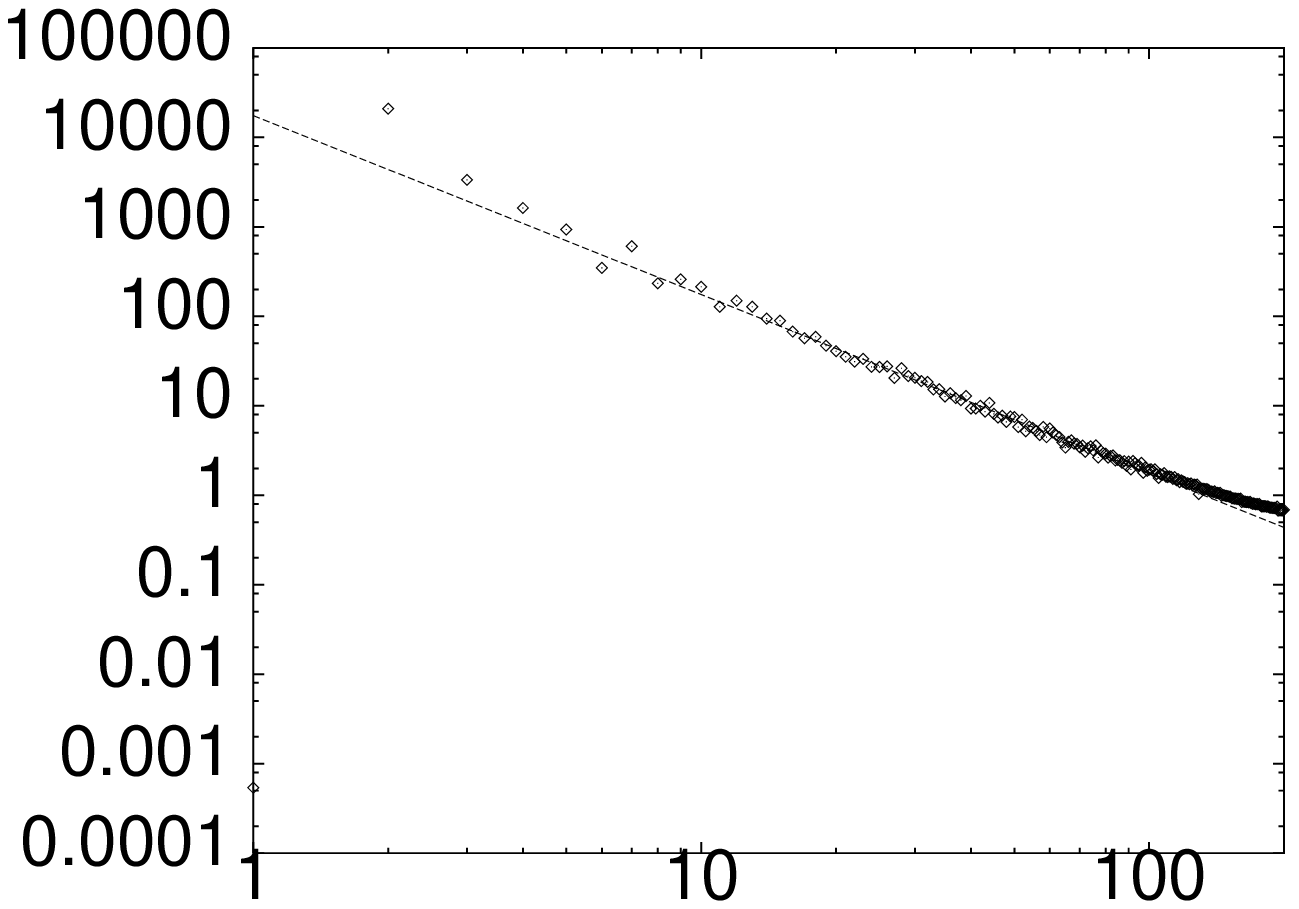}
        \epsfysize=5cm
        \epsfbox[81 50 410 302]{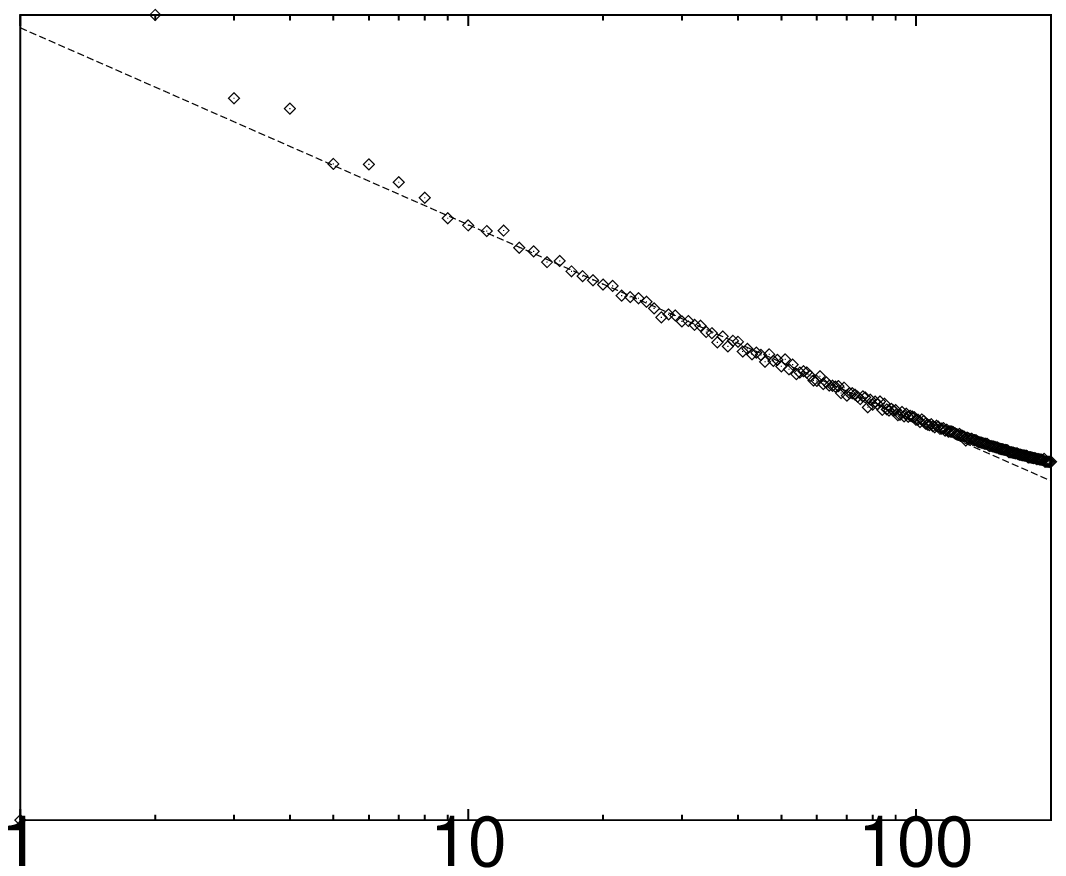}
}
\hbox{
        \makebox[0.25cm]{}
        \makebox[10cm]{\hspace*{3cm}$f$\hspace*{6cm}$f$}
}
\end{center}
\caption{
log-log plot of the power spectrum along the 
a) horizontal and 
b) vertical directions for a $512 \times 512$ image produced with 
the generative model.
The straight lines correspond to $1/f^2$ fits; the horizontal axes are given in
cycles {\it per} image.}
\label{fig:power_spectrum}
\end{figure}

\indent
We have still to check that this model has the correct second order
statistics \cite{Fi87}. Using a $512 \times 512$ image produced with
the generative model we have verified that the power spectrum is
indeed of the form $S(\vec{f})\sim 1/f^2$, for an appropriately chosen
$\Psi$'s. The plot and the fit are exhibited in
Fig.~\ref{fig:power_spectrum}.


\subsection{Geometrical interpretation of the generative model}
\label{subsection:generative_geometrical}

\indent

Table \ref{fig:synthetic_catalogue} presents a catalogue of images
produced with the generative model for different values of the
parameters $\beta$ and $\bar s$. The images have been generated in
such a way that those with identical $\bar s$ (shown in the same
column) contains the same number of modulations for each
$\alpha_{j \vec{k}}$. That is, a given site $j \vec{k}$ has a random
choice of $n_{j \vec{k}}$ modulations which is the same for all the
images in the column.  There is also a continuous change for fixed
$\beta$ as $\bar s$ varies, which is obtained keeping fixed the
cumulative probabilities used to generate the number of modulations at
a given site.

\begin{table}[htb]
\begin{center}
\vspace*{-2cm}
\begin{tabular}{||c|c|c|c|c|c||} \hline\hline
 & $\bar{s}=0.0$ & $\bar{s}=0.5$ & $\bar{s}=1.0$ & $\bar{s}=1.5$ &
$\bar{s}=2.0$ \\
 & ($D_{\infty}=2.0$) & ($D_{\infty}=1.5$) & ($D_{\infty}=1.0$) &
($D_{\infty}=0.5$) & ($D_{\infty}=0.0$)  
\\\hline
&&&&& \\
$\beta=0.0$ 
& \epsfxsize=2.5cm \epsffile{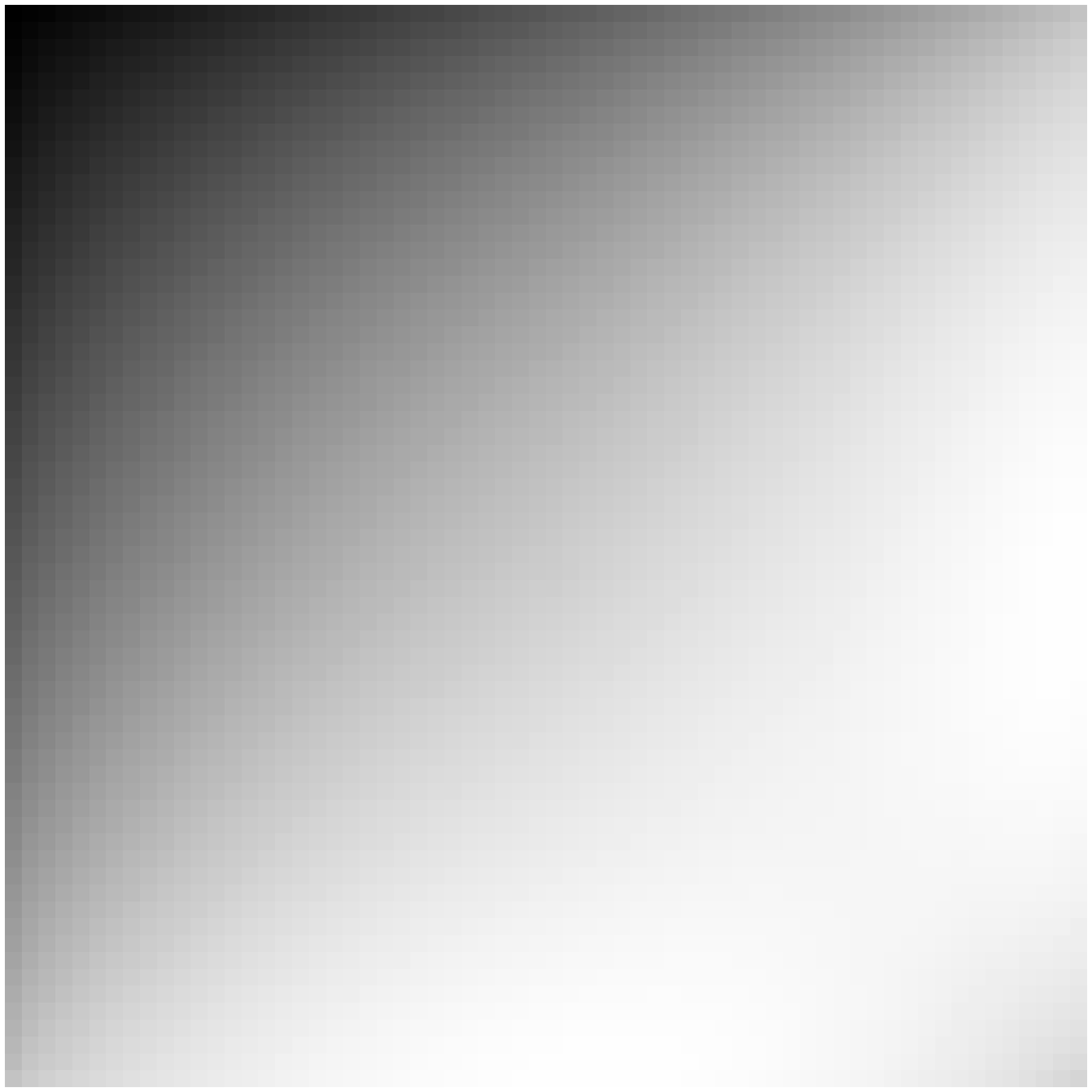}
& \epsfxsize=2.5cm \epsffile{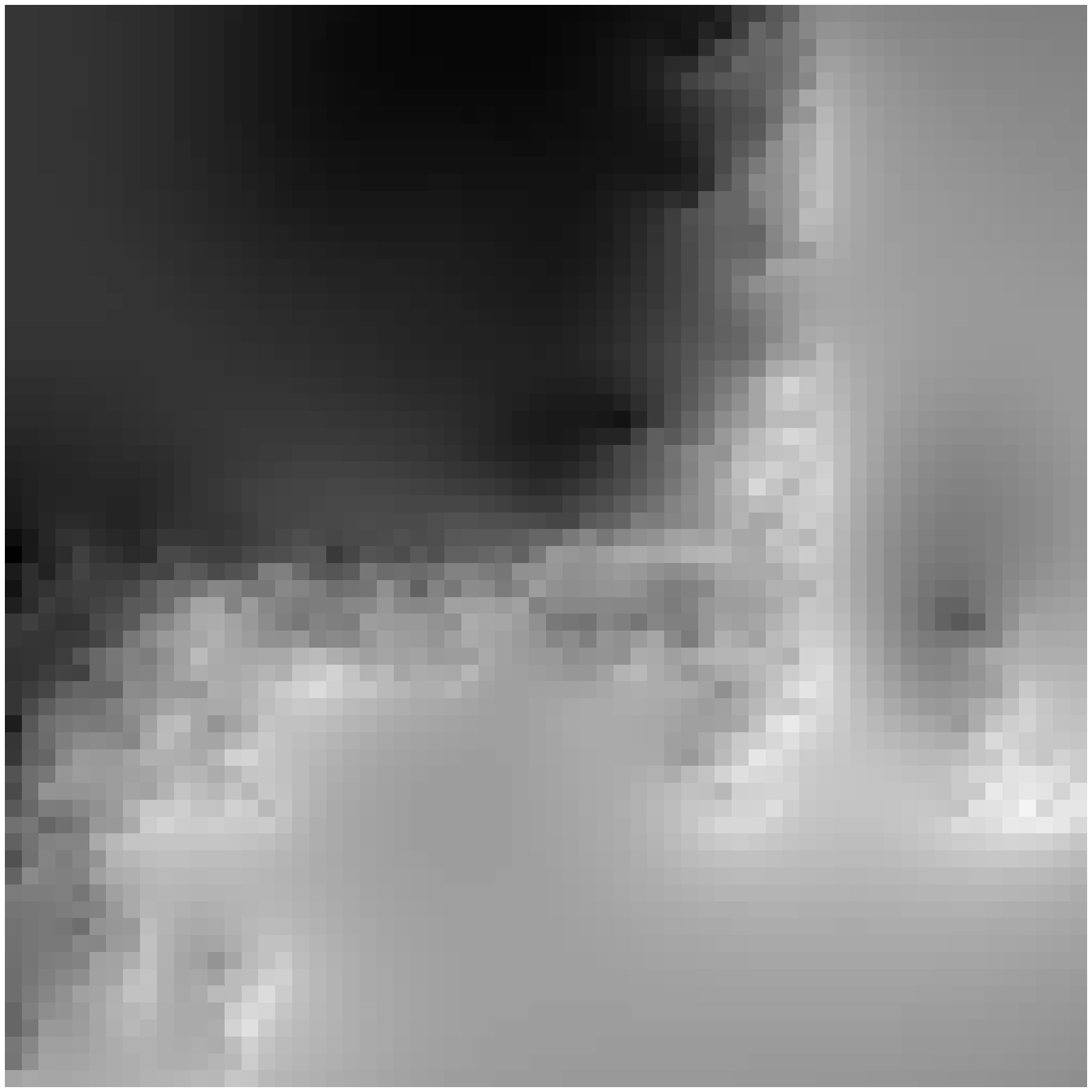}
& \epsfxsize=2.5cm \epsffile{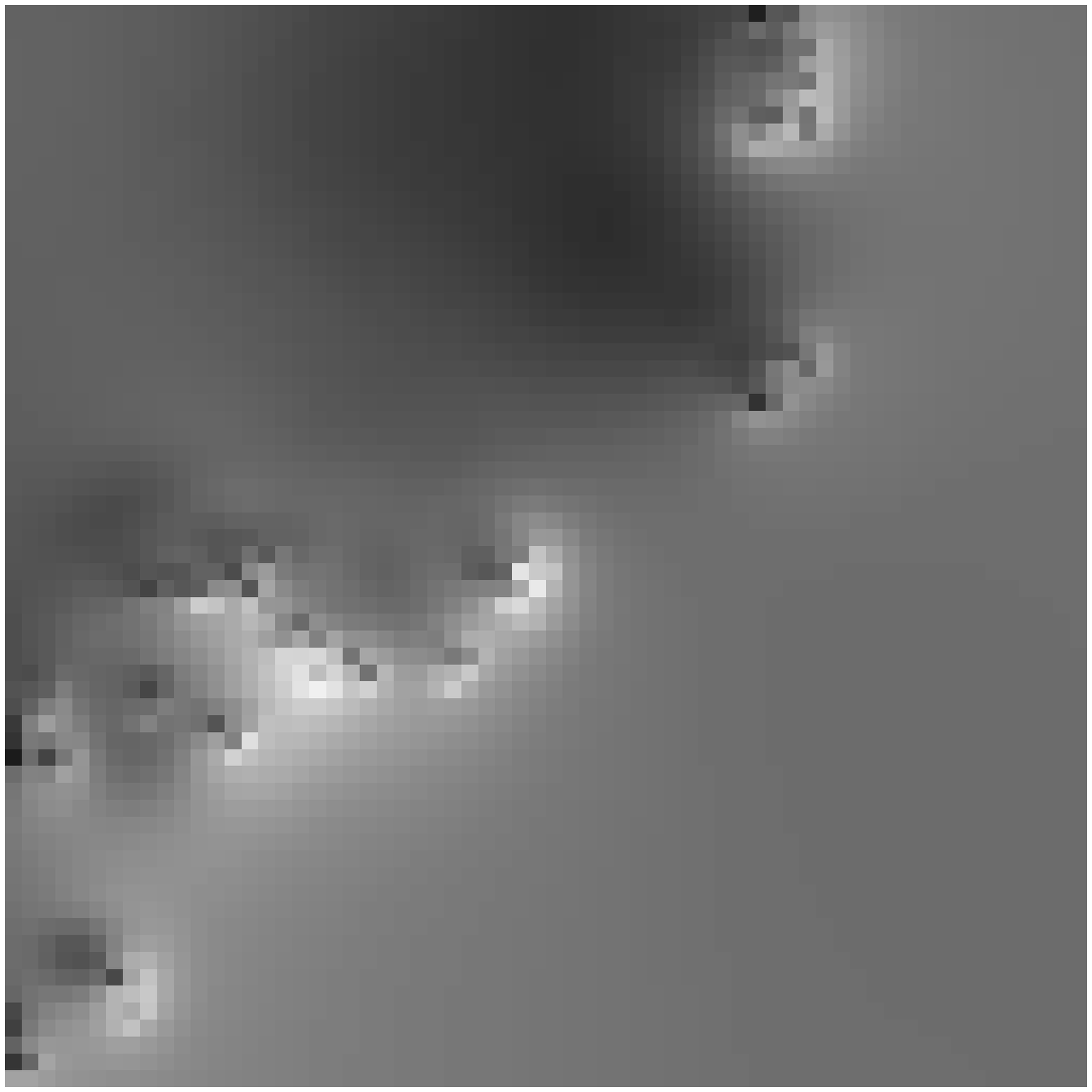}
& \epsfxsize=2.5cm \epsffile{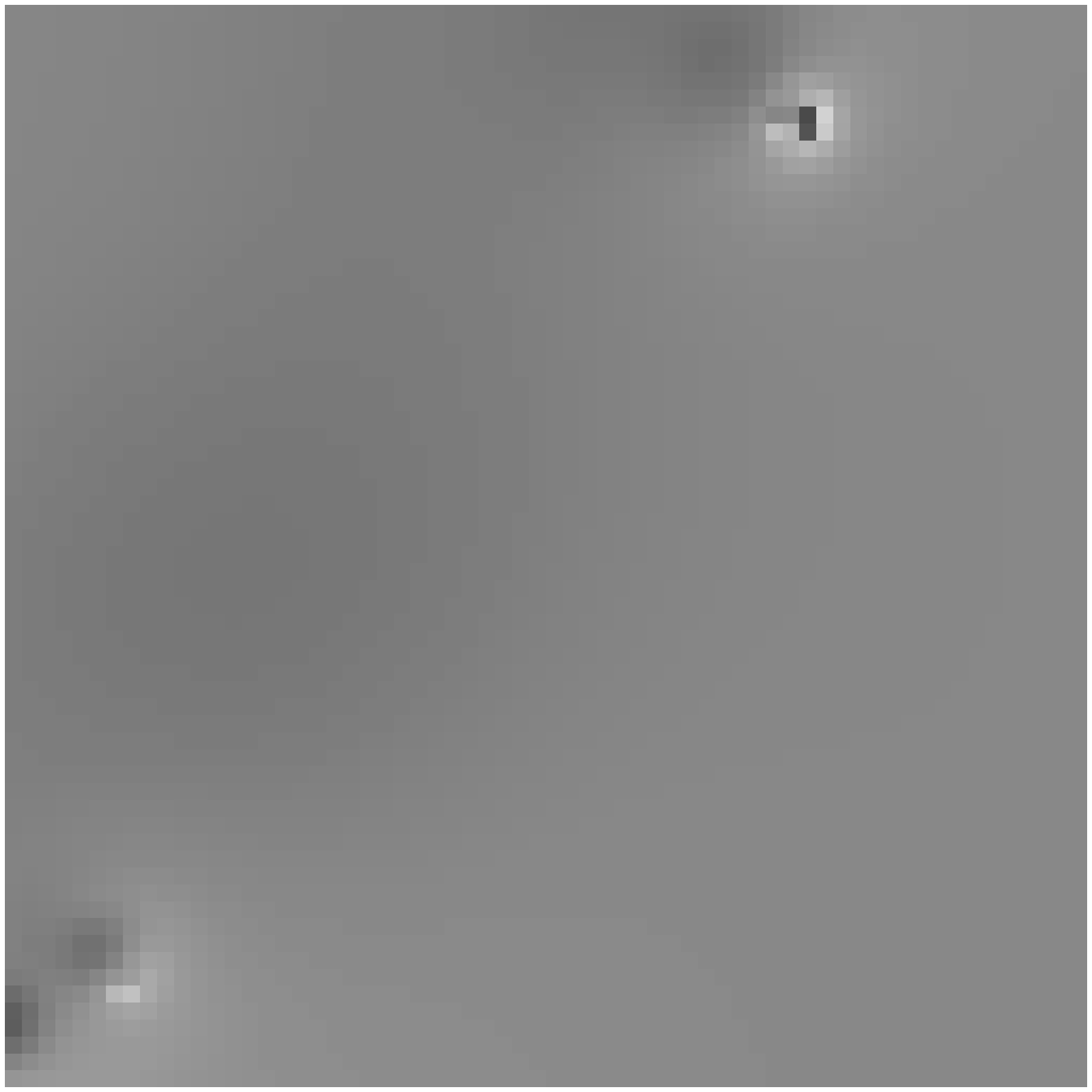}
& \epsfxsize=2.5cm \epsffile{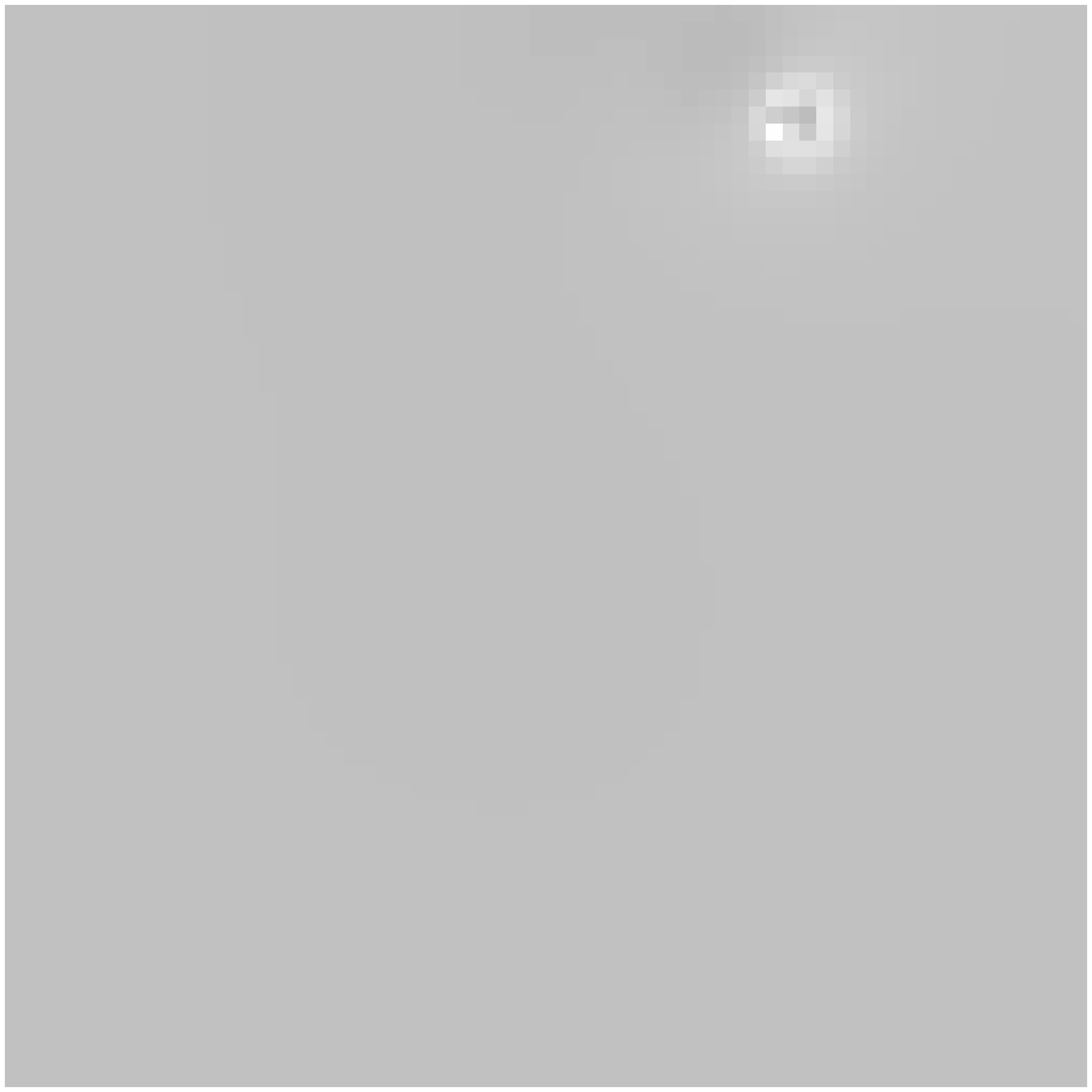}
\\&&&&& 
\\\hline
&&&&& \\
$\beta=0.2$ 
& \epsfxsize=2.5cm \epsffile{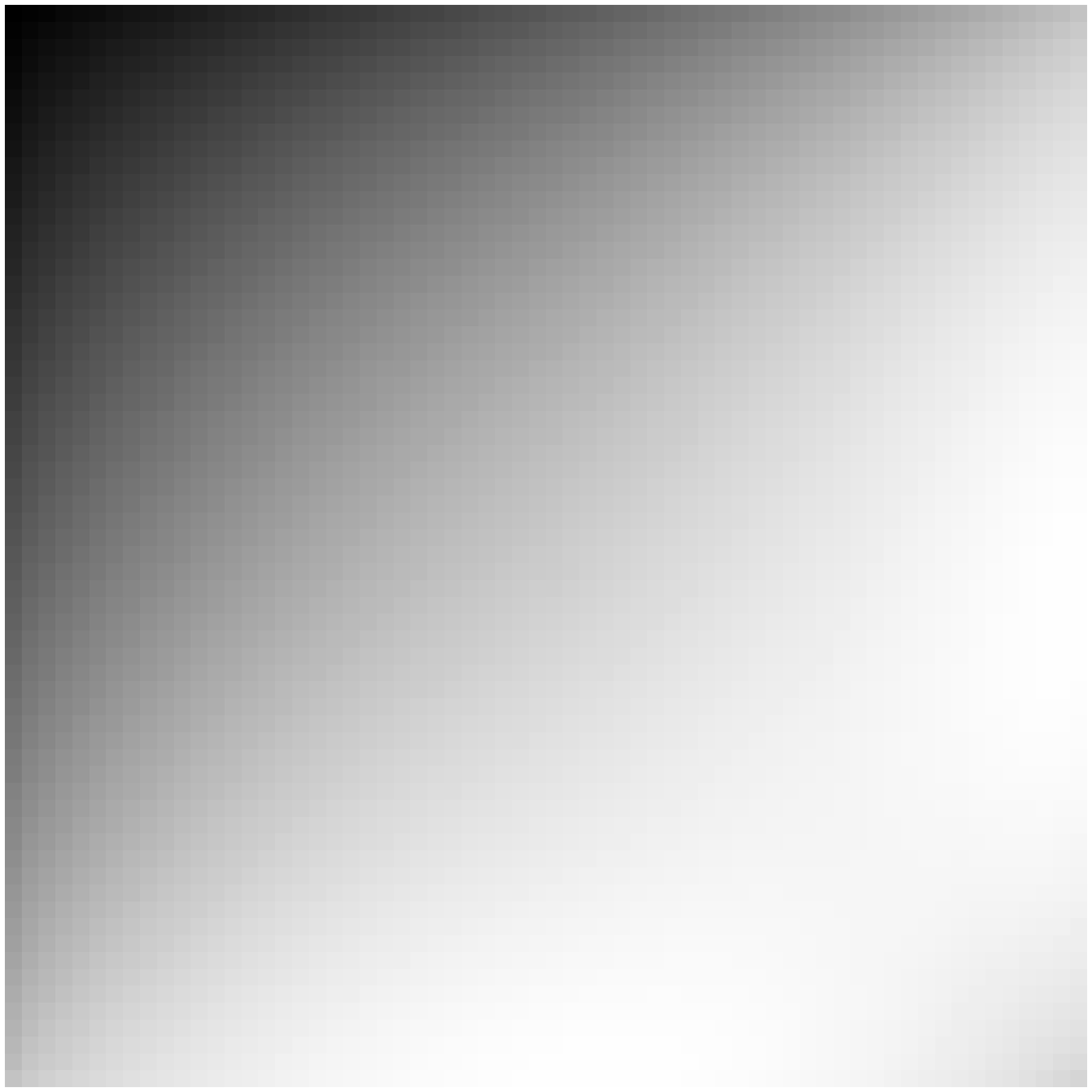}
& \epsfxsize=2.5cm \epsffile{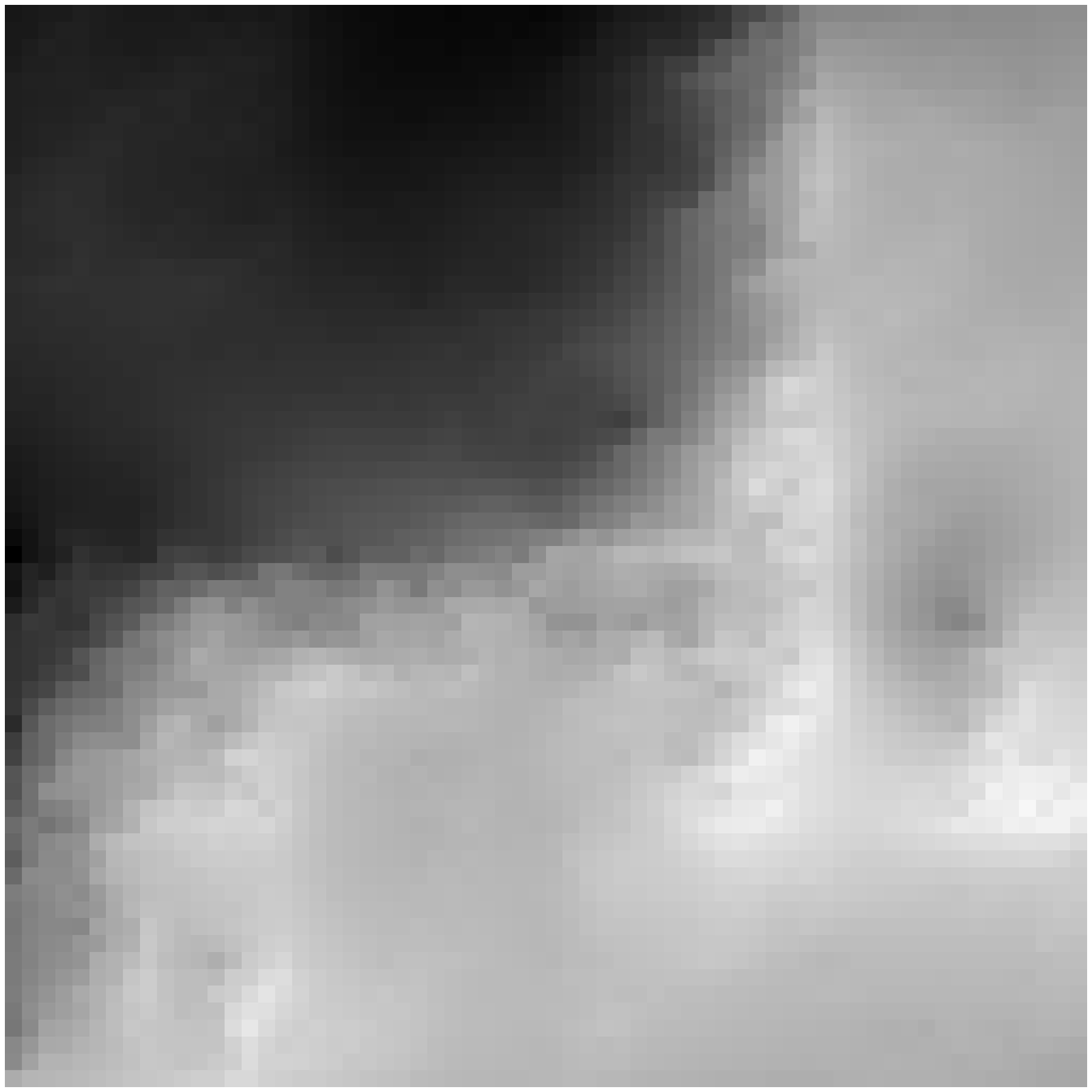}
& \epsfxsize=2.5cm \epsffile{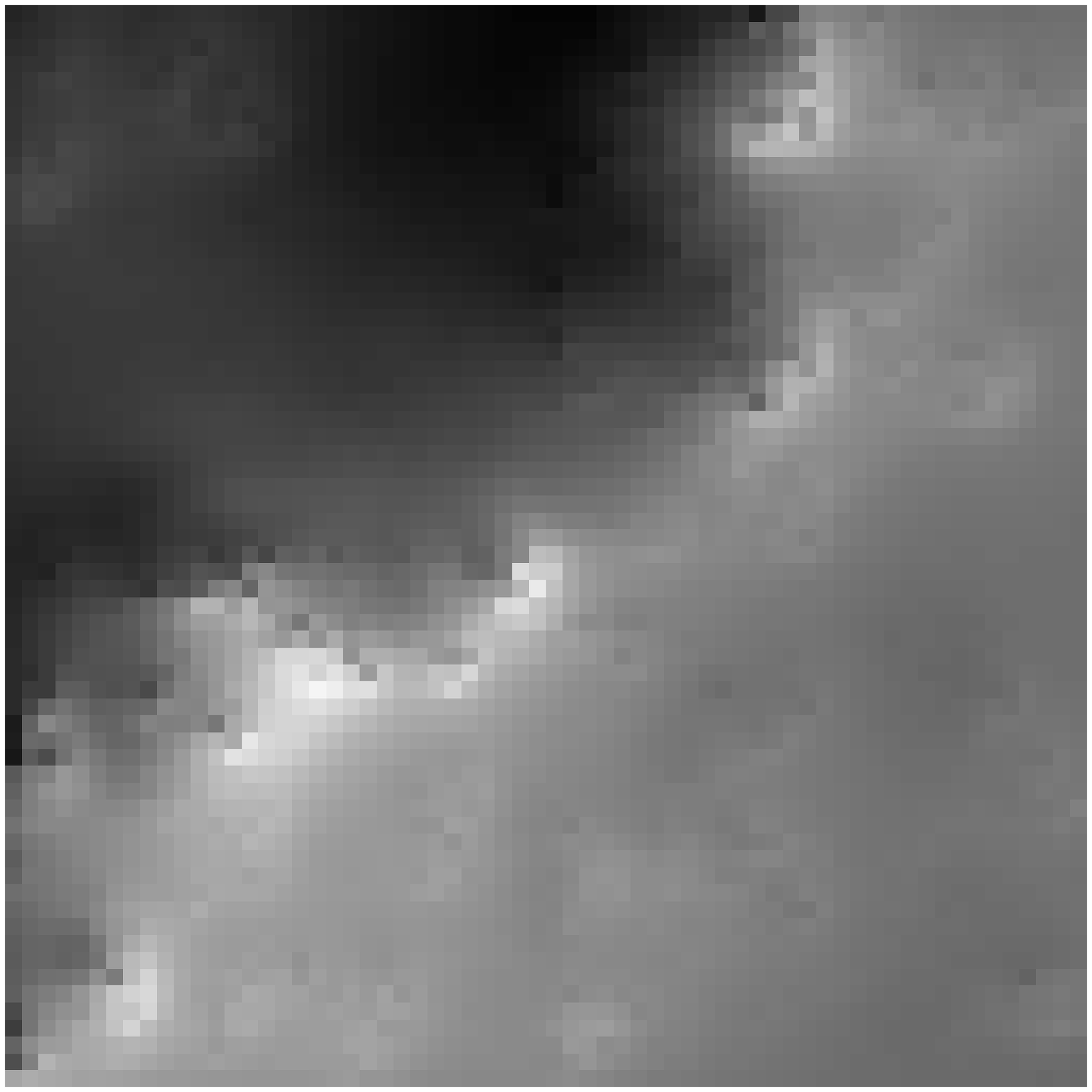}
& \epsfxsize=2.5cm \epsffile{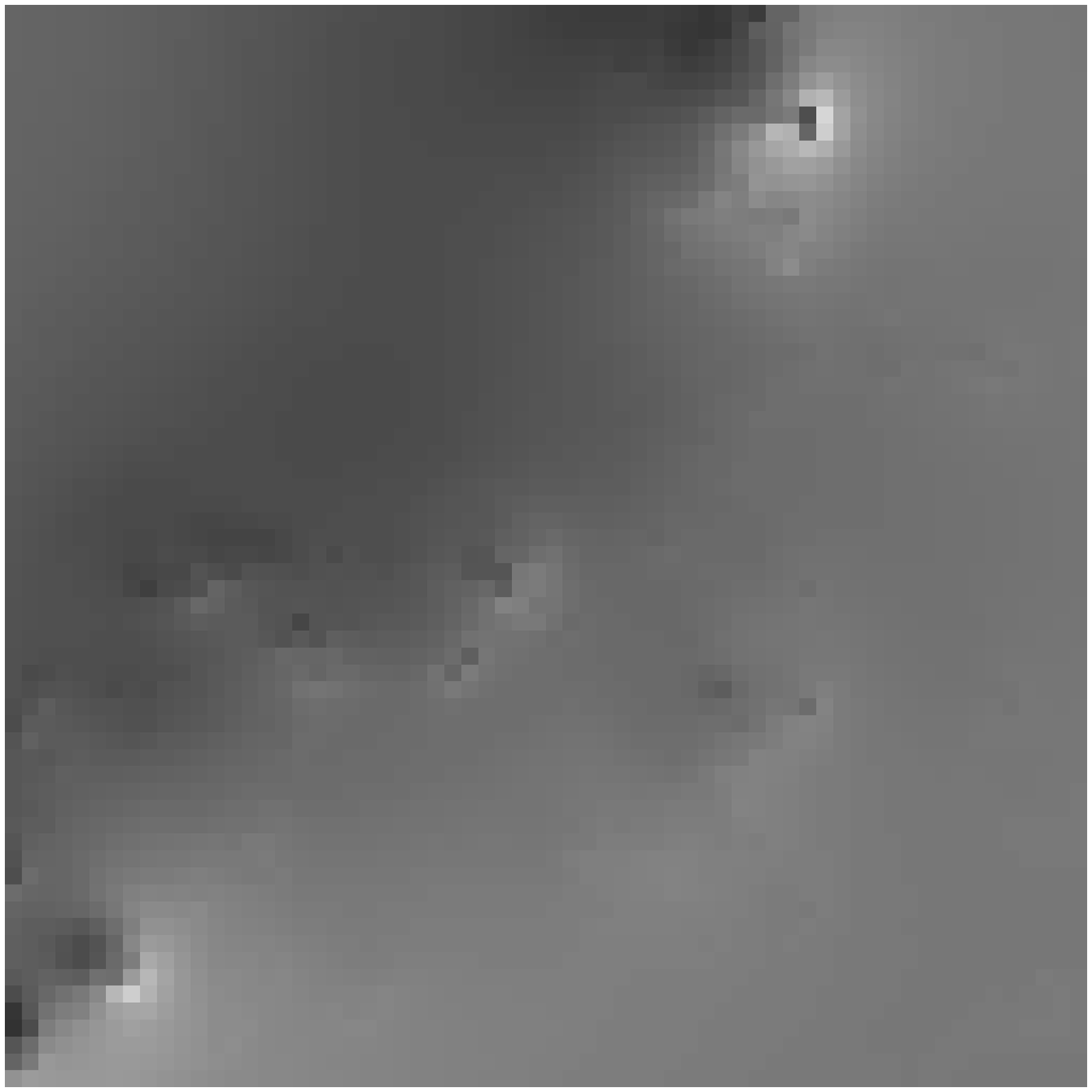}
& \epsfxsize=2.5cm \epsffile{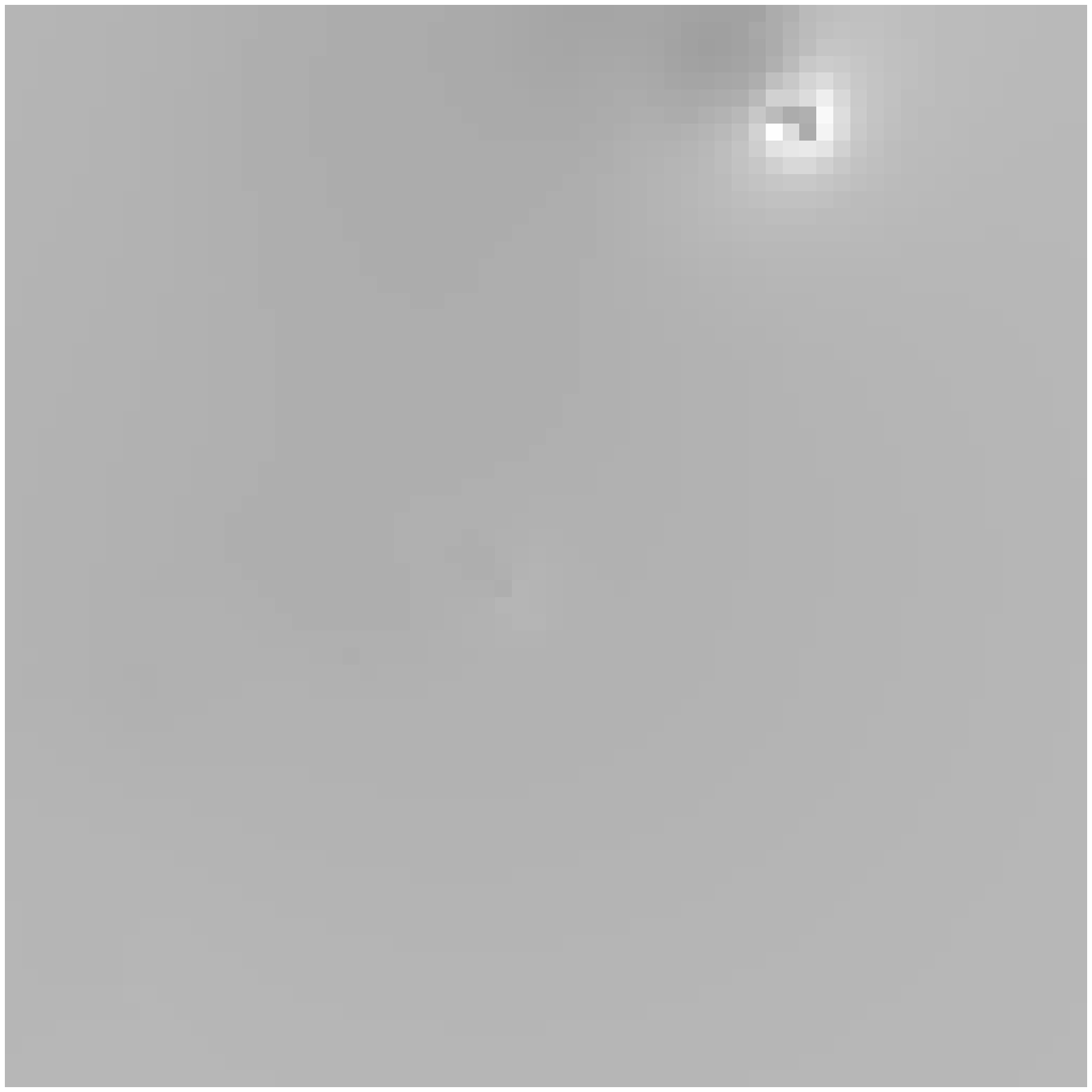}
\\&&&&& 
\\\hline
&&&&& \\
$\beta=0.4$ 
& \epsfxsize=2.5cm \epsffile{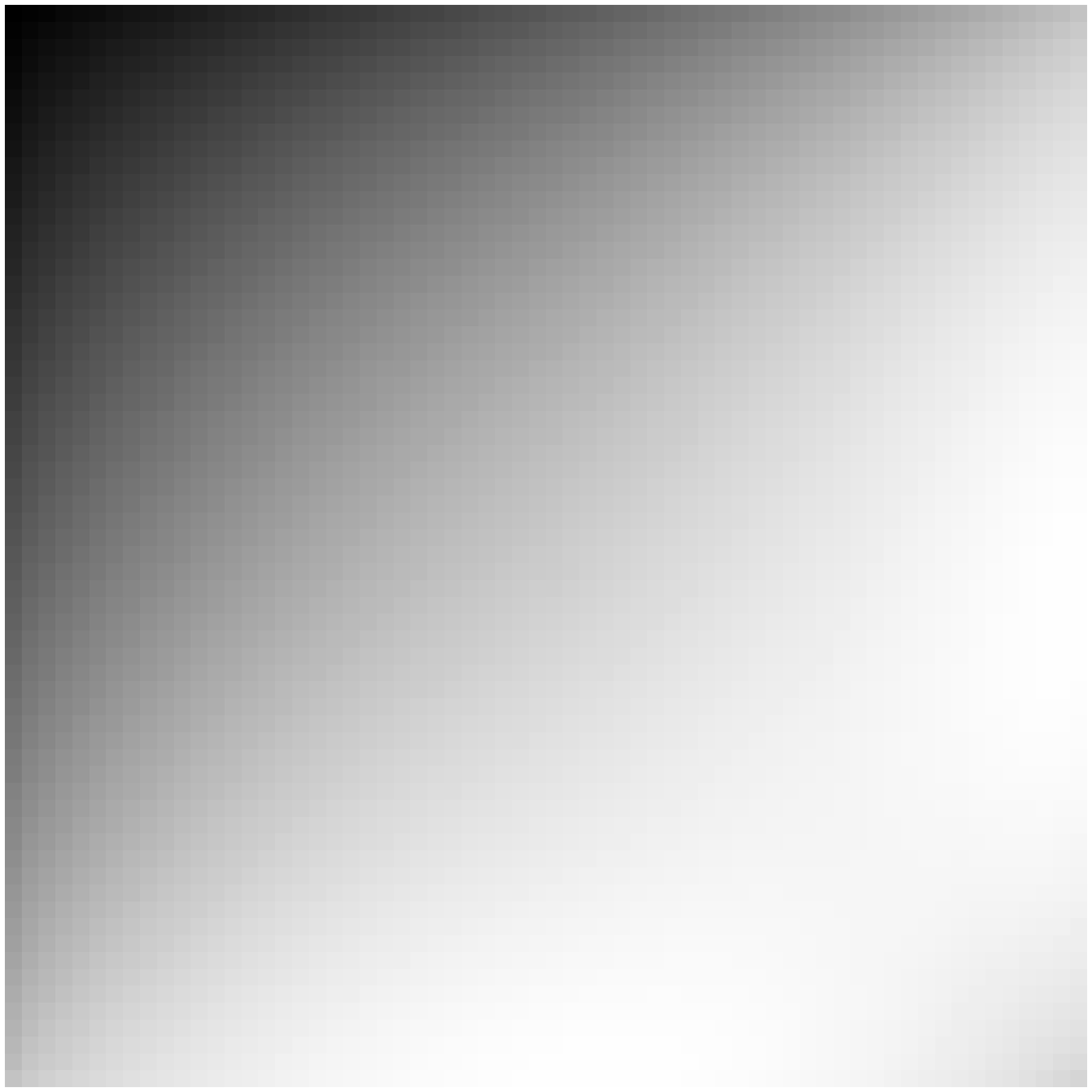}
& \epsfxsize=2.5cm \epsffile{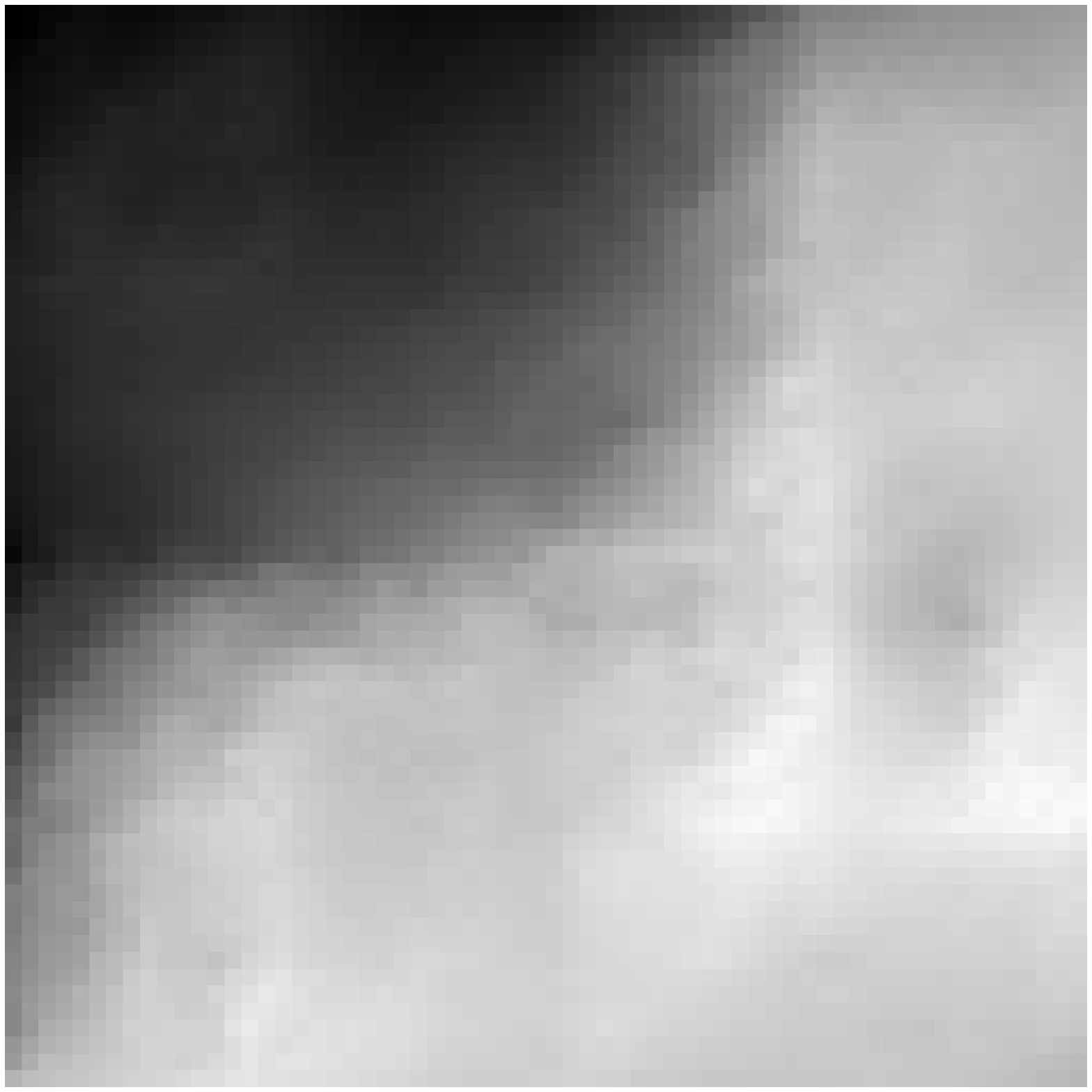}
& \epsfxsize=2.5cm \epsffile{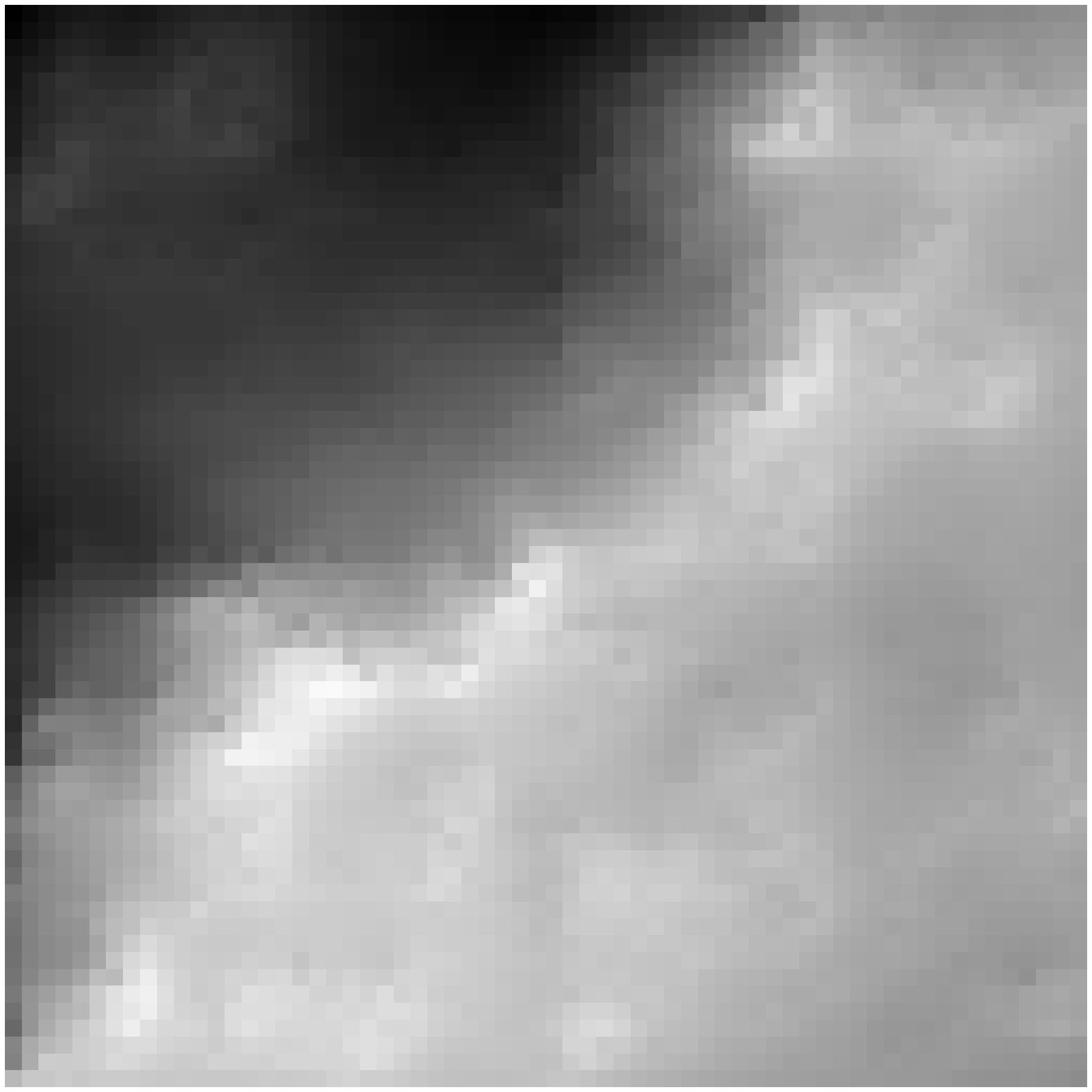}
& \epsfxsize=2.5cm \epsffile{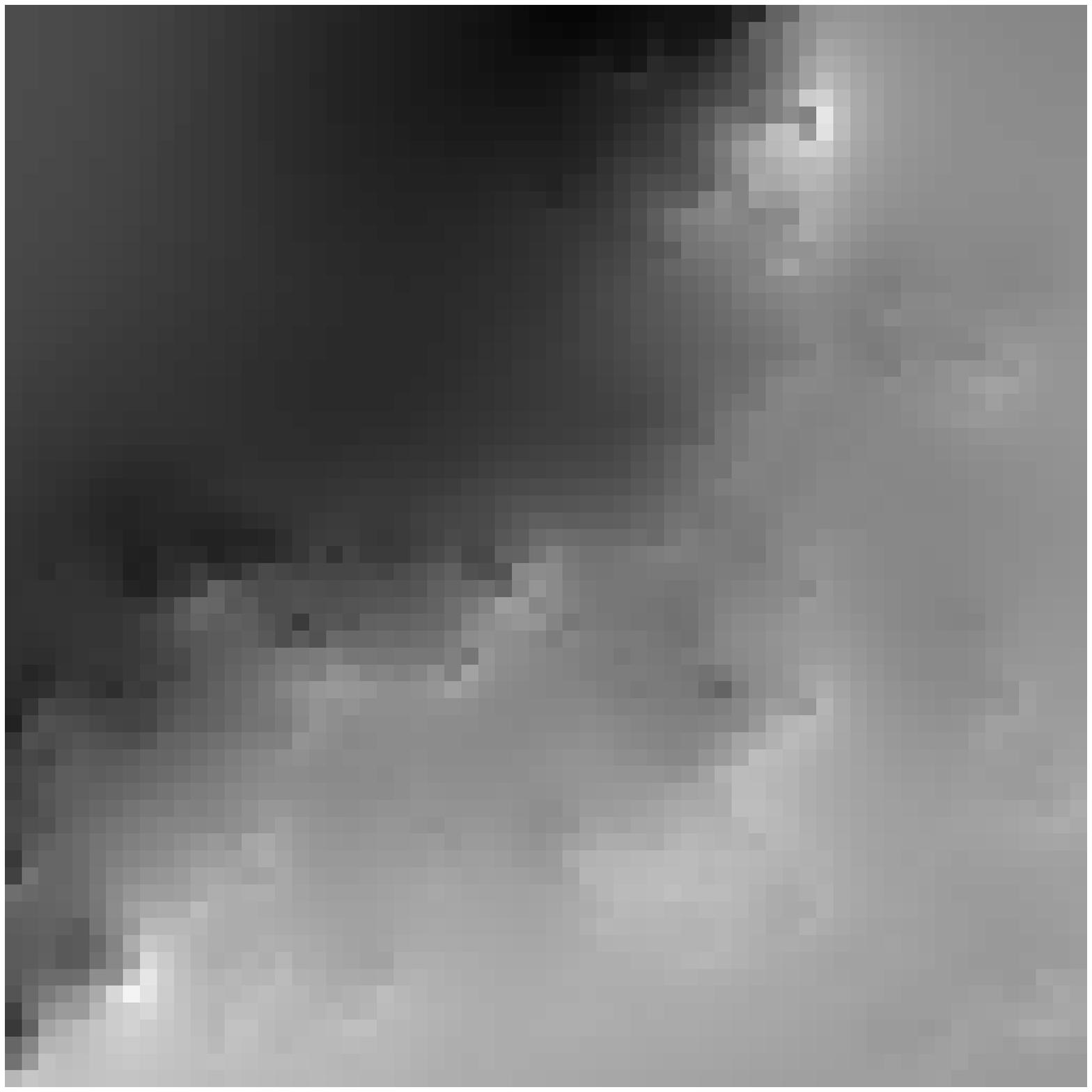}
& \epsfxsize=2.5cm \epsffile{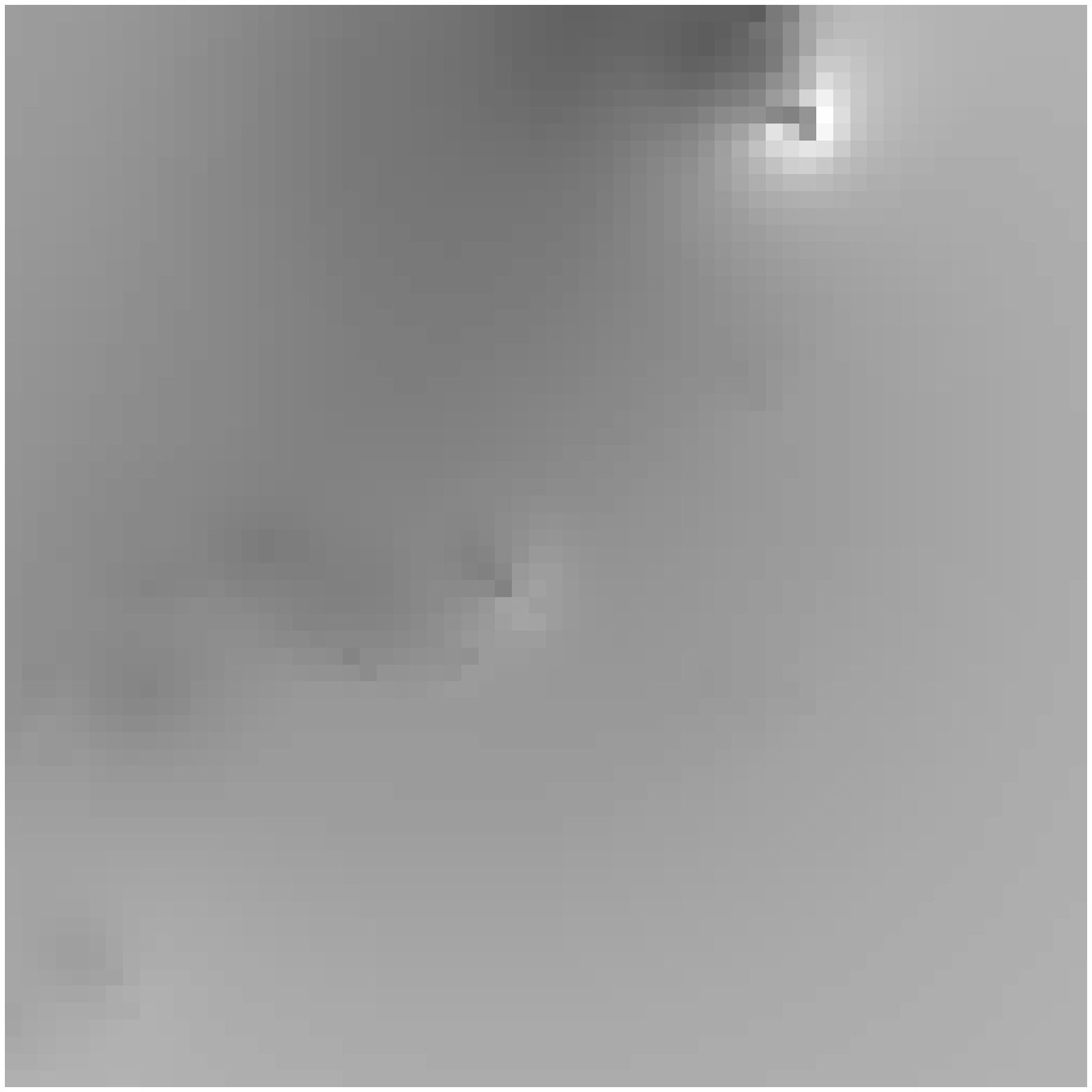}
\\&&&&& 
\\\hline
&&&&& \\
$\beta=0.6$ 
& \epsfxsize=2.5cm \epsffile{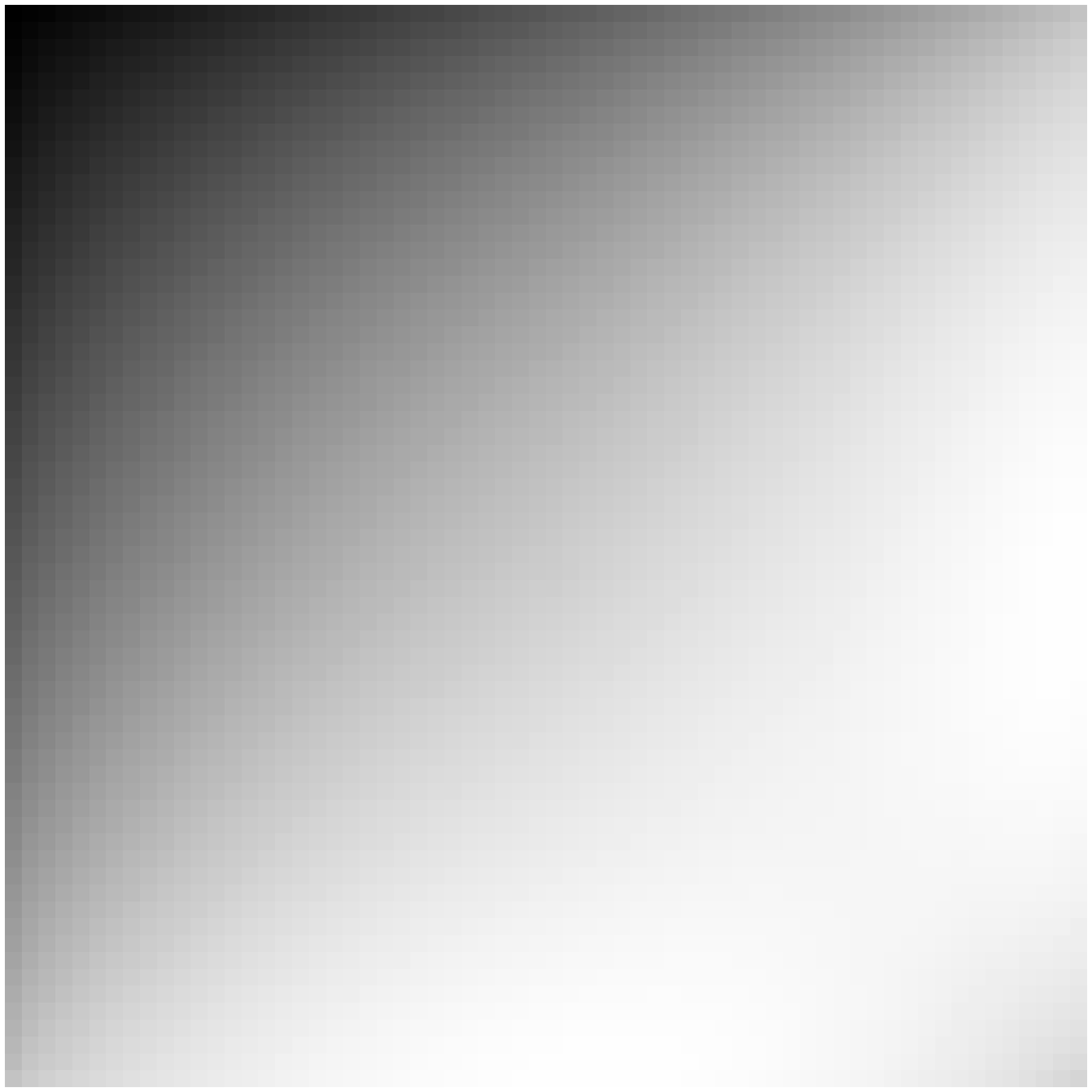}
& \epsfxsize=2.5cm \epsffile{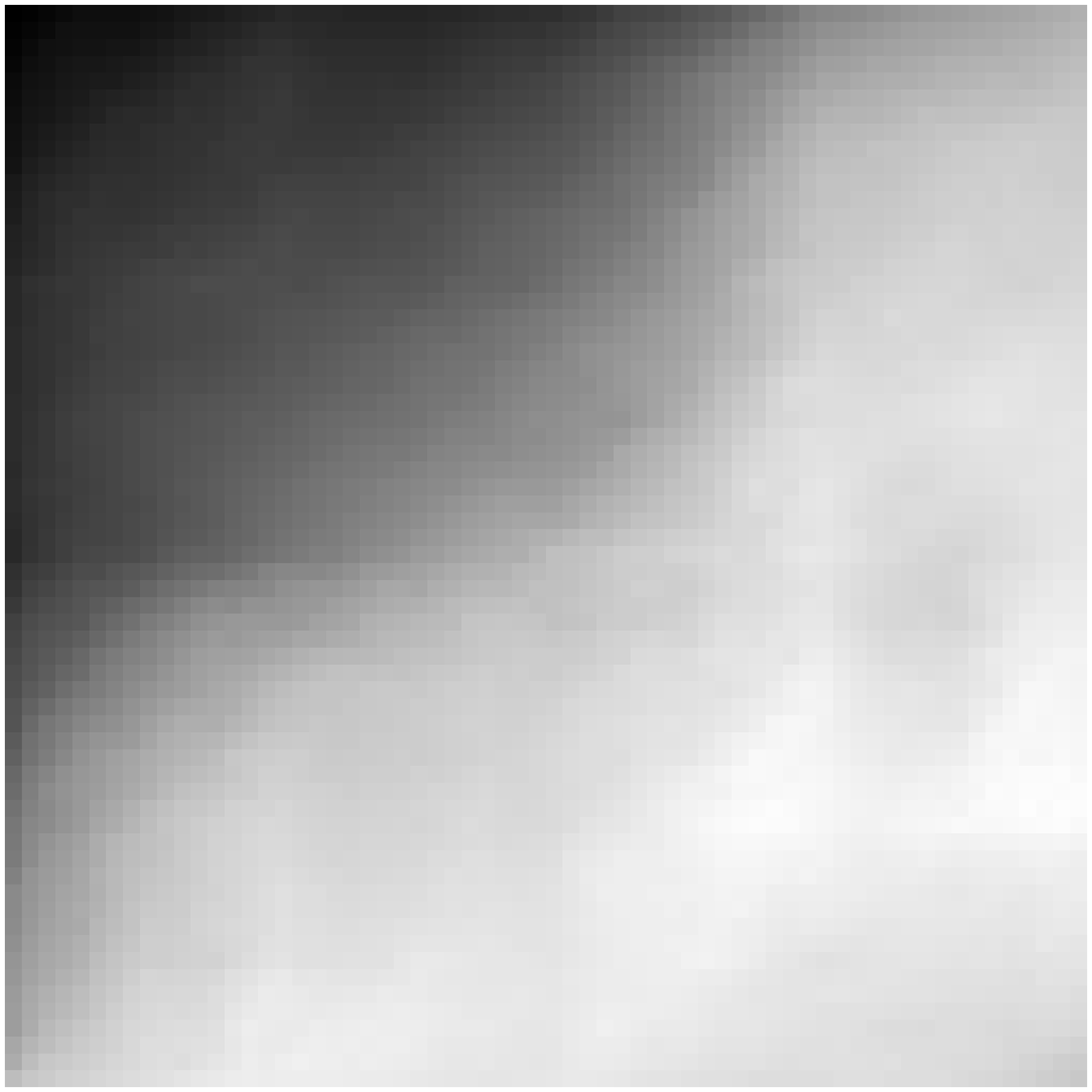}
& \epsfxsize=2.5cm \epsffile{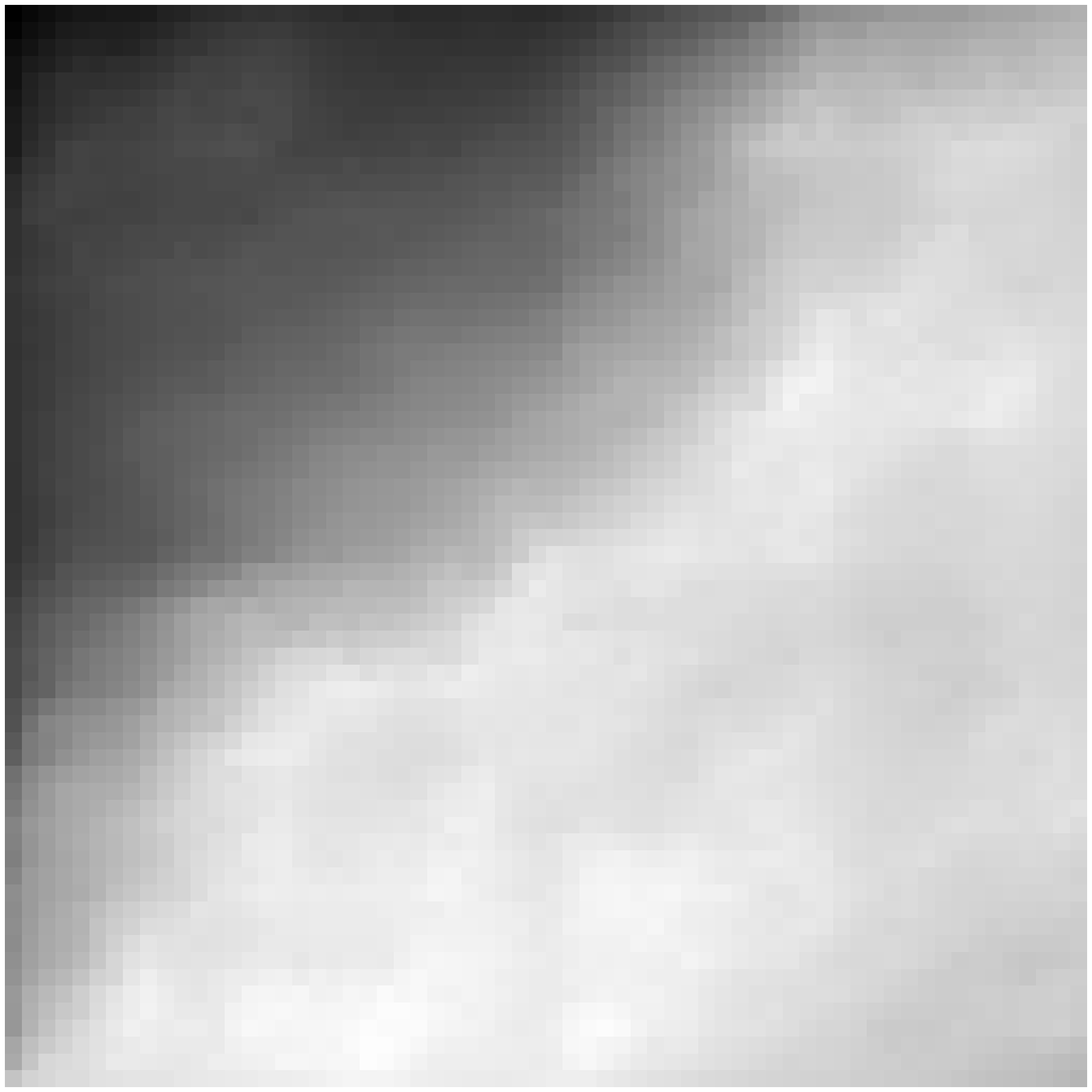}
& \epsfxsize=2.5cm \epsffile{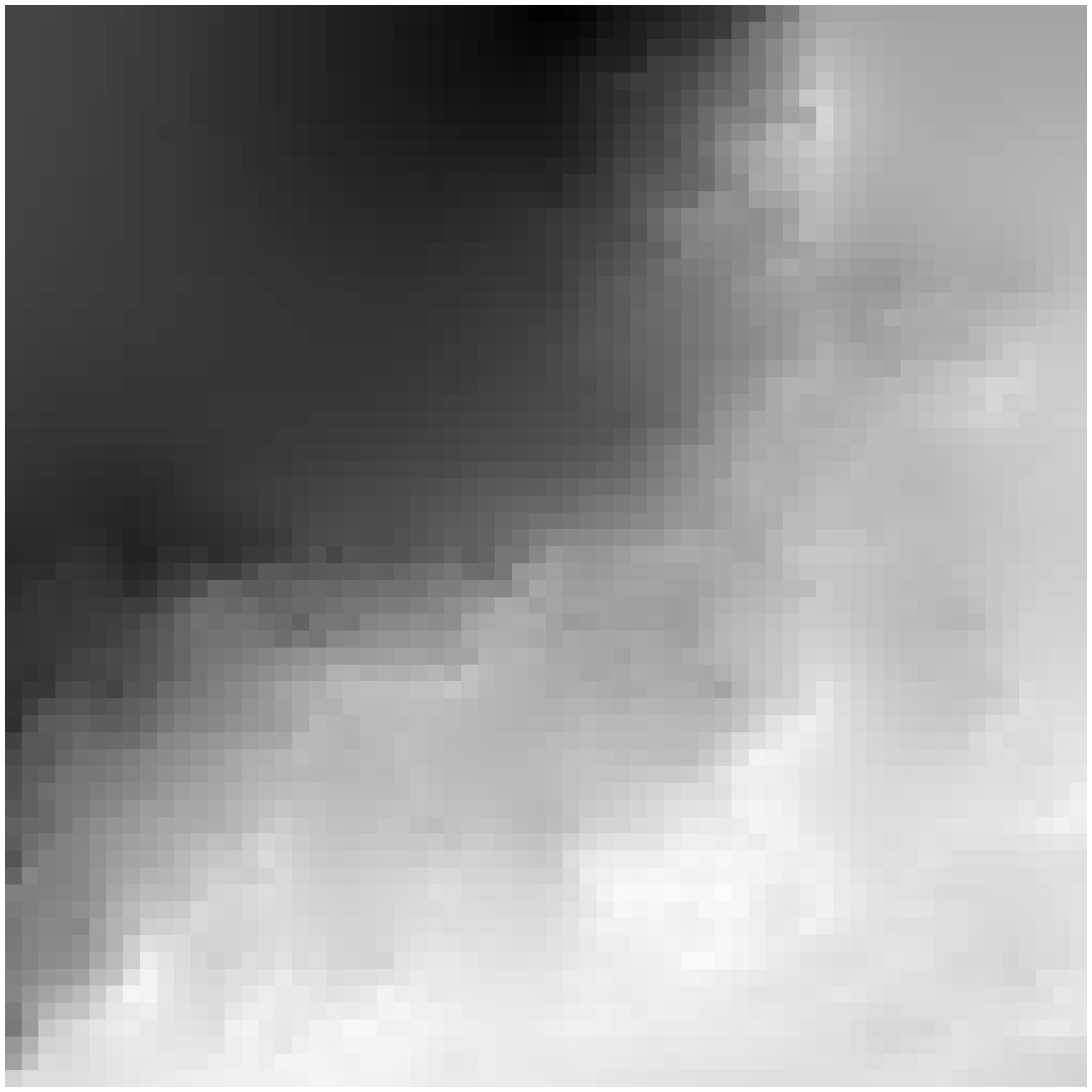}
& \epsfxsize=2.5cm \epsffile{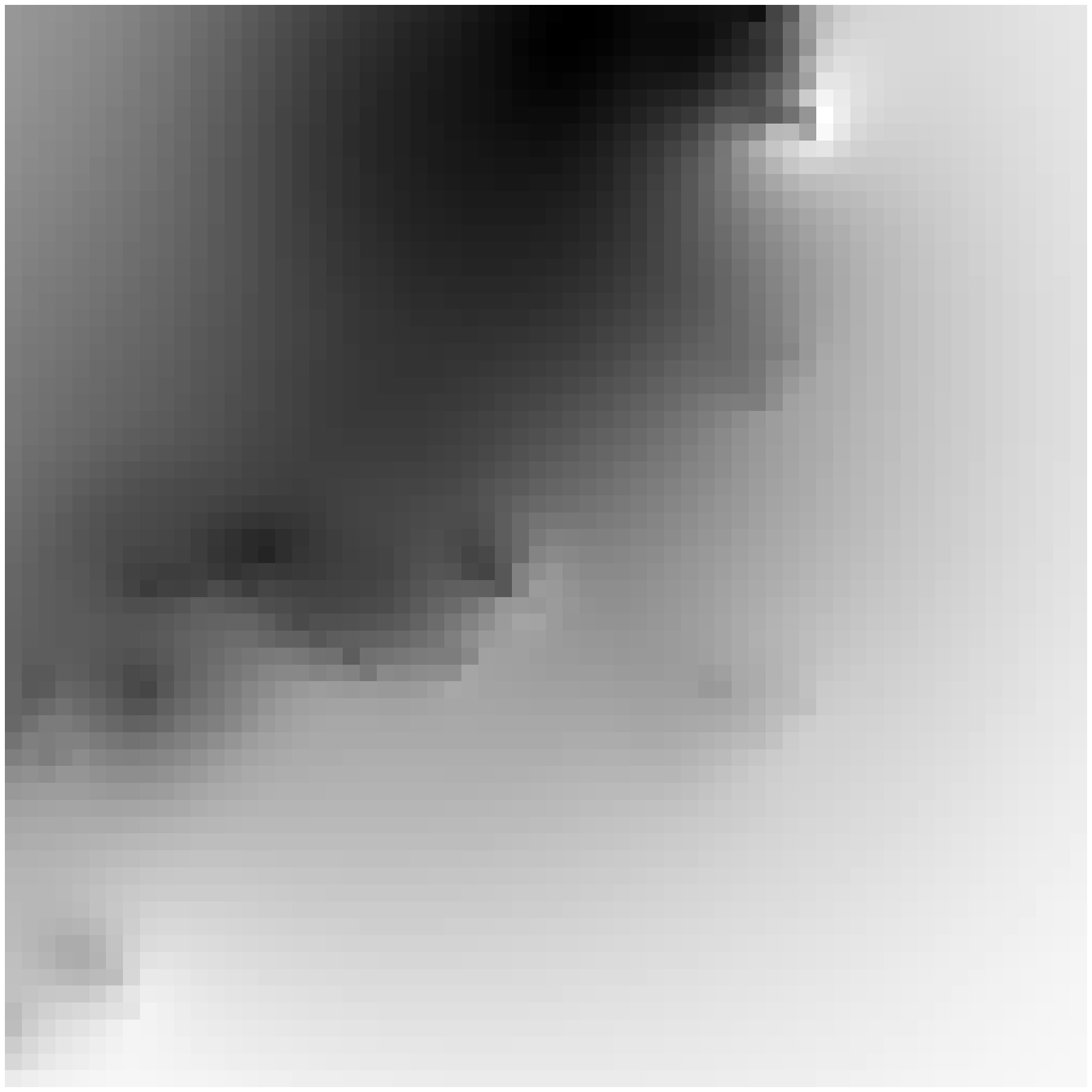}
\\&&&&& 
\\\hline
&&&&& \\
$\beta=0.8$ 
& \epsfxsize=2.5cm \epsffile{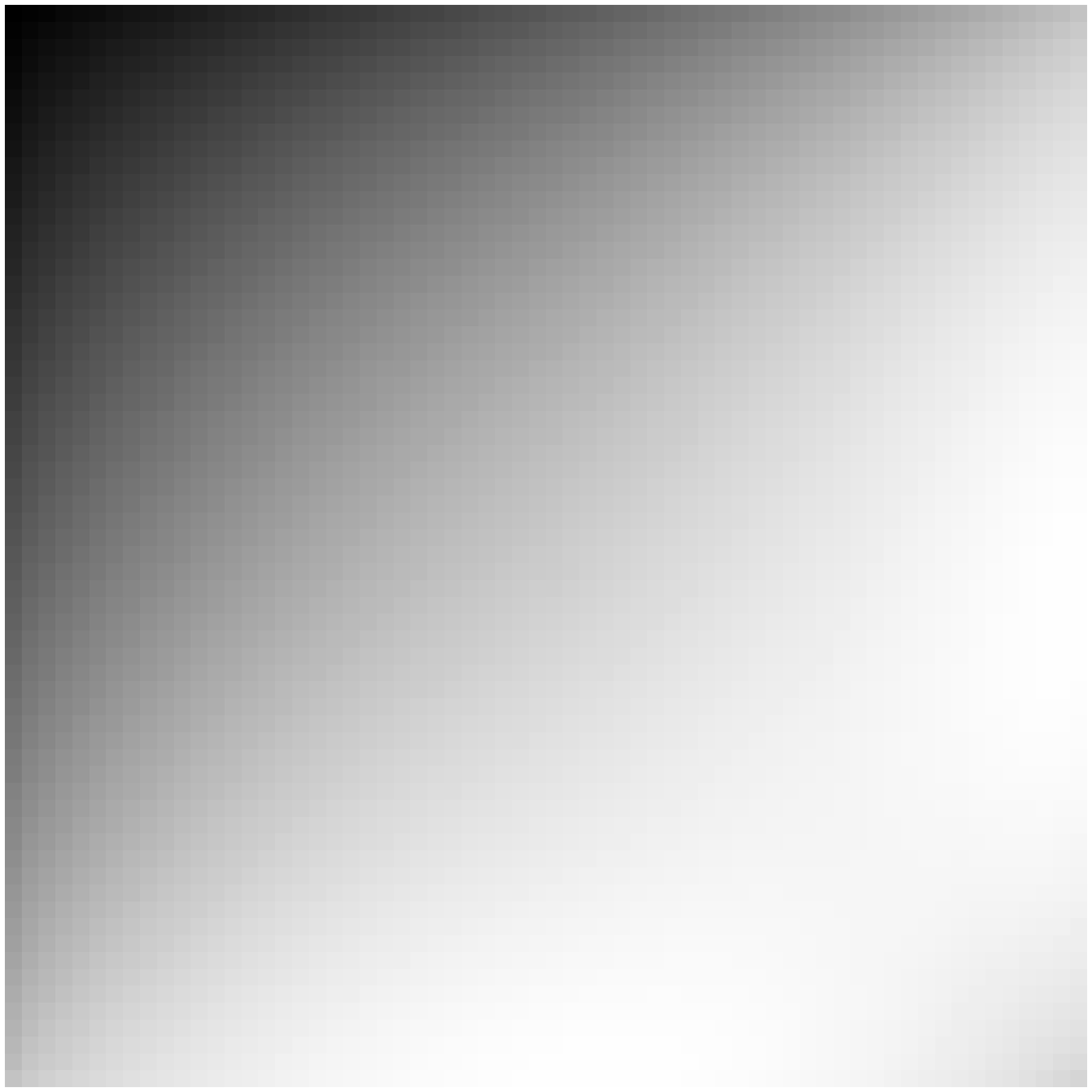}
& \epsfxsize=2.5cm \epsffile{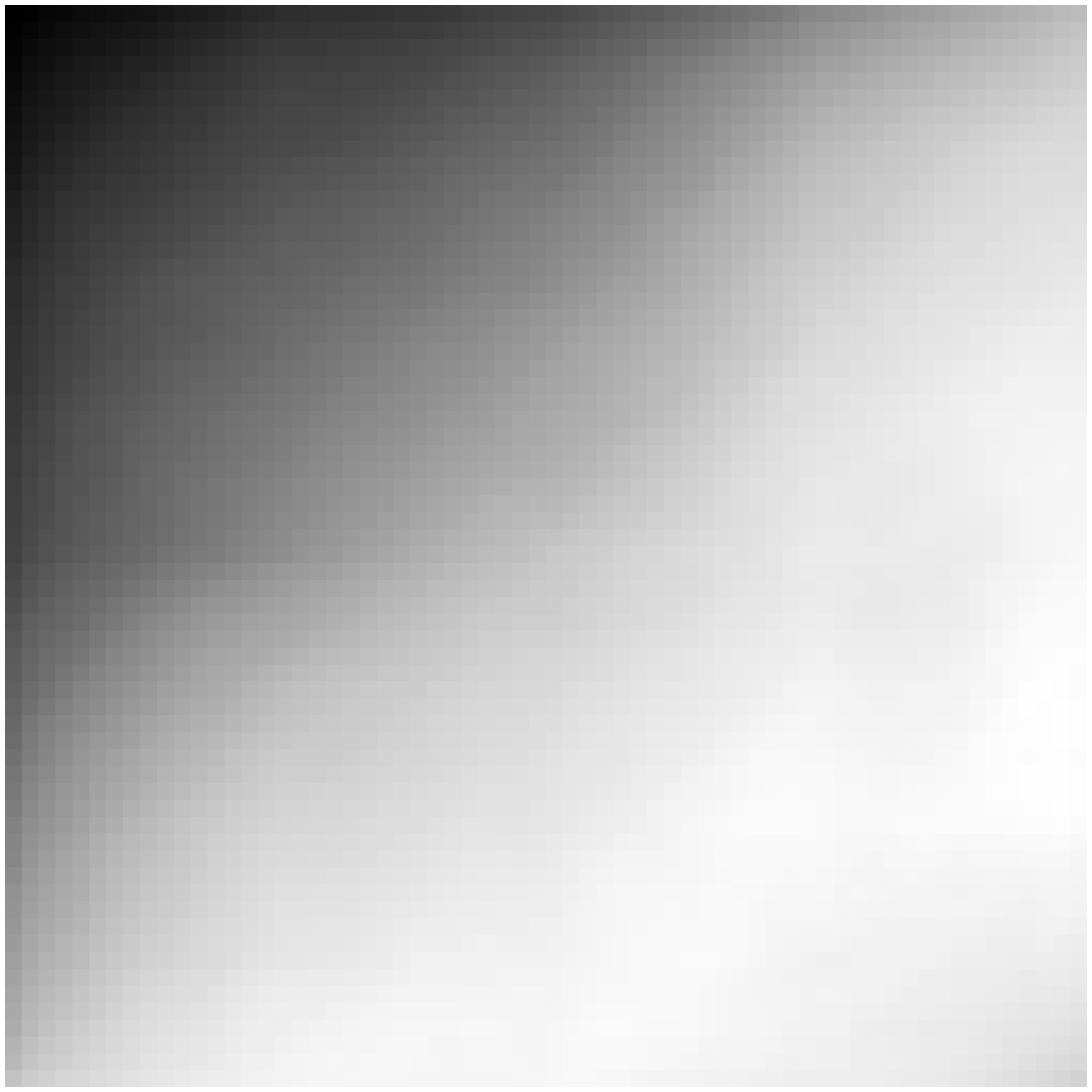}
& \epsfxsize=2.5cm \epsffile{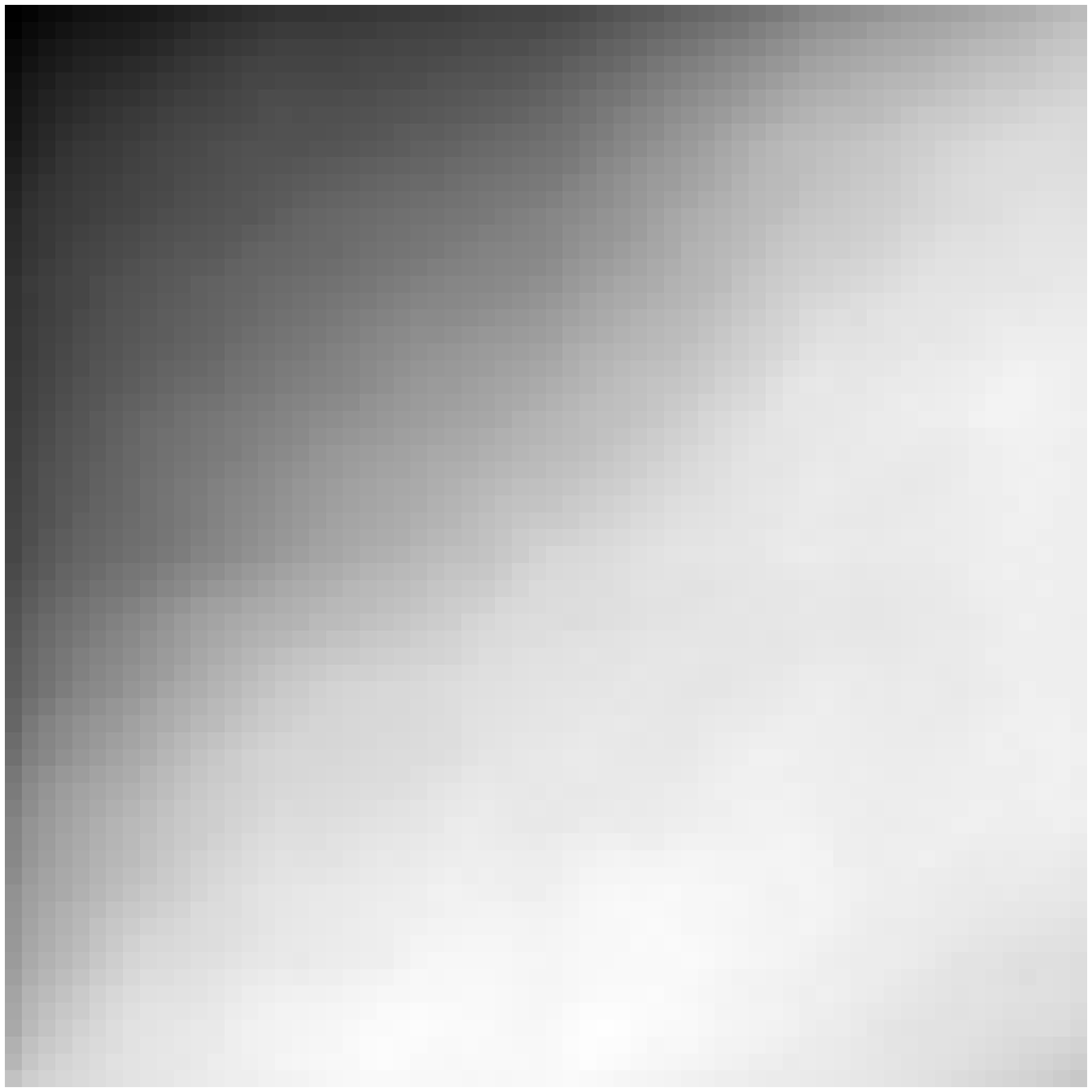}
& \epsfxsize=2.5cm \epsffile{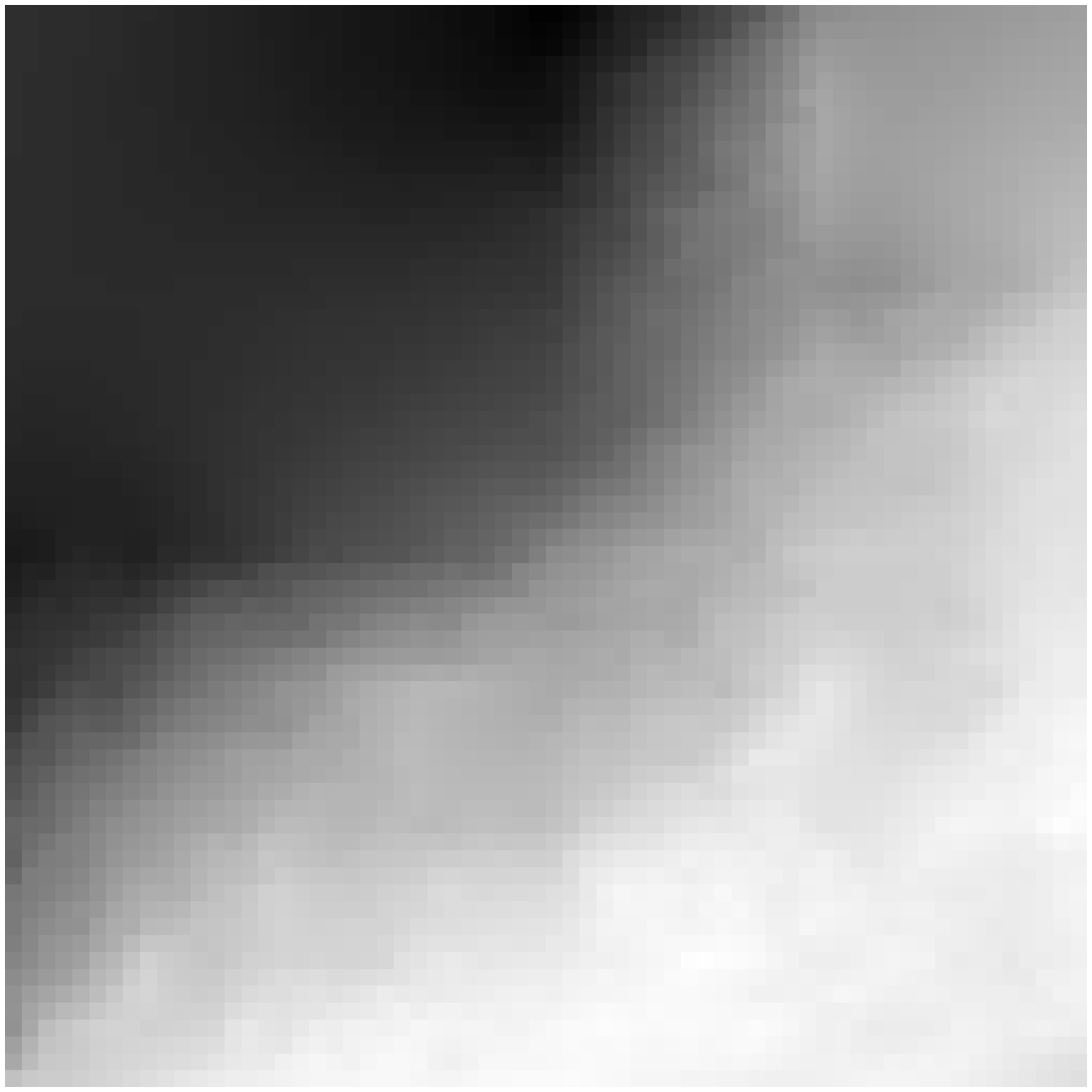}
& \epsfxsize=2.5cm \epsffile{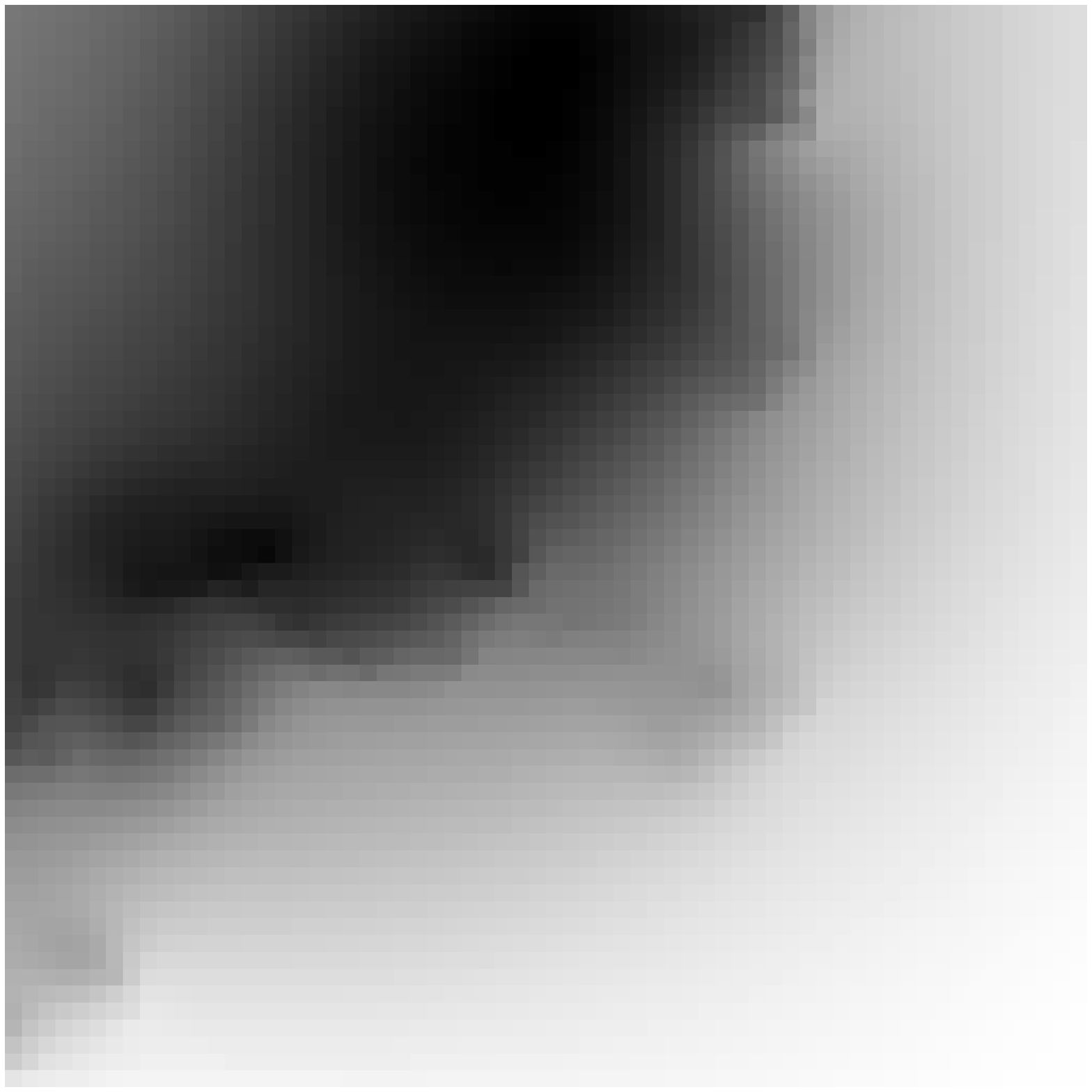}
\\&&&&& 
\\\hline
\end{tabular}
\end{center}
\caption{Catalogue of synthetic images. The case $\beta=1$ is not 
shown because the images are like those for $\bar s=0$ (see text)}
\label{fig:synthetic_catalogue}
\end{table}


\indent
The geometrical characteristics of the images are clear from 
visual inspection of the figures in Table
\ref{fig:synthetic_catalogue}. The images in the column $\bar s=0$ are
all the same, as the possibility of having even one modulation is
neglected. These images are also obtained as the limit
$\beta\rightarrow 1$ (this is because for $\beta=1$ the modulations
become irrelevant, as they do not change the value of $\gamma_{j
\vec{k}}$). These two cases should correspond to a uniform contrast
field (the non-uniformity in the borders is due to finite size
effects).  For non-trivial values of $\bar s$ and $\beta$ the
following properties are observed:

\begin{itemize}

\item
$D_{\infty}$ is directly related to $\bar{s}$ by
eq.(\ref{eq:beta_vs_delta}): $D_{\infty}=2-\bar{s}$. Therefore, for
fixed $\beta$, as $\bar{s}$ increases $D_{\infty}$ decreases, as can
be seen in Table \ref{fig:synthetic_catalogue} going from left to
right.

\item
As $\beta$ increases at fixed $\bar s$, $\Delta$ decreases. Visually
it can be appreciated that the singularity exponents decrease in
Table \ref{fig:synthetic_catalogue} from top to bottom.
\end{itemize}


\subsection{Numerical analysis of the generative model}
\label{subsection:generative_numerical}

\begin{figure}[hbt]
\begin{center}  
\hspace*{0.0cm}
\makebox[7cm]{\makebox[1cm]{$P(\beta)$}\makebox[6cm]{}}
\\
\vspace*{.5cm}
\makebox[7cm]{\makebox[1cm]{}
\makebox[6cm]{
\epsfysize=6cm
\epsfbox[50 50 410 302]{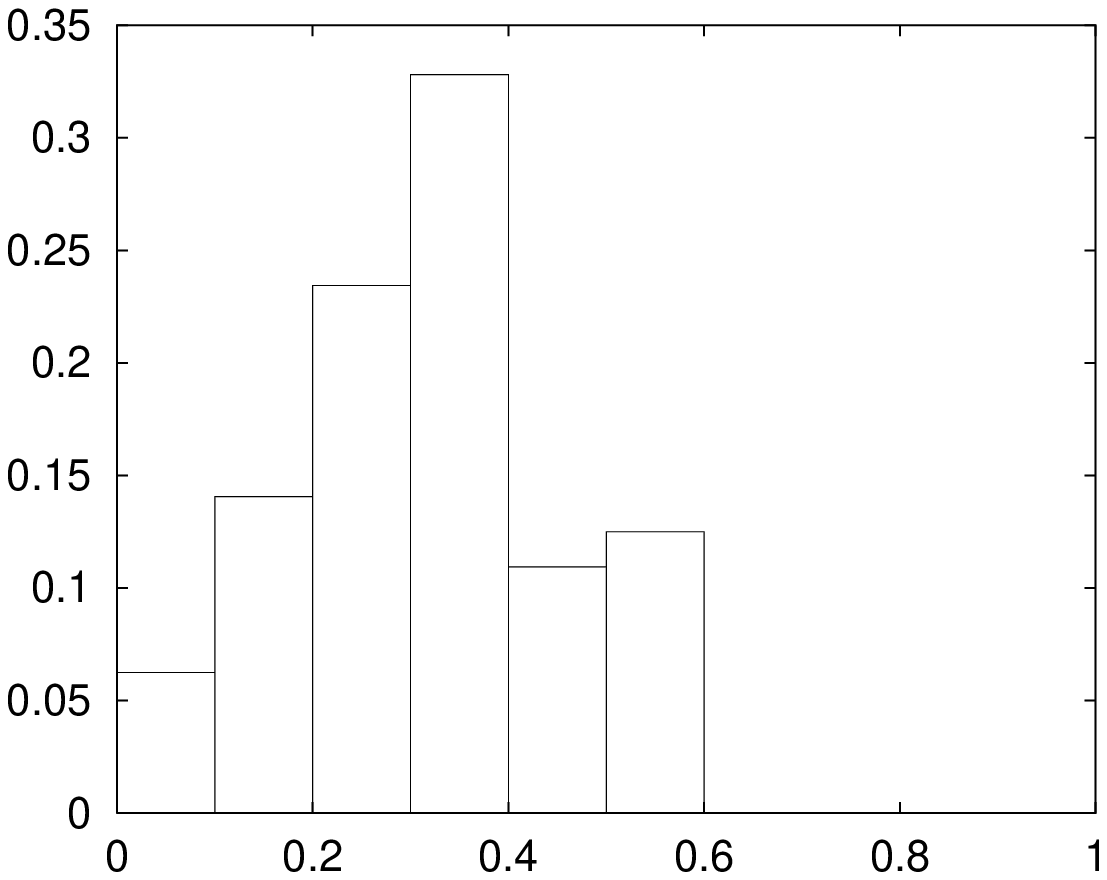}
}}
\\
\makebox[7cm]{\makebox[1cm]{}\makebox[6cm]{\hspace*{2cm}$\beta$}}  
\end{center} 
\caption{ Generative model.  Histogram of $\beta$ for the $32$ $256\times256$
synthetic images (and two directions) with a theoretical $\beta =0.40$}
\label{fig:generative_histogram}
\end{figure}

The artificial images are constructed to have a certain theoretical
$\beta$ and $\bar s$. We study here the properties of $32$ generated
images of $256 \times 256$ pixels with $\beta = 0.40$. The histogram
in Fig. \ref{fig:generative_histogram} shows the distribution of the
$64$ computed $\beta's$ for the horizontal and vertical direction of
each of the $32$ images. The mean of the histogram is $\beta=0.31$
with standard deviation $\sigma_{\beta}=0.13$. The fluctuations around
the mean and its bias with respect to the theoretical $\beta$ are
mainly due to the finite size of the images and not to the
size of the data set.  To verify this we have randomly divided
the data set in two sets of $12$ and $20$ images and have checked that
neither their means nor their standard deviations change appreciably.

\begin{figure}[hbt]
\hspace*{5cm}$\bf{a}$ \hspace*{6cm} $\bf{b}$ \\
\hspace*{0.5cm} ${\rho}(p, 2)$ \hspace*{5.5cm}$\bar{\rho'}(p, 2)$
\begin{center}
\leavevmode
\epsfysize=4cm
\epsfbox[50 61 388 291]{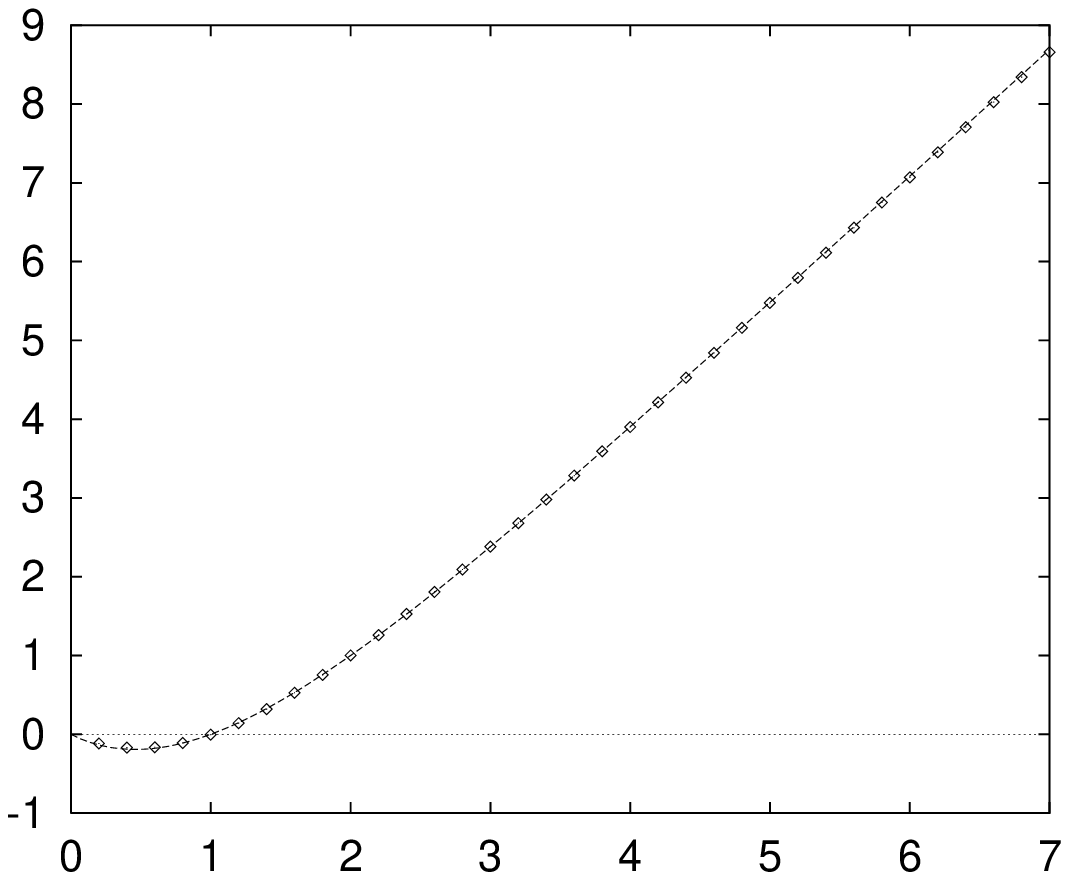}
\hspace*{1cm}
\epsfysize=4cm
\epsfbox[72 61 410 291]{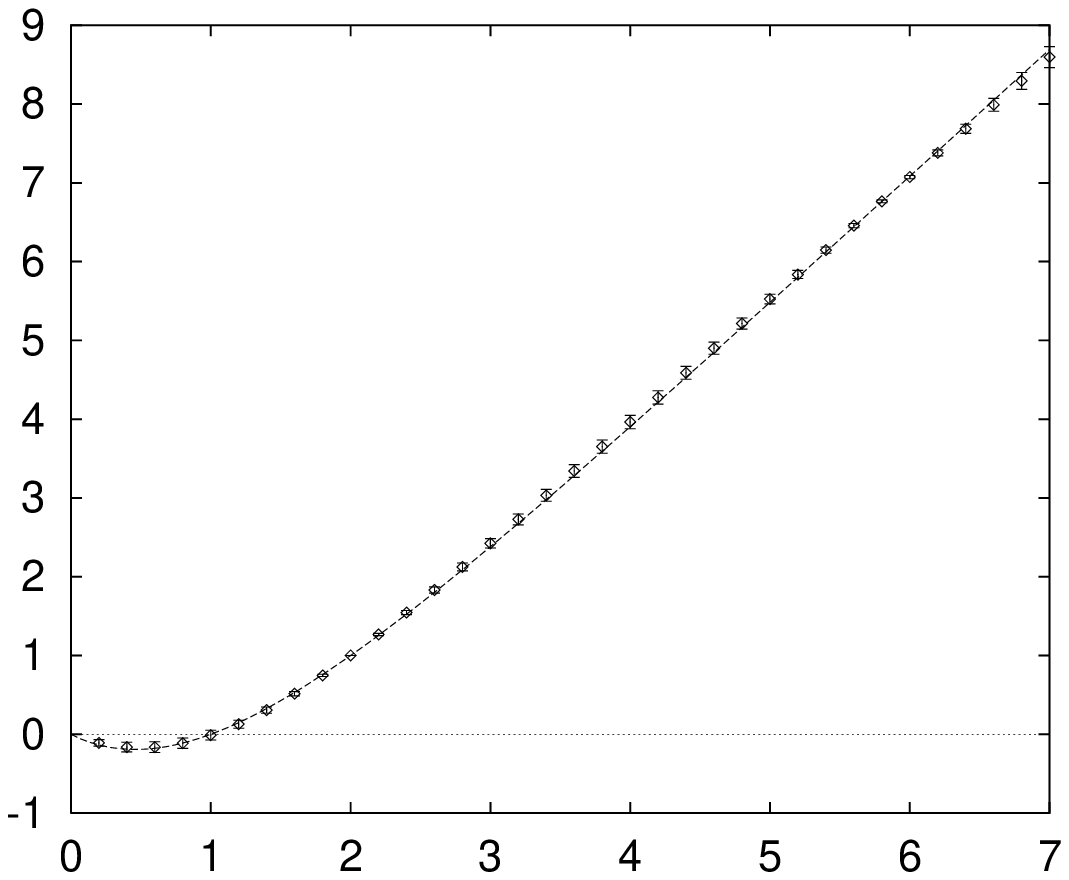}
\\
\hspace*{1cm}$p$
\hspace*{6.5cm}$p$
\\
\end{center}

\caption{ 
Test of the log-Poisson multiplicative process for the generative
model.  
a) $\rho(p,2)$ for the averaged moments of the whole ensemble
of $32$ $256\times256$ images. The dashed line corresponds to the ESS
exponents predicted by the log-Poisson model for $\beta = 0.38$.  
b) The mean of $\rho'(p, 2)$ over the $32$ synthetic images and the two
directions, where now $\rho'(p, 2)$ has been normalized to $\beta=0.38$.
The dashed line represents $\rho^{0.38}(p,2)$, the
log-Poisson $\rho(p,2)$ with $\beta=0.38$.  The error bars are of the
size of the data points.}
\label{fig:generative_rho}
\end{figure}

On the other hand, averaging the moments of the $64$ cases (which
approximates very well the moments computed over a single $2048
\times 2048$ image) yields  $\beta = 0.38$ (Fig.~\ref{fig:generative_rho}a), 
rather close to the theoretical $0.40$.

What is interesting is that even when these fluctuations in $\beta$
are due to finite size effects the multiplicative process for each
image is preserved. This can be seen in
Fig.~\ref{fig:generative_rho}b where the mean and variance of the
normalized ESS exponent $\rho'(p, 2)$ (averaged over the $64$
cases) is shown.

One can wonder whether the fluctuations in $\beta$ observed in the true
images could be thought of as finite size effects. This does not seem to
be the case. As we have seen in Sec.  \ref{subsection:results_whole} even
the whole set of images follows the log-Poisson model with a given
$\beta$. Considering this set as a single large image, each of its pieces
(i.e. the single images) should follow the log-Poisson model with
$\beta$'s distributed as in Fig.~\ref{fig:generative_histogram}. However,
if the fluctuations of the artificial images were a good model for the
fluctuations of the true ones, then the dispersion of the histogram in
Fig.~\ref{fig:generative_histogram} should be larger than that of the
histogram in Fig.~\ref{fig:400_imag_rho}III (because the synthetic images
are smaller than the true ones).



\section{Discussion}
\label{section:discussion}

\indent
One of the main conclusions of this work is that the multiscaling properties
of changes in contrast are robust in natural scenes: Both SS and ESS are 
present in almost every single image considered, which are of very different
type. In addition, a vast number of the images in our data set ($94\%$) can
be described in terms of a multiplicative process of the log-Poisson type.
The modulation parameter $\beta$ is distributed over its admissible range,
although the probability distribution decreases with increasing $\beta$.

\indent
Four different ensembles of images (described in
Sec.~\ref{section:environments}) were considered in order to
understand how the presence or absence of certain features affects the
edge statistics. The moments of the relevant variables $\epsilon_r$
were then computed by averaging over all the images in the ensemble. It
was found that SS, ESS and the same multiplicative process are also
present in each of the four ensembles and for both vertical and
horizontal statistics.  Again, these eight cases differ only in the
value of $\beta$.  

\indent
The same properties (SS, ESS and log-Poisson) hold when the whole data
set is taken as the ensemble. We can conclude from our experimental
study of the three categories of ensembles that SS, ESS and the
log-Poisson model are robust features of natural images. It is very
remarkable that ensembles following log-Poisson laws with different
values of the parameter $\beta$ can be combined together giving again
a log-Poisson model.

\indent
One point that we want to emphasize is that the existence of a
multiplicative process of a given type (the log-Poisson model) is very
robust. It is present in images with rather different aspect, what
supports the expectation that it has actually been detected by the
early visual system in the course of evolution.  If so, it should be
contained in the structure of the receptive fields. On the other hand,
there is a high variability in the value of the model parameter
$\beta$. Then, when the image ensemble changes, the cell should adapt
to the new value of $\beta$. This is not an unusual situation in the
visual system: it adapts, for instance, to the average contrast
\cite{AhAlDoMa97,CaFe97}.


\indent
The generative model studied in this paper provides a simple way to
produce images with the multiscaling properties of natural scenes. It
could be used as a better description of the contrast distribution that
contains not only the correct second order statistics but also a
correct description of the non-gaussian aspects of changes in
contrast. The model could then be used to study the consequences that
these properties have on the early stages of the visual system
\cite{TuPa99}.

\indent
The model is not based on objects. A quick glance to
Table~\ref{fig:synthetic_catalogue} shows cloudy structures that have
little to do with natural images (except e.g. those of the sky).  In
particular their appearance is rather fractionary, lacking real
contours.  However the model could provide an indication of which
aspects of the visual processing depend only on the multiscaling
properties and which ones are more dependent on spatial correlations.

\indent
By analyzing the finite size effects of synthetic images, we have
concluded that the value of $\beta$ is an essential feature of each
natural image, which cannot be regarded as a fluctuation from a
universal value. The generative model was also used to illustrate the
meaning of the parameters $\beta$ and $\bar{s}$.

\indent
The wavelet used in the generative model studied in this work was
somewhat arbitrary. A better way to represent the multiscaling
properties of natural images is to derive a wavelet basis from the
image data set itself. When this is done, it is found that the
coefficients of the wavelet expansion are scale-independent and weakly
space-dependent, thus providing a quasi-optimal code for the image
\cite{TuPa99}.  Their distribution for a given image is log-Poisson
(eq.~(\ref{eq:Log-Poisson})), with exactly the same parameter $\beta$
that would be obtained from the statistics of $\epsilon_r$.  This is a
remarkable result, and supports our proposal that the visual system
has adapted to the multiscaling properties of natural scenes and that
its architecture contains, in particular, information about the
log-Poisson multiplicative process.

\indent
Further equalization of these quasi-independent coefficients that
follow the distribution given in eq.~(\ref{eq:Log-Poisson}) could be
an advantageous strategy for the visual system since it optimizes the
mutual information between the source (visual stimuli) and the
internal representation \cite{NaPa94,NaBuPa98}. Since the probability
distributions of the quasi-independent coefficients depend on $\beta$
this proposed equalization depends on the image being processed.

\indent
To conclude, in this work we have established that the multiscaling
properties of changes in contrast are a robust feature in natural
images.  The existence of this statistical regularity for single
images has several implications for visual processing.  This result is
relevant because it implies a reduction of the entropy of the
distribution probability of the ensemble of natural images. As a new
statistical regularity is found, the set of images that can be
considered as natural becomes more restricted. The fact that this
property is present not only in ensembles but also in single images
greatly reduces the entropy of the ensemble of natural
images. Therefore its existence should be taken into account when
modeling visual processing in the brain.

\section*{Acknowledgments}

We are thankful to Angela del Pozo for her help during the initial
stages of the preparation of this work.  We are grateful to Hans van
Hateren, some of whose images we used for the statistical
analysis.  Antonio Turiel is financially supported by an FPI grant from
the Comunidad Aut{\'o}noma de Madrid, Spain.  This work was funded by
a Spanish grant PB96-0047.

\end{document}